\documentclass[12pt,preprint]{aastex}
\shorttitle{Wolf-Rayet Stars in M31}

\shortauthors{Neugent et al.}

\begin{document}

\title{The Wolf-Rayet Content of M31\altaffilmark{1}}

\author{Kathryn F.\ Neugent and Philip Massey\altaffilmark{2}}
\affil{Lowell Observatory, 1400 W Mars Hill Road, Flagstaff, AZ 86001;\\ kneugent@lowell.edu; phil.massey@lowell.edu}

\author{Cyril Georgy}
\affil{Centre de Recherche Astrophysique de Lyon, \'{E}cole Normale Sup\'{e}rieure de Lyon,\\ 46 all\'{e}e d`Italie, 69384, Lyon Cedex 07, France; Cyril.Georgy@ens-lyon.fr}

\altaffiltext{1}{The spectroscopic observations reported here were obtained at the MMT Observatory, a joint facility of the University of Arizona and the Smithsonian Institution. MMT telescope time was granted by NOAO, through the Telescope System Instrumentation Program (TSIP). TSIP is funded by the National Science Foundation. This paper uses data products produced by the OIR Telescope Data Center, supported by the Smithsonian Astrophysical Observatory.}
\altaffiltext{2}{Visiting Astronomer, Kitt Peak National Observatory, National Optical Astronomy Observatory, which is operated by the Association of Universities for Research in Astronomy, Inc.\ (AURA) under cooperative agreement with the National Science Foundation.}

\begin{abstract}
Wolf-Rayet stars are evolved massive stars, and the relative number of WC-type and WN-type WRs should vary with the metallicity of the host galaxy, providing a sensitive test of stellar evolutionary theory. However, past studies of the WR content of M31 have been biased towards detecting WC stars, as their emission line signatures are much stronger than those of WNs. Here we present the results of a survey covering all of M31's optical disk (2.2 deg$^2$), with sufficient sensitivity to detect the weaker-lined WN-types. We identify 107 newly found WR stars, mostly of WN-type. This brings the total number of spectroscopically confirmed WRs in M31 to 154, a number we argue is complete to $\sim95\%$, except in regions of unusually high reddening. This number is consistent with what we expect from the integrated H$\alpha$ luminosity compared to that of M33. The majority of these WRs formed in OB associations around the Population I ring, although 5\% are truly isolated. Both the relative number of WC to WN-type stars as well as the WC subtype distribution suggest that most WRs exist in environments with higher-than-solar metallicities, which is consistent with studies of M31's metallicity. Although the WC-to-WN ratio we find for M31 is much lower than that found by previous studies, it is still higher than what the Geneva evolutionary models predict. This may suggest that Roche-lobe overflow produces the excess of WC stars observed at high metallicity, or that the assumed rotational velocities in the models are too high.
\end{abstract}


\section{Introduction}
The evolution of massive stars can be likened to peeling an onion layer by layer, with strong stellar winds doing the work. When the hydrogen-rich outer layers are gone, a Wolf-Rayet (WR) star is formed.  The type of WR star (WN - nitrogen rich, or WC - carbon rich) then depends upon which layer is visible. Stellar winds first peel away enough material to reveal the star's H-burning products, helium and nitrogen (WN stage) before peeling away more to reveal the He-burning products, carbon and oxygen (WC stage). These winds also create the WR's characteristically broad emission lines (Cassinelli \& Hartmann 1975, Conti \& de Loore 1979). As the ``Conti scenario" describes, the amount of mass lost by a massive star dictates whether a star will evolve only to the WN stage, or whether it will evolve first to a WN and then to a WC (Conti 1976; Maeder \& Conti 1994). Since this mass-loss is driven by radiation pressure on highly ionized metal atoms, a massive star born in a higher metallicity\footnote{In this paper we use ``metallicity" to mean the oxygen abundance, as tabulated by Massey (2003) and references therein. This is well justified, as oxygen is one of the main drivers of the stellar winds for hot O stars (Abbott 1982), and its abundance can be directly measured from studies of HII regions, unlike the case for iron group elements, which are also important in driving stellar winds (see, for example, Puls et al.\ 2000).} environment will have a higher mass loss rate. It follows that WC stars will be more common relative to WN stars in high metallicity galaxies while low metallicity galaxies will have just a few or even zero WCs. Determining the relative number of WC-type and WN-type WRs (the WC to WN ratio) allows us to test stellar evolutionary models by comparing what we see observationally to what the models predict. Reliable evolutionary tracks affect not only the studies of massive stars, but the usefulness of population synthesis codes such as STARBURST99 (Leitherer et al.\ 1999), used to interpret the spectra of distant galaxies. For example, the inferred properties of the host galaxies of gamma-ray bursts depend upon exactly which set of stellar evolutionary models are included (Levesque \& Kewley 2007). It is also important for improving our knowledge of the impact of massive stars on nucleosynthesis and hence the chemical enrichment of galaxies (see, e.g., Meynet 2008). Thus, determining an accurate ratio of WC to WN stars in a galaxy turns out to have its uses far beyond the massive star community, as discussed by Maeder et al.\ (1980).

This strong dependence between the relative number of WCs to WNs and an environment's metallicity has previously been compared to evolutionary model predictions by Meynet \& Maeder (2005), who find good agreement at low metallicities but poor agreement at higher metallicities where the models predict too small a ratio. As discussed in Neugent \& Massey (2011), it is not clear whether the models or observations are at fault since the strongest emission feature in WCs (C III $\lambda4650$) is nearly 4$\times$ stronger than the strongest line in WNs (He II $\lambda4686$) (Conti \& Massey 1989). This makes WNs much more difficult to detect by surveys. However the results of Neugent \& Massey's (2011) galaxy-wide complete survey of M33 suggests that the model predictions continue to under-estimate the WC to WN ratio at high metallicities, such as those in the center of M33. Still, the observational problem likely remains for the higher metallicity Milky Way and M31.

Here we examine the WR content of the Local Group galaxy with the highest metallicity, M31. Before this galaxy-wide survey, only small areas of M31 had been studied with sufficient sensitivity to detect even the weakest-lined WNs such as we do here (see, for example, Massey et al.\ 1986). Earlier wide-area surveys simply lacked this sensitivity and thus failed to detect the majority of WNs (see, for example, Moffat \& Shara 1983, 1987), as discussed in Massey \& Johnson (1998, hereafter MJ98). But here, we present the first galaxy-wide, unbiased survey of Wolf-Rayet stars in M31.

The metallicity of M31 is slightly controversial, as we will discuss in \S5.0. Analysis of HII region data (Zaritsky et al.\ 1994, Sanders et al.\ 2010) suggests that Population I ring's metallicity, where the most active star-forming regions of M31 are located, is about 2$\times$ solar in the inner disk, and solar in the outermost regions. Analysis of a few supergiants by Venn et al.\ (2000) and Smartt et al.\ (2001) suggests that the metallicity is approximately solar instead (but see discussion in Massey et al.\ 2009). Regardless, M31 provides the highest metallicity sample in the Local Group, as interstellar reddening makes it difficult to observe a complete sample of WRs in the Milky Way\footnote{Note, though, the progress being made in the Milky Way by the near infra-red survey of Shara et al.\ 2009, 2012.}. Here, we assume a metallicity of log$\frac{O}{H} + 12 \sim 8.9$ for M31 (Zartisky et al.\ 1994). This metallicity is slightly higher, yet comparable to, the inner metallicity of M33 (log$\frac{O}{H} + 12 = 8.7$; Magrini et al.\ 2007, see discussion in Neugent \& Massey 2011) where Neugent \& Massey (2011) determined the WC to WN ratio to be $0.58\pm0.09$. Thus, we expect to find a similar or higher WC to WN ratio here in M31.

In \S2 we describe how our candidates were selected while in \S3 we describe how they were observed and the data reduced. In \S4 we present a census of WR stars in M31 followed by a discussion of our results in \S5. Finally, in \S6 we summarize our results, discuss our conclusions and highlight where to go from here.

\section{Candidate Identification}
To identify potential M31 WR candidates, we used a method almost identical to that used by Neugent \& Massey (2011) when surveying M33, with the notable exception that we did not use photometric criteria in this study as the M33 study showed that the additional effort involved was not beneficial. Here we present an overview of both the on- and off-band imaging and image subtraction methods. A more detailed description can be found in \S2 of Neugent \& Massey (2011).

\subsection{On- and Off-Band Imaging}
The strong emission lines characteristic of WRs allow for their relatively easy detection. For this survey, we used the filter system first introduced by Armandroff \& Massey (1985). This system is comprised of three filters: one is centered on WCs' strongest line, C III $\lambda$4650, another on WNs' strongest line, He II $\lambda$4686, while the third is centered on the continuum, $\lambda$4750. Each filter has a bandwidth of $\sim$50~\AA.

We obtained images of M31 through each of these three filters using the Kitt Peak 4-meter Mayall telescope and the Mosaic CCD camera on (UT) 2001 November 3-6. The entire optical disk of M31 was imaged using ten overlapping 36$\arcmin$ $\times$ 36$\arcmin$ fields covering a 2.2 deg$^2$ area. Each field was imaged three times for 300s each with a slight amount of dithering between the exposures to fill the CCD camera's gaps. We additionally took bias frames every night as well as exposures of an illuminated uniform screen attached to the dome for flat fielding. Image reductions were identical to those described in Neugent \& Massey (2011) for M33.

\subsection{Image Subtraction}
With these on- and off-band images in hand, we used image subtraction to identify potential WR candidates. While this sounds like a simple task, seeing variability drastically increases the complexity.  When identifying WRs in M33, we used the Astronomical Image Subtraction by Cross-Convolution program which uses point-spread fitting and cross-convolution techniques to simplify the process (Yuan \& Akerlof 2008). We found that this image subtraction method was more reliable than our previous method of doing photometry on every star, as we demonstrated in Neugent \& Massey (2011). The image subtraction program was first run on each of the three dithered images in each field. We then took the median of the three outputted images to decrease the effect of bad pixels, cosmic rays and erroneous results caused by bright foreground stars. Even though the three images had been dithered to fill in the CCD camera's gaps, we only examined the area covered by two pairs (or more) of images when searching for WR candidates. Thus, small gaps still existed. Visually blinking the images and comparing the stars on the WC or WN image with the stars on the continuum image allowed us to remove even more false positives.

An examination of all ten M31 fields yielded 216 WR candidates. Of these, 28 were removed because their colors indicated they were red ($B-V = 0.65 - 1.7$); as Neugent \& Massey (2011) found that late-type stars could show up as spurious WR candidates due to an absorption band in the continuum filter. This left 188 WR candidates. MJ98 list 48 spectroscopically confirmed WRs in M31\footnote{Note that here we reclassify one of these 48 as an O7 Iaf star based upon a better spectrum.}, plus an additional 6 candidates. Of the 48 confirmed WRs, we recovered 46 (96\%), and one of the six candidates. Of the two spectroscopically confirmed WRs we missed, one fell into a gap between the CCDs, and the other was simply overlooked. In designing the spectroscopic observations described in the next section we started with 190 targets, the 188 we found here (including the previously known ones) plus two of the three spectroscopically confirmed WRs that we missed.  (The third one, J004628.51+421127.6, previously called OB 102 WR1, was excluded from the target list due to an oversight by the second author.)

\section{Spectroscopy}
To confirm new M31 WR candidates and reclassify previously known WRs, we used the multi-object fiber-fed spectrograph Hectospec (Fabricant et al.\ 2005) on the 6.5-m MMT. The large field of view (1$^\circ$ in diameter), multitude of fibers (300), and their allowed close spacing (20\arcsec) made it practical to obtain observations of our M31 WR candidates using just 3 pointings. The data were all taken with the 270 line mm$^{-1}$ grating, resulting in spectral coverage from 3700-9000~\AA, at a resolution of 6~\AA\ (5 pixels).  Beyond $\sim$7500~\AA\ we expect the spectra to be contaminated by second-order blue light, but this should have no effect on our program, as all the principle WR classification lines are found between 4000-6000~\AA.

The equivalent of one night of dark time was granted through NOAO (2011B-0083), although the observations were carried out in a self-staffed ``queue" mode over the course of several nights. When designing the fiber configurations, we gave highest priority to our new WR candidates, and lower priority to the known WRs. We were able to assign 182 of our 190 targets (96\%) using just three configurations, two of which were observed on (UT) 2011 Sept 23 and one on Sept 26, with the second author the observer for the latter. Each of these configurations were observed for 7200s (2 hrs). Additionally, some of the unobserved targets were inserted into four additional configurations designed for other programs, one of which was observed for 3600s on 2011 Sept 26, two observed for 3600s on 2011 Oct 21\footnote{The two 2011 Oct 21 fields belonged to Maria Drout, and we thank her for allowing us to insert some of our unobserved candidates into her configurations. We additionally thank her for bringing our attention to three new WRs hidden within her own target list.}, and one observed for 5400s on 2011 Nov 17. Only a small percentage of our targets were observed in these shorter exposures, and although they were not observed for as long as our primary exposures, the data are still of good quality. Reductions were carried out through the standard Hectospec pipeline (Mink et al.\ 2007) by Susan Tokarz of the OIR Telescope Data Center, and a short description can be found in Neugent \& Massey (2011). In the end, we successfully obtained spectra of 182 of the 190 target stars (96\%), missing only four of our own newly found candidates, the one previous WR candidate, and four previously spectroscopically confirmed WR stars (one of which had not been included among the list of 190 target stars). These nine stars are listed in Table~\ref{tab:Whinners}\footnote{Throughout this paper we identify stars using the Local Group Galaxies Survey (LGGS) catalog of M31 from Massey et al.\ (2006), with minor revisions given in Massey et al.\ (2011).}.

Of the 182 assigned target stars, 147 (81\%) turned out to be Wolf-Rayet stars. Their classifications were guided by the criteria listed by van der Hucht (2001), and based upon those listed by Smith (1968a), with extension to the WC4, WN2, and the WN9-11 subclasses.  (We have not used the intermediate WN2.5 type described by van der Hucht 2001 to allow easier comparison with earlier works.)  We list the spectral types in Table~\ref{tab:Winners}, along with the measured equivalent widths (EWs) of the He II $\lambda 4686$ (WN stars) or C III $\lambda 4650$ lines (WC stars) and that of C IV $\lambda 5803$, if present.  Additionally included are the full widths at half maxima (FWHM) of these lines.   

Spectroscopy carried out as part of another project identified eight additional M31 WRs which the imaging survey had missed, invariably because the star fell into a gap or on a bad column of the original CCD frames. All eight of these were observed because they were identified as stellar H$\alpha$ sources by Massey et al.\ (2007a). One of these, J004130.37+410500.9, is either a late-type WN star (WN7?) or an Of star; it is sufficiently embedded in nebulosity that it is hard to tell. Another, J003857.19+403132.2, is also swamped by its HII region, and only He II $\lambda 4686$ and N III $\lambda \lambda 4634,42$ show up as broad stellar lines. The other six we also classify as late-type WNs: J004024.33+405016.2 (WN9), J004242.33+413922.7\footnote{Called a P Cyg type Luminous Blue Variable (LBV) candidate by Massey et al.\ (2007a).} (WN10), J004337.10+414237.1 (WN10), J004357.31+414846.2 (WN8)\footnote{The star J004357.31+414846.2 was also independently found to be a late-type WN star by Nelson Caldwell (2011, private communication), and we are grateful for his calling it to our attention.}, J004511.21+420521.7 (WN11), and J004542.26+414510.1 (WN10). The WN10 and WN11 stars show a broad blend of N II emission with He II $\lambda 4686$, as well as P Cygni lines of He I, typical of these ``transitional" objects (Walborn 1977 and Bohannan \& Walborn 1989; see discussion in Crowther et al.\ 1995, but also Neugent \& Massey 2011). In the case of J004242.33+413922.7, we detected the star in our survey, but incorrectly eliminated it as the subtracted image was characteristic of a bright star.  J004542.26+414510.1 was missed simply because it barely shows up as enhanced on the WN frame; its equivalent width (-4~\AA) is quite small, as are those of the other WN9-11 stars. Our suspicion is that even though the majority of the missed stars fell in gaps on some of the pairs, they would have been correctly identified as viable candidates from the other pairs if they had been stronger lined. This suggests a bias against the latest-type WN stars, but we expect these to be very rare, given what we know of the WN content of other nearby galaxies (see discussion below).

After our spectroscopy we were somewhat surprised to find that three of our newly confirmed WRs had $B-V$ values approaching our color cutoff of $B-V=0.65$, presumably because there are regions within M31 that are more highly reddened than the small areas previously surveyed for massive stars.  However, when we reexamined the stars we eliminated as likely red stars (and hence found through absorption in the continuum band), we found that several were quite bright ($V\sim 16.2-18.0$), and thus very unlikely to be heavily reddened WRs. We list in Table~\ref{tab:toored} five candidates that are a bit redder than our original cutoff. The faintest of these five stars, J003944.96 +402038.2 is located in an OB association, and thus is a possible WR star. However, while one of the remaining stars is located within the Population I ring (discussed in \S4.1), and another near an OB association, we frankly do not expect any of these remaining four to be real WRs, but we include them here for completeness. We hope to later follow up on J003944.96 +402038.2 to determine if it is a genuine WR star.

In Figure~\ref{fig:wns}, we show representative spectra of M31 WNs, and in Figure~\ref{fig:wcs} we do the same for WCs. Although we have included a range of types, we will discuss in \S 5.1 how the distribution of WC subtypes is skewed towards later types in M31 compared to in other well-studied galaxies, and its significance. We make a few notes below.

\subsection{WN Classification}
{\it WN2:} We classified only one star, J004455.82+412919.2, as WN2, and it is a peculiar star at that. He~II $\lambda 4686$ is very broad, and there's no sign of N~III $\lambda \lambda$4634,42, N~V $\lambda \lambda 4603$, 19 or N~IV $\lambda 4058$, leading us to the WN2 designation. But, WN2s should have strong He~II lines, and as Figure~\ref{fig:wns} shows, there is no He~II $\lambda 4542$ at all. Furthermore, N~IV $\lambda 7110$ {\it is} present, along with strong C~IV $\lambda 5806$.  The lack of any C~III $\lambda 4650$ would suggest that the presence of strong C~IV is primarily an ionization effect, and not indicative of a transition WN/C type (Conti \& Massey 1989).

{\it WN4 -- 6:} The classification of WN4 -- 6 stars is more complex than one might expect. As listed by van der Hucht et al.\ (1981), a WN4 star has N~III $\lambda \lambda$4634,42 ``weak or absent", and N~IV $\lambda 4058$ ``about equal to" N~V $\lambda\lambda$4503,19 in strength.  With a WN6, N~IV is about equal to N~III in strength, and N~V is ``present but weak." We found those two extremes easy to classify. However, the WN4.5 intermediate class has N~IV $>$ N~V, and N~III ``weak or absent", while WN5 stars have N~V $\approx$ N~IV $\approx$ N~III. Instead, our spectra show that between WN4 and WN6, the stated criteria are somewhat misleading. We found N~IV $\lambda$4058 to be the strongest of the three nitrogen lines, and either N~V a little stronger than N~III or the reverse is true. As long as N~III is a reasonable fraction of N~V, we continued to classify the star as a WN5.  We illustrate two examples in Figure~\ref{fig:wns}, an ``early" WN5 star, J004445.90+415803.8, with N~V $>$ N~III and a ``late" WN5 star with N~V $<$ N~III, J004426.32+413419.8. To determine if this is somehow unique to M31, we checked our old spectrophotometry (the basis for the measurements given in Conti \& Massey 1989) and conclude that it is not. For instance, the WN5 Galactic star HD 193077 (WR 138) is a ``late" WN5 star with N III $>$ N~V but N~IV stronger than either. By contrast, the WN5 Galactic star V444 Cyg (WR 139) is an ``early WN5", also with N~IV stronger than either N~III or N~V.  Oddly, the WN4.5 star HD 219460 would also be a ``late WN5" by our reckoning, with N~IV stronger than either N~III or N~V, and N~III stronger than N~V. The WN4.5 HD 65865 is similarly an ``early WN5."  Our point is simply to note that the stated classification criteria, while adequate to the photographic error, could be made more quantitative in the digital era, and that the problem is most severe for the WN4.5-WN5 subtypes.  
 
{\it WN7 -- 8:} Notice that the WN7 and WN8 spectra we illustrate contain hydrogen, as is typical with these late-type WNs. This can be seen by comparing the even-n Pickering series of He II (e.g., $\lambda \lambda$ 4100, 4339) with the odd-n Pickering series of He II (e.g., $\lambda \lambda$ 4200, 4542). The even series is coincident with the hydrogen Balmer lines so when the even-n lines are stronger than the intervening odd-n lines, hydrogen must be present.
 
{\it WN9 -- 11:} Seven of our M31 stars would traditionally be classified as ``slash" stars, with spectral type ``Ofpe/WN9". This class was introduced by Bohannan \& Walborn (1989), who suggested that these stars represented a transitional phase between the Of and WN stages (see Conti 1976), and who noted that some such stars are known to undergo eruptions similar to those of LBVs. Crowther et al.\ (1995) argued that since the absorption lines in these stars are always blue shifted, they are formed in the stellar winds, and not in the (presumed) static photosphere, and thus these stars should be classified as pure WRs. We disagree with that argument, as evidence suggests that the outflow of material has already begun at the photospheric level in Of stars (Conti et al.\ 1977; see Hutchings 1979, and discussion in Massey et al.\ 2012). Nevertheless, it is useful to distinguish among these stars, and here we have adopted the scheme proposed by Crowther et al.\ (1995), who replace the Ofpe/WN9 designation with the WN9, WN10, and WN11 types. The distinctions are based primarily on the relative strengths of N II and N III, not too dissimilar to what was originally proposed by Walborn (1977).  Of the seven very late WN stars in M31, three are classified as WN9 according to this scheme (N III $>$ N II), three are classified as WN10 (N III $\approx$ N II) and one is classified as WN11 (little or no N III). We illustrate examples of all three types in Figure~\ref{fig:wns}. Note that Massey (1998) compares an earlier spectrum of one of these stars, J004344.48+411142.0 (OB 69 WR2) to that of the prototypical Ofpe/WN9 star HDE 269927c, which is now classified as WN9 (Crowther et al.\ 1995).
 
Such stars are quite rare, even among WRs. In the LMC, roughly 12 similar stars (WN9 -- WN11) are known out of a sample of 134 (9\%) according to Breysacher et al.\ (1999). In M33, only 10 of the 206 (5\%) known WRs are Ofpe/WN9 stars (Neugent \& Massey 2011). For M31, we find a similar percentage, with only seven WN9 -- WN11 stars known out of 154 (5\%). The fact that such stars are more prevalent in the LMC suggests a metallicity effect, although none of the 11 SMC WNs are of late-type (Massey et al.\ 2003), and the SMC is lower in abundance than that of the LMC.

\subsection{WC Classification}
{\it WC4 -- WC8:} The classification of the WC stars was straightforward; we see the entire range from WC4 to WC8 type. However, we did not find any WC9s, which are distinguished simply by the ``presence" of C II $\lambda$4267. We carefully examined our four WC8 star spectra and see no sign of this line despite our good signal-to-noise (100 -- 150 per 6~\AA\ spectral resolution in this wavelength region). Still, the criterion would be better defined if there was a minimum equivalent width requirement for the C II line.

\subsection{Interesting Losers}
Our detection method was designed to find objects with a stronger flux near $4650\pm25$~\AA\ (WC filter) or $4690\pm25$~\AA\ (WN filter) that in the continuum (CT filter) near $4750\pm25$~\AA. Most of our non-WR detections turned out to be cases where bright stars caused the image subtraction program (mentioned in \S2.2) to give spurious detections. These non-WRs were often interesting in their own right, with our spectra identifying them as B or A supergiants. We plan on discussing these as part of a larger study of newly found early-type stars in M31 and M33 (see brief summary in Smart et al.\ 2012). However, in other cases, the non-WRs had legitimate emission in the WC or WN filter, and these too are quite interesting. We list these five objects in Table~\ref{tab:Losers}.

Two of these objects turn out to be massive stars in M31. One, J004427.95+412101.4, had previously been called ``WNL/Of" by Armandroff \& Massey (1991), and its spectrum is shown in Figure~\ref{fig:Of}. Absorption lines are clearly visible in our higher quality spectrum (compare to that of Armandroff \& Massey 1991), and we classify the star as roughly O7. The strong N III $\lambda 4634, 42$ and He II $\lambda 4686$ lines lead to an ``Iaf" luminosity classification. The other M31 member, J004221.78+410013.4, was first identified by Massey et al.\ (2007a, their Table 8) as a probable H$\alpha$ emission star. Its spectrum is shown in Figure~\ref{fig:LBV}, and we identify the star here as a ``hot LBV candidate", a star whose spectrum resembles that of other known LBVs in M31 and M33. The emission is considerably stronger in this star than of others shown in Figure 10 of Massey et al.\ (2007a). A sea of emission, primarily of [Fe II], likely lead to the excess flux in the WR filter.

The remaining three objects are all high-redshifted background quasars! Their spectra are shown in Figure~\ref{fig:qsos}. The strongest emission lines in quasars are found at rest wavelengths of 1216~\AA\ (Ly$\alpha$), 1549~\AA\ (C IV), and 1909~\AA\ (C III]). These will show up in our detection filters at redshifts of $z\sim 2.85$, $z\sim 2.00$, $z\sim 1.45$, respectively. In M33 we found two quasars among our WR candidates, one with $z=2.85$ and one with $z=1.99$.  Here we find a $z=2.87$ QSO, as well as two $z=1.45$ specimens. One of these, J003856.68+403446.8, was also misidentified as a probable WR star in the objective prism survey of Meyssonnier et al.\ (1993). All three are X-ray sources, appearing in the second ROSTAT point source catalog (Supper et al.\ 2001) and the XMM-Newton point source catalog (Stiele et al.\ 2011). J003856.68+403446.8 is also associated with a radio source (Flesch 2010). The faintness of these objects ($V=19.8-20.8$) probably precludes their use as probes of M31's interstellar media in the immediate future, although they may provide astrometric fiducials with truly negligible proper motions. 

\section{A Census of M31's WRs}
Of the 150 WR stars for which we obtained spectroscopy, 107 are newly discovered WRs, either as part of our survey (99 stars) or because of followup spectroscopy of stellar H$\alpha$ sources (8 stars), as described in the previous section. We did not obtain new spectroscopy for another 4 previously known WRs (see Table 1). This brings the total number of WRs known in M31 to 154. Our discoveries were skewed towards finding more WNs (79) than WCs (28), which was expected given the previous completeness bias.

\subsection{The Catalog}
We list in Table~\ref{tab:BigTable} all of the 154 currently known Wolf-Rayet stars in M31. Included are the LGGS designations along with the alternate IDs for any of the previously known WRs. Two stars, X004256.05+413543.7 and X004308.25+413736.3, lacked a counterpart in the LGGS survey and thus are referred to as ``X00...". The spectral types are from our new spectroscopy, except as indicated.

Also included in Table~\ref{tab:BigTable} is the value $\rho$, the galactocentric distance within the plane of M31, normalized the a radius of 95\farcs3, corresponding to a surface brightness of $\mu_B=25.0$ (de Vaucouleurs et al.\ 1991).  We computed $\rho$ by adopting a position angle of the major axis of 35$^\circ$, an inclination of 77$^\circ$, and the coordinates for M31's center of $\alpha_{2000}=00^h42^m44.4^s$ and $\delta_{2000}=41^\circ16^\prime08\farcs0$ (de Vaucouleurs et al.\ 1991, van den Bergh 2000). A value of $\rho=1.0$ corresponds to 21 kpc, taking the distance to M31 as 0.76 Mpc (van den Bergh 2000).

We include an estimate of the absolute visual magnitude $M_V$ in Table 5 for each star.  To obtain these, we adopted $E(B-V)=0.2$ ($A_V=0.6$), based on the median $E(B-V)=0.13$ found for early-type stars by Massey et al.\ (2007b), where we have rounded up to allow for higher reddening.  However, for stars redder than $B-V=0.2$ we list no $M_V$. This color would roughly correspond to a reddening of $E(B-V)=0.35$ ($A_V=1.1$), based upon a compromise $(B-V)_0=-0.15$ found by Feinstein (1964). We would of course have preferred to de-redden the photometry star-by-star, but the uncertainties in the intrinsic broad-band colors (due to the presence of emission lines) precluded that, and we were unable to measure reliable narrow-band photometry from our fiber spectroscopy, as discussed below. The $M_V$'s listed using an average value are thus good to $\sim0.5$~mag.

Using van den Bergh (1964), we determined which M31 WR stars are located within OB associations. These results are also included in Table~\ref{tab:BigTable}. Overall, 133 (86\%) of the confirmed M31 WR stars are located in or very near an OB association. Further more, 123 (80\%) of the WRs are located within M31's Population I ring located between 9 kpc ($\rho = 0.43$) and 15 kpc ($\rho = 0.71$). This ring is characterized by the most active regions of star formation within the galaxy (van den Bergh 1964). However, there are still 7 (5\%) WRs that are neither located within an OB association or the Population I ring. These we believe to be truly isolated and are marked as such in Table~\ref{tab:BigTable}.

Figure~\ref{fig:WRlocations} shows the locations of all known WR stars within M31 with the blue $\times$ representing WN stars and the red $+$ representing WCs. The two black ellipses denote the inner and outer boundaries of the Population I ring discussed above. Note that the relative number of WCs to WNs doesn't appear to change with respect to $\rho$ suggesting that the metallicity in the regions of M31's disk where WRs are found is fairly constant.

We are making our M31 and M33 WR spectra available via a tar file accessible through the electronic edition. Such data may prove interesting to other researchers, and provides an archival record if one of these stars explodes as an SNe within the next few decades (see Shara et al.\ 2009, 2012), a prospect we consider statistically unlikely\footnote{The duration of the WR phase is half a million years, $5\times 10^5$ years. Thus a population of 350 WRs will suffer a spectacular loss on average once every $\sim$1,500 years.}, but certainly exciting. We have included all of the WRs in M31 for which we have new data (i.e., those listed in Table 2 in the present paper) as well as the WRs and Of-type stars in M33 listed in Table 3 of Neugent \& Massey (2011). The spectra are unnormalized.  If more than one spectrum of a particular star was available, we included the second version denoted with an ``alt" on the name. This data set includes spectra of 150 of the 154 WRs known in M31 plus 84 of the 206 WRs known in M33, as well as 12 M33 stars we consider to be of Of-type.

\subsection{Completeness}
Several factors may affect the completeness of this survey. Our biggest limitation deals with differential reddening within the galaxy. The unfavorable inclination combined with numerous dust clouds results in areas of high extinction. However, the area completely obscured by dust is small: using Hodge (1980) we find only $\sim$330 arcmin$^2$, or 4\% of our survey area, should be covered by dust clouds, and only a tiny fraction of that area is in the Population I ring. As for the relatively unobscured regions, Massey et al.\ (2007a) measured the average (typical) extinction of the OB stars, finding $E(B-V) = 0.13$, where half is due to foreground reddening and half is due to internal extinction. Yet, there are clearly regions of higher (and lower) reddening; for instance, early-type stars in the OB associations OB 48 and OB 8 -- 10 show $E(B-V) = 0.24$, while OB 120 shows $E(B-V) = 0.08$ (Massey et al.\ 1986). In areas of high reddening we are understandably not as complete as in areas of low reddening. However, by removing the highly reddened stars from our sample (as we discuss below), we can test our completeness throughout the ``reasonably reddened" areas of M31. We perform this test using the emission-line fluxes of our WN stars (since, as discussed in \S1, the emission lines in WNs are much fainter than those in WCs). As discussed in MJ98, if two stars are equally bright, it will be harder to detect the star with the smaller He II $\lambda$4686 EW. However, if two stars have the same He II $\lambda$4686 EW but different luminosities, it will be easier to then detect the brighter star. This is because surveys such as ours are primarily flux-limited in emission bands (see discussion in Neugent \& Massey 2011)\footnote{The presence of an OB companion will lead to a brighter magnitude but a lower line EW, as the line will be partially washed out by the continuum of the other star.  However, as we show below, we would have detected even the weakest lined WRs known in the LMC, and this includes many WRs with OB companions.  The most likely circumstance where we might miss a few WRs would be if companion is an evolved B supergiant, which is visually brighter than most O stars.  But even so, we detect several such systems, as shown in Table 5.}.

In previous papers (MJ98, Neugent \& Massey 2011), the V-band flux times the EW was used to act as a surrogate for intrinsic line fluxes. This was straight-forward for the SMC, LMC, and M33, where the reddening is uniform. However, as discussed above, this is not the case in M31. We can, of course, use the LGGS $B-V$ and $U-B$ measurements of these stars to estimate their color excesses ($E(U-B)/E(B-V)=0.72$), but because of their strong emission lines, the intrinsic broad-band colors of WR stars are not well defined. The ``proper" way to correct for WR star extinction would be to use narrow-band {\it b} and {\it v} fluxes set on the continuum (see, for example, Torres-Dodgen et al.\ 1988 and Massey 1984). Here we attempted to flux-calibrate our spectra to compute synthetic {\it b} and {\it v} values, but found a $V$-band magnitude-dependent difference between $b-v$ and the LGGS $B-V$, which we attribute to imperfect sky subtraction. With fiber spectrometers, sky subtraction is never local, and therefore it does not properly take into account the underlying galaxy background, unlike in the case of the broad-band LGGS {\it UBV} values. So, instead we use the $M_V$ values, derived as stated earlier, and ignore the stars for which $B-V>0.2$.

In Figure~\ref{fig:complete} we show a test of our current survey's sensitivity. By comparing the locations of the M31 WNs (open circles) to the WNs found in M33 (green boxes, from Neugent \& Massey 2011), the LMC (blue $\times$s, from Conti \& Massey 1989), and the SMC (red asterisks, from Massey et al.\ 2003), it is clear that our survey has detected WNs (and thus WR stars in general) that are as intrinsically faint and as weak-lined as those in other nearby galaxies, where the WR content is known completely. However, because of the differential reddening in this galaxy, we can only claim completeness in the portions of M31 that aren't highly reddened. We stress that the WN9 -- 11 stars (Ofpe/WN9 ``slash" stars) do not appear on this diagram, as they were not found as part of our on-band, off-band survey. These stars have weaker lines than the -10~\AA\ EW we were aiming for in our survey. We thus would not be surprised if additional stars of this type are discovered in the future. Still, given their scant numbers in other galaxies, we expect the missing number of WNs to be a few percent of the total number of WRs.

Given that our survey is sufficiently sensitive to detect all but the weak-lined WN9 -- 11 stars, our remaining incompleteness is due to stars falling in gaps or simply being overlooked. For M31, our fields had quite a bit of overlap, and we estimate the missed area covered to be around 5\% of our total survey area. This is consistent with our recovery of 96\% of the known WRs, and the fact that our extensive other spectroscopy revealed only 3 additional WRs (excluding the WN9 -- WN11s), or about 2\%. We suspect we have additional incompleteness for WN9 -- WN11s, but due to our complementary spectroscopy, we expect that our total number of 154 WRs is complete to 95\%, similar to what we found in M33 (Neugent \& Massey 2011). Thus, the total number of WRs in M31 is approximately 160 -- 170, although some additional stars may be hidden in regions of very high reddening, such as behind dust lanes.

\section{Results} 
As discussed in Section 1, we found that the relative number of WC and WN stars is a sensitive function of metallicity among the galaxies of the Local Group that have been well surveyed. For instance, in the SMC (log$\frac{O}{H}+12 = 8.13$, Russell \& Dopita 1990) this ratio is 0.09 (Massey et al.\ 2003). In the LMC, where the metallicity is 0.24~dex higher (log$\frac{O}{H}+12=8.37$, Russell \& Dopita 1990), the ratio is 0.23 (based upon Breysacher et al.\ 1999; see Neugent \& Massey 2011), and in M33 where there is a metallicity gradient, the WC to WN ratio changes from 0.22 in the outer portions (log$\frac{O}{H}+12=8.29$) to 0.58 in the inner region (log$\frac{O}{H}+12=8.72$).  

Zaritsky et al.\ (1994) used the $R_{23}$ method to determine the oxygen abundances of M31's HII regions, finding log$\frac{O}{H}+12=9.1$ at a radius of 3.3~kpc and a very slight gradient, -0.020 dex kpc$^{-1}$, where both numbers have been adjusted to the slightly larger distance we adopt here (0.76~Mpc vs 0.70~Mpc).  The average galactocentric radius of our entire sample is 11~pc ($\rho=0.53$), corresponding to log$\frac{O}{H}+12=8.93$ (1.7$\times$ solar). Note that over the 9 -- 15~kpc Population I ring where the vast majority of our WRs are found (\S4.1) we expect the oxygen abundance to only change by $\pm0.05$~dex. Therefore, we cannot look for gradients in WR properties as we did in M33. Venn et al.\ (2000) and Smartt et al.\ (2001) suggest that M31's metallicity is closer to solar, and this controversy is discussed extensively by Massey et al.\ (2009). Since then, Sanders et al.\ (2010) have obtained new data on the HII regions; their analysis agrees well with the higher abundances found by Zaritsky et al.\ (1994), and with the metallicity reaching about 2$\times$ solar in the inner disk, and 1$\times$ solar at 20~kpc. Thus we adopt log$\frac{O}{H}+12 \sim 8.9$.  Fortunately, the evolutionary models of Meynet \& Maeder (2005) predict very similar WC to WN ratios at both solar and 2$\times$ solar metallicity, and so our comparison of this ratio with the evolutionary models below in \S5.3 is not very sensitive to whichever value we adopt.

\subsection{WC Subtype Distribution}
Besides influencing the WC to WN ratio, the metallicity of a galaxy also affects its WC subtype distribution. This was first noted by Smith (1968b), after finding that WCs found inwards of the solar circle are of later type than those found outwards.  Smith (1968b) suggested that this difference was due to the Galactic metallicity gradient.  We now understand that high metallicity environments lead to increased stellar wind densities. This then affects the strength of C III $\lambda$5696 which is used to classify WCs (Crowther et al.\ 2002). It also affects the appearance of a WC star's spectrum where late-type WCs have weaker (small EWs) and skinnier (small FWHMs) CIV $\lambda$5806 lines than early-type WCs. Thus, a figure showing FWHM vs.\ EW demonstrates an environment's average metallicity even if the spectral subtype hasn't been well determined. Figure~\ref{fig:hockeystick} shows diagrams for M31 as well as the three different metallicity environments within M33. Since the spectral type variation within M31 is small compared to that across all of M33, we can infer that the metallicity variations throughout the regions of M31 where WRs are found are small. In M33 WRs are found throughout the entire disk, but in M31 the WRs are mostly found in the Population I ring. Additionally, a comparison of the M31 figure with the three M33 figures shows that the subtype distribution of M31 WCs is similar to the subtype distribution in the inner region ($\rho<0.25$) of M33. Note though, that there are more WC7's than in M33, and also note the presence of WC8's, which are nonexistent in M33. This suggests that the metallicity of M31 is higher than that of the inner portion of M33, which is thought to be solar. We can also see this by comparing the median FWHMs of the WCs in the two samples where the median in the inner portion of M33 is 49~\AA, while the median in M31 is 53~\AA.

The relationship between the WC subtype and metallicity can be shown by creating histograms based on subtype, as shown in Figure~\ref{fig:histograms}. As the metallicity increases (starting with outer portion of M33 with a metallicity of log$\frac{O}{H} + 12 = 8.29$ and ending with the inner portion of the Milky Way with a metallicity of log$\frac{O}{H} + 12 = 9.07$), the number of early-type WCs decreases and the number of later-type WCs increases. When examining the Milky Way, it is important to remember that interstellar reddening makes a complete census impossible. However, here we use the data from van der Hucht (2001) to create one histogram for the area inwards of the solar circle ($\le 8.4$ kpc) and another for outwards of the solar circle ($> 8.4$ kpc). We estimate the range in metallicity for these regions using the metallicity gradient (for oxygen) of $-0.043\pm0.01$ dex/kpc (Shaver et al.\ 1983, Deharveng et al.\ 2000, Esteban et al.\ 2005). 

\subsection{WC to WN Ratio}
Using the data from Table~\ref{tab:BigTable}, we can finally compute a meaningful WC to WN ratio for M31. With 62 WCs and 92 WNs, we get a ratio of 0.67. We can also compute the ratio after leaving out the highly reddened stars, which might reside in areas where we have missed WNs. In this case, we get a WC to WN ratio of 0.63. Since these two values are quite similar, it suggests our survey is not overly biased towards WCs even in areas of high reddening (i.e., we may be missing WCs and WNs nearly equally).

Before this survey, there were only 48 known WRs in M31: 33 WCs and 15 WNs (see Table 10 of MJ98) so the WC to WN ratio was 2.2. This is quite a significant difference from the value of 0.67 we present here! MJ98 additionally restricted their sample to only 9 small regions that had been well surveyed with CCDs and found 27 WRs (13 WCs and 14 WNs). While the WC to WN ratio of 0.93 they found is much closer to our value of 0.67, the difference is still significant. However, as we'll see in the next section, 0.67 still isn't low enough to match the evolutionary model predictions.

We can now compare this ratio to the value we might expect to find based on M31's metallicity. In the inner region of M33, at a metallicity of log$\frac{O}{H}+12=8.72$ (roughly solar; though, this value itself is controversial), Neugent \& Massey (2011) found a WC to WN ratio of 0.58. Here in M31, at a metallicity of log$\frac{O}{H} +12 \sim 8.9$, we expect the ratio to be comparable if not a bit higher. Therefore, this value of 0.67 is consistent with what we expect to find if the metallicity were higher than solar. We compare the WC to WN ratio with the ratios of other Local Group galaxies in Table 9.

We have also improved our estimate of the errors on the WC/WN ratio compared to that given in Neugent \& Massey (2011). There are two approaches we've taken to estimating this uncertainty. The first of these assumes that the surveys for WRs are 95\% complete in each of these galaxies. Neugent \& Massey (2011) argued this was the case for their M33 survey, and we've argued the same here for M31. But, that number is probably pretty good for the SMC and LMC as well, which were surveyed by different means. Massey \& Duffy (2001) surveyed essentially all of the SMC using similar filters, finding 2 additional WN-type WRs, bringing the total to 11, and stated that despite their efforts they could not preclude the possibility that they had missed one or two, particularly in crowded regions. And indeed the next year, Massey et al.\ (2003) found a 12th SMC WR (also of WN-type) accidentally while obtaining spectra of previously unobserved blue SMC stars. The discovery of another WR in the SMC would be surprising, but not terribly so, consistent with a 5\% uncertainty, which would correspond to 0.6 stars. In calculating the effect of the 5\% incompleteness on the WC/WN ratio, we have made the two extreme assumptions; either that {\it all} of the unobserved stars are of WN type or {\it all} of the unobserved stars are of WC type. This leads to asymmetrical errors, as the WC/WN ratio is $<$ 1 for each of these galaxies. This makes the observed amounts almost lower limits, as the discovery of additional WCs would have a disproportionately large effect on the ratios, particularly in the SMC where the number of stars involved is so small. (In computing the errors we have rounded to the nearest integer number of stars.) Of course, given that WCs are so much easier to find than WNs, we would argue that this may significantly exaggerate the errors for the SMC, given the small numbers. Still, the LMC, which has been surveyed extensively by Azzopardi \& Breysacher (1979, 1980), and Morgan \& Good (1985, 1990), still reveals the occasional new WR star, including one of WC type just discovered accidentally by the first two authors with Nidia Morrell, details of which will be published later.   

While these five-percent error bars represent the uncertainties in the current population of WR stars in these galaxies, what would we have observed one million years ago? Or what will we observe one million years in the future? (Note that the WR lifetime is shorter than a million years.) The number of massive stars would still be about the same, and the metallicity of the galaxy would be the same, but the ``snapshot" of the current WR population would look a bit different. Thus, for our second estimate of the error, we used a stochastic approach. Although stars are not photons, it is not unreasonable to estimate the expected variation using a similar method; root-$N$ statistics (i.e., if we observe 9 WC stars today, the statistical fluctuation is of order 3). The error estimates produced by this assumption are, coincidentally, almost the same as the upper limit error based on the 5\% assumption. This is most likely the more appropriate error for comparing with stellar evolutionary theory\footnote{Note that MJ98 assumed root-$N$ statistics, while Neugent \& Massey (2011) used the 5\% incompleteness but applied the upper error symmetrically.}.

\subsection{Comparison with Theory}
Now that we have an unbiased sample of the WRs within the reasonably reddened regions of M31, we can compare our observed results with the predictions of the latest Geneva evolutionary models. Neugent \& Massey (2011) previously made this comparison for the LMC, SMC and M33 but here we extend it to M31, which is at a higher metallicity, and also compare the predictions of the older and newer Geneva models. Figure~\ref{fig:wcwn} shows both the expected ratio of WCs to WNs as computed by the Geneva evolutionary models as well as what we observe, with the data taken from Table~\ref{tab:WCWN}.

In Figure~\ref{fig:wcwn}, the dotted curve shows the predicted WC/WN ratio as computed by Meynet \& Maeder (2005), and we see that despite the considerable lowering of the value for M31, there is still a serious disagreement with the model predictions at solar and above metallicities, as noted by Neugent \& Massey (2011) for the inner portion of M33. However, since the Meynet \& Maeder (2005) study, a new generation of evolutionary models are under construction by the Geneva group incorporating a number of improvements, including revising the prescription for mass-loss during the red supergiant phase, changing the shear diffusion coefficient, and adopting a more realistic assumption for the initial rotation velocities as a function of mass. A full set of these models are currently only available for $z=0.014$ (solar metallicity, Ekstr\"{o}m et al.\ 2012) and for $z=0.006$ (LMC-like metallicity, Chomienne et al., in prep.). Georgy et al.\ (2012) computes the WC/WN ratio for the $z=0.014$ models, and we have additionally computed it for the $z=0.006$ models following the same procedure. The predictions are shown in the figure as solid blue lines\footnote{In the evolutionary models, the WC phase is recognized by the predicted surface composition of carbon relative to nitrogen. Georgy et al.\ (2012) begin this phase when the mass surface abundance of carbon becomes more than that of nitrogen and when the hydrogen mass fraction is lower then 30\% and log$_{\rm Teff} > 4$. Meynet \& Maeder (2005) start when the mass surface abundance of carbon becomes more than 10$\times$ that of nitrogen. The result is highly insensitive to the exact value used.}. The lower blue line corresponds to the realistic case of the initial rotational velocity being 40\% of the critical (breakup) velocity, while the upper blue line corresponds to the unrealistic case of no initial rotation. Clearly the new models offer a dramatic improvement, and it will be most interesting to see the results for metallicities that are lower (SMC-like) and higher (M31-like).  

Georgy et al.\ (2012) argue that some significant fraction of WR stars may be formed through Roche-lobe overflow (RLOF) in binaries, based upon (a) the fact that the models predict a lower relative number of WR per O star than what appears to be observed in the Milky Way, and (b) that adopting a non-zero fraction for the number of RLOF-formed WRs would make the WC/WN ratio larger.  We might argue instead that the formation of {\it some} WRs have been helped by RLOF, we note that we would then expect to find a higher percentage of WR binaries at low metallicities if this fraction was significant.  After all, stellar wind induced mass-loss will be less important at low metallicities, while we would naively expect that the percentage of close binary O-type progenitor stars will be the same. However, the percentage of close binary WRs in the Magellanic Clouds appears to be statistically indistinguishable from that of the Milky Way (Bartzakos et al.\ 2001; Foellmi et al.\ 2003a, 2003b; Schnurr et al.\ 2008; see Moffat 2008 for a summary).  An alternative explanation may simply be that 40\% of the breakup velocity may be slightly too high a value. It is clear that a much higher value is predicted by the models with no rotation, suggesting that the prediction is sensitive to the exact assumed rotation velocity. The 40\% value is chosen as it appears a good match to solar-metallicity B-type stars (see discussion in Ekstr\"{o}m et al.\ 2012) but this should represent a sort of average behavior.  Eventually one would want to make the predictions based upon a realistic distribution function, since some stars will have faster and some with slower rotational velocities. It will also be interesting to see what the predicted WC/WN ratio is for new models at metallicities higher than solar. Nevertheless, it should be noted that at lower rotational velocities there is an even smaller relative number of WRs to O stars predicted (Georgy et al.\ 2012), which may run into conflict with observations, although the number of O stars is quite uncertain in nearby galaxies (Massey 2010).

\section{Summary and Conclusions}
M31 contains 154 confirmed WR stars, 62 of WC type, 92 of WN type and zero of WO type. Out of these 154 WR stars, 107 are presented for the first time in this paper. Before this survey, results were heavily biased towards WC stars which are easier to identify due to their very strong lines. Our new survey covered the entire disk of M31 (2.2 deg$^2$), and was sensitive enough to detect even the weakest-lined WN stars in regions of normal reddening. We believe that the total number of WR stars found is complete to roughly 5\% ($\sim$8 stars), except in regions of unusually high reddening, and except for possibly a few missed weak-lined WN9 -- 11 stars. Our spectroscopy shows WN types ranging from WN2 through WN11 with WC types (which are believed to be directly related to the environment's metallicity) ranging from WC4 through WC9. These findings are consistent with M31's high metallicity since amongst the Local Group galaxies, only the inner part of the Milky Way has WC types as late as WC9. We also found that most of the WRs are located in the Population I ring between 9 and 15~kpc, where the available HII region data suggest an oxygen abundance of 1.5 -- 2.0$\times$ solar.

The ratio of WC to WN stars found in M31 (0.63 in the normally reddened regions) is much lower than previously thought, but is still well above that predicted by the Meynet \& Maeder (2005) models at high metallicities. However, it is in accord with what Neugent \& Massey (2001) found when comparing the inner portion of M33 with the models. While the latest Geneva evolutionary models do a good job predicting the observed WC to WN ratio at metallicities lower than solar, there is still work to be done, especially at the higher metallicities such as in M31 and the inner portion of M33. At this point, models with metallicities above solar are not yet available. We currently cannot be sure what this problem is caused by: either there is a deficiency yet unresolved in the models (such as adopting too high an initial rotation rate), or single-star models are simply not sufficient and require Roche-lobe overflow in close binaries to produce the ``extra" WC stars at higher metallicities. However, this would require a higher close binary fraction at higher metallicities and at first glance, there is no strong theoretical argument as to why binary frequencies of massive stars should be metallicity-dependent. Still, there have been observational hints that this dependence may be the cause (Zinnecker 2003) and sufficiently little is known about massive binary formation that this can't be ruled out. We are planning to pursue this question in the next observing season. 

Our survey suggests that the total number of WRs in M31 is about 160 -- 170, with the possibility that a few additional WRs may be hidden behind dust lanes or in other regions of high reddening.  This is only about 75\% of the number of WRs in M33 (Neugent \& Massey 2011). Compared to M33, M31 is about 30$\times$ more massive, and 9$\times$ more luminous in $V$ (van den Bergh 2000 and references therein), but M31 is an Sb galaxy while M33 is an Sc galaxy, and it is well-known that Sc galaxies have higher current star-formation rates (SFR). A good tracer of the current SFR is the integrated H$\alpha$ luminosity, and indeed M31's H$\alpha$ luminosity is about 80\% that of M33 (Kennicutt et al.\ 2008), consistent with the WR content, and lending additional credence to the argument that our survey of M31 is close to complete.

\acknowledgements
We would like to thank Perry Berlind for his help observing, Susan Tokarz for reducing the spectra as described in \S3.1, Nelson Caldwell for correspondence on the issue of M31's oxygen abundance plus help in observing the Hectospec configurations, Georges Meynet for his helpful suggestions and ideas, and Maria Drout for bringing three previously undiscovered WRs to our attention. This work was supported by the National Science Foundation under AST-1008020.

\begin{figure}
\epsscale{0.33}
\plotone{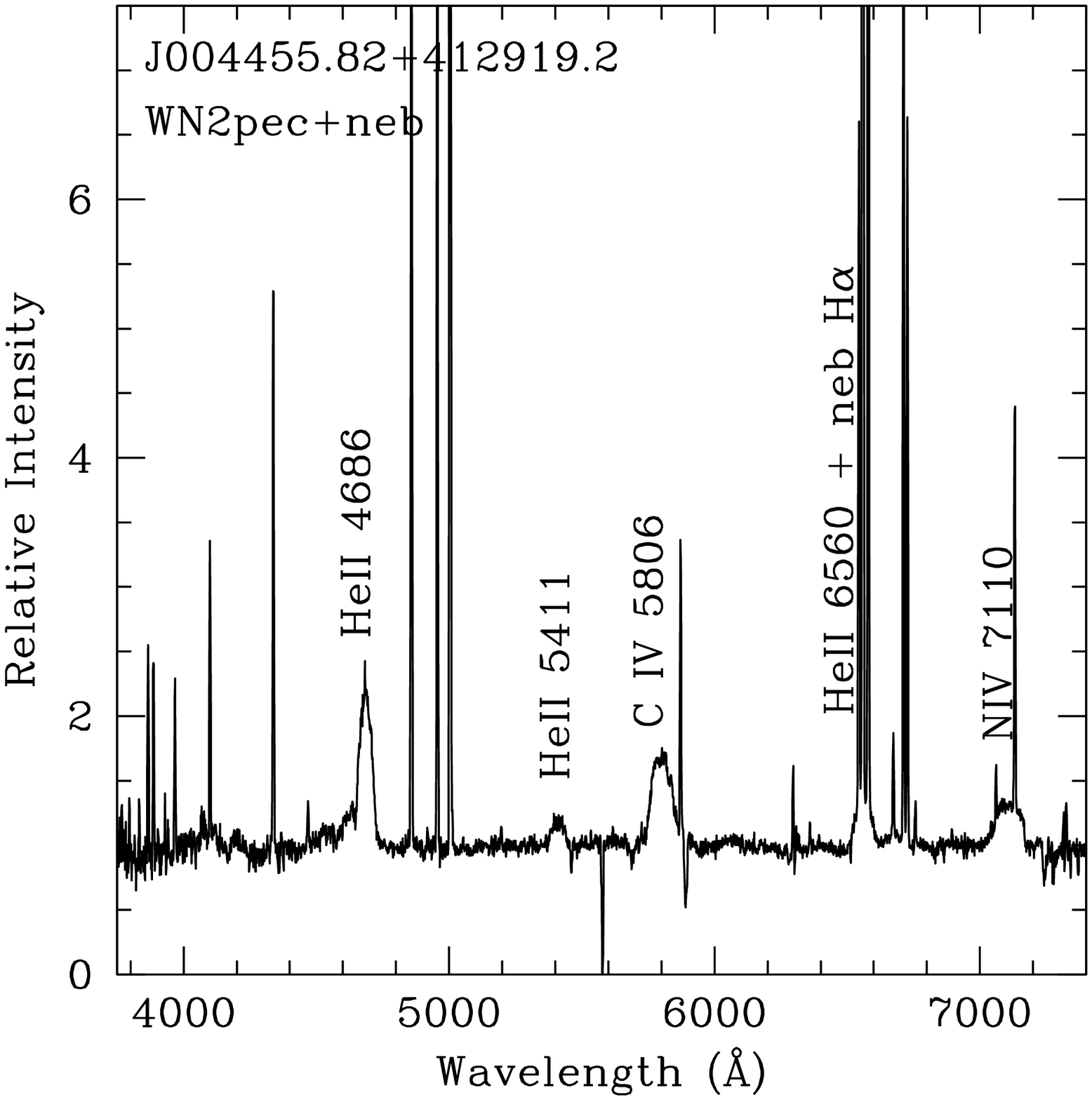}
\plotone{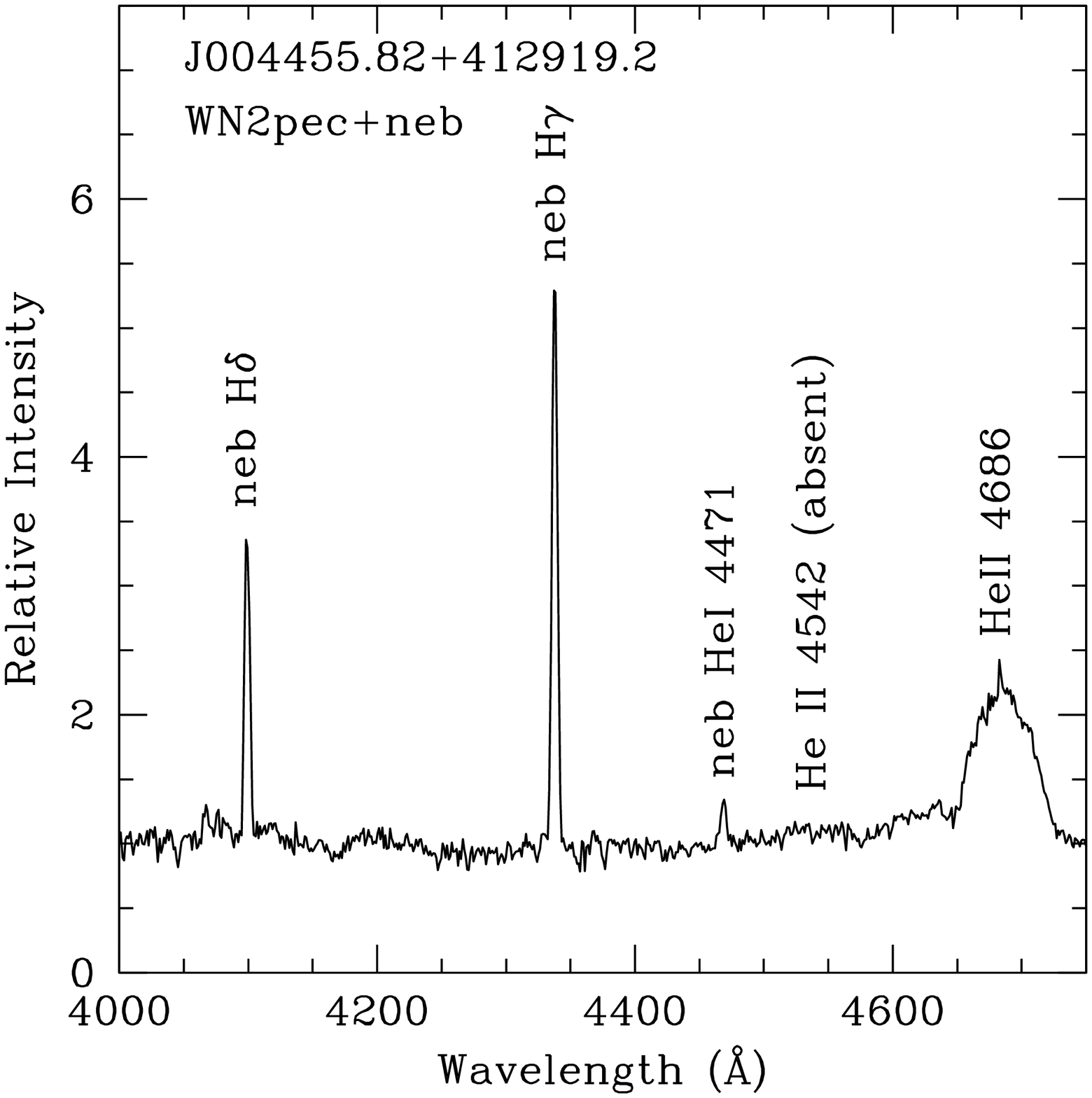}
\plotone{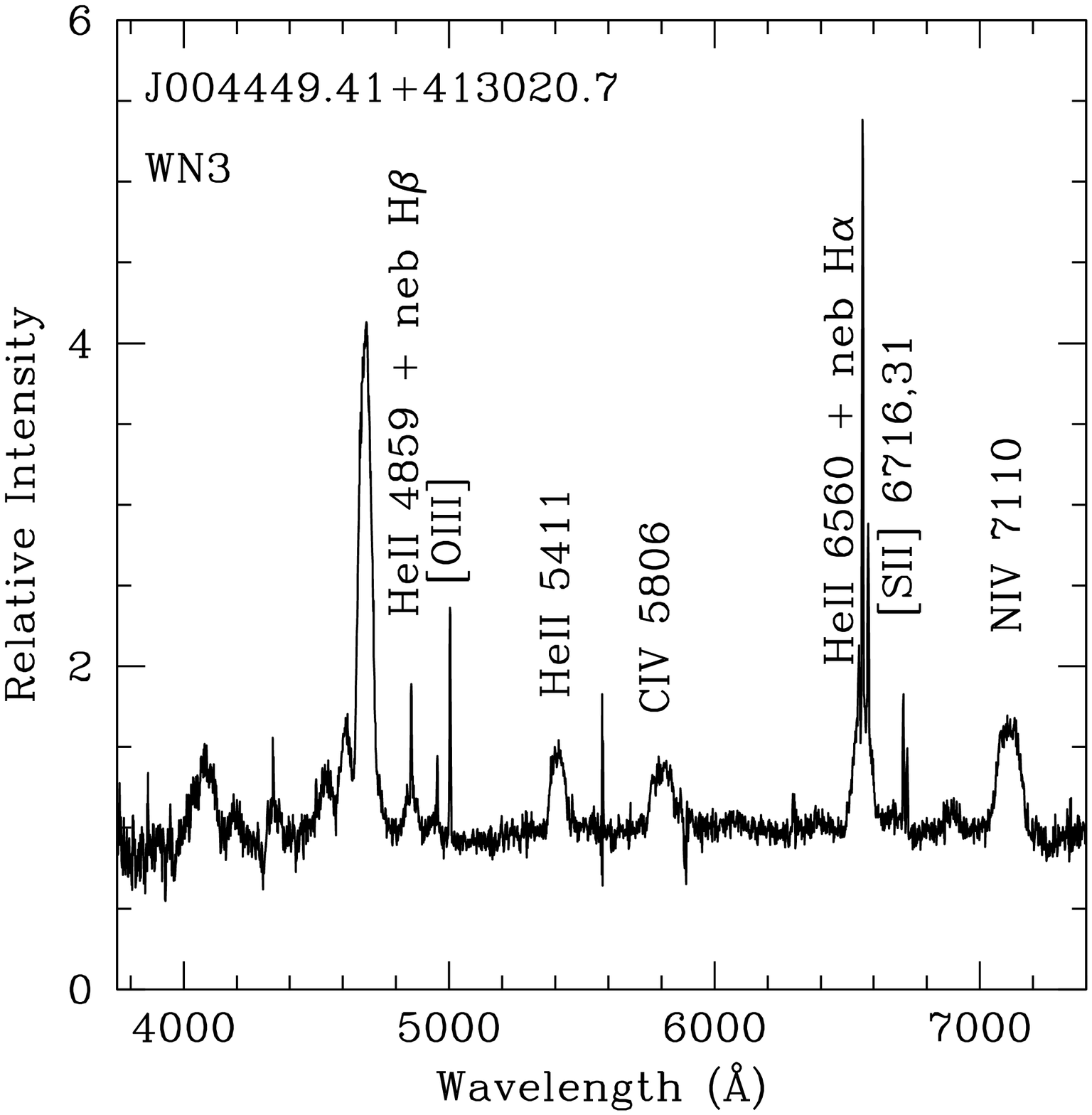}
\plotone{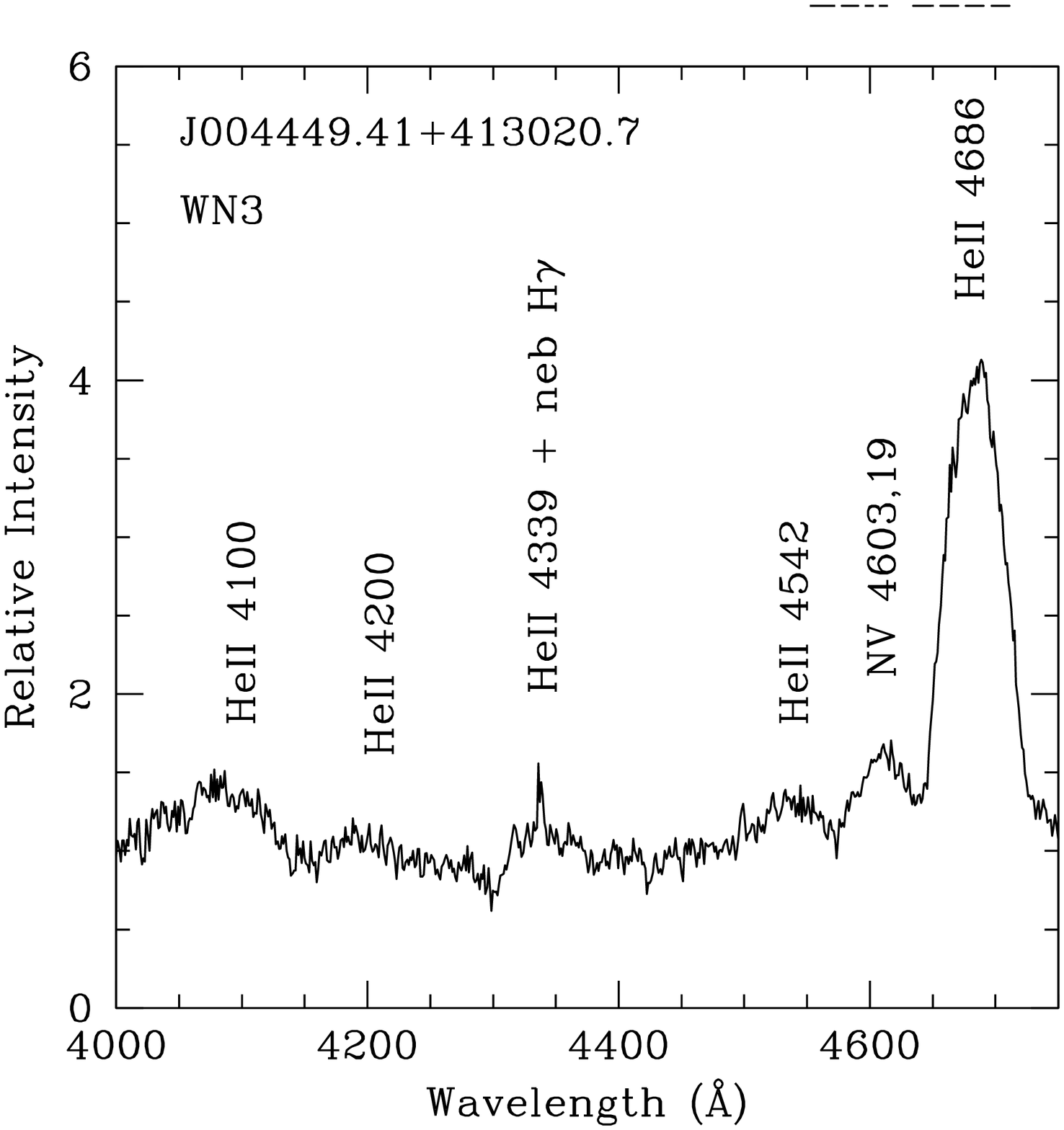}
\plotone{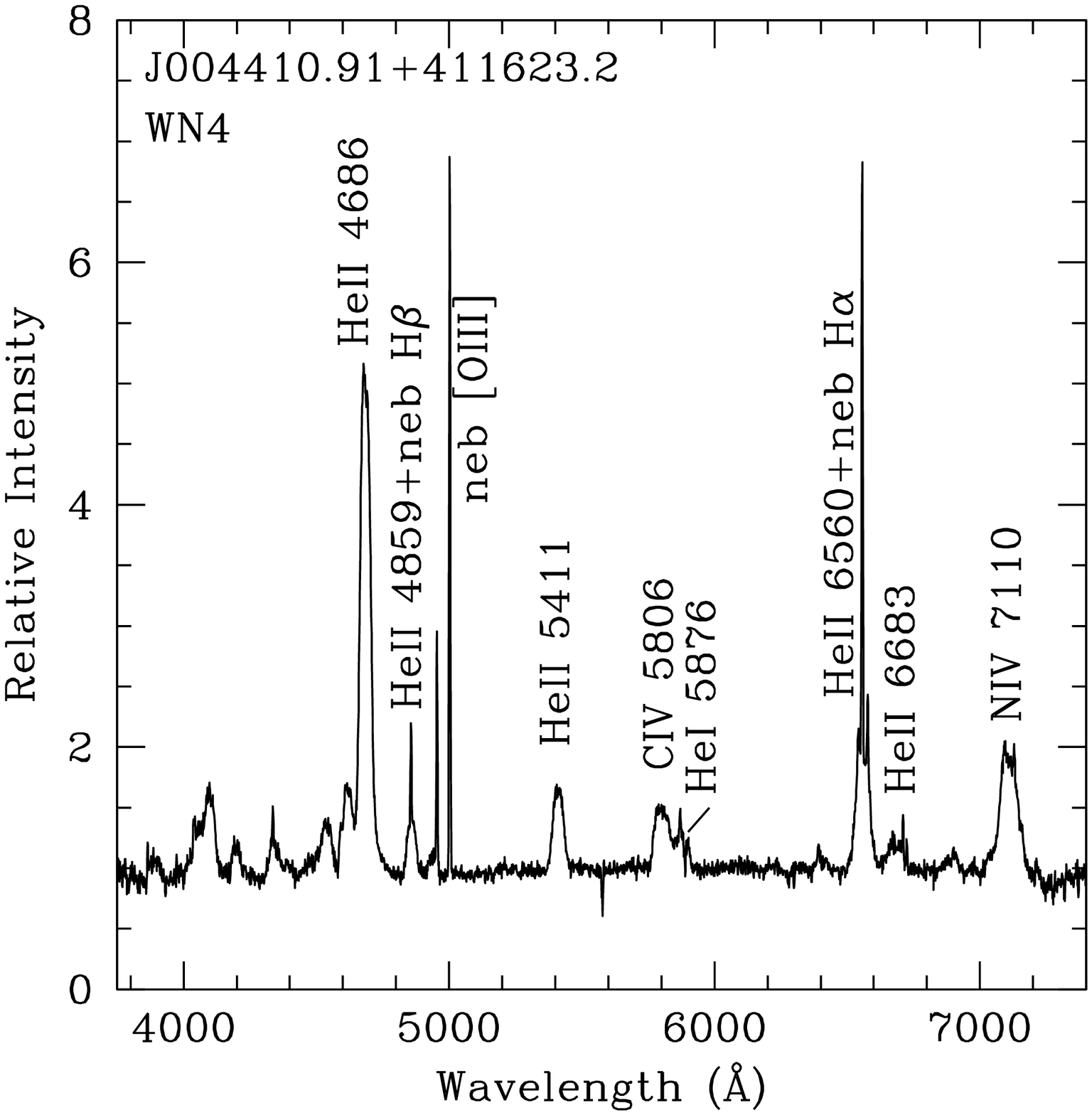}
\plotone{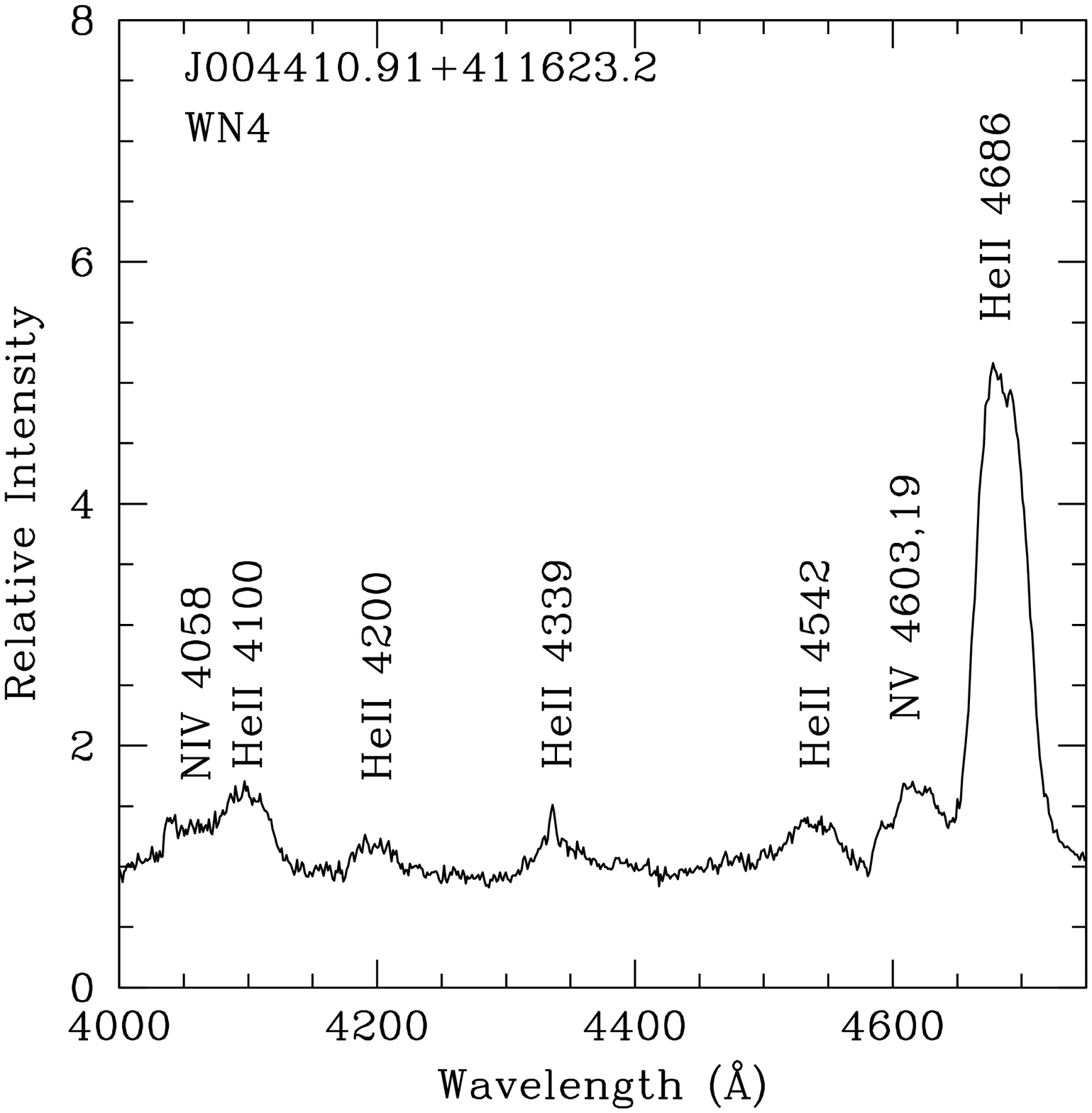}
\caption{\label{fig:wns} The spectra of representative WN stars.  On the left we show the 3750 -- 7400\AA\ region of the spectrum, while on the right we show an expansion of just the blue region. The latter contains the N III $\lambda 4634, 42$, N IV $\lambda 4058$, and NV $\lambda\lambda 4603, 19$ lines which
form the primary basis for the classification.}
\end{figure}

\addtocounter{figure}{-1}
\begin{figure}
\epsscale{0.33}
\plotone{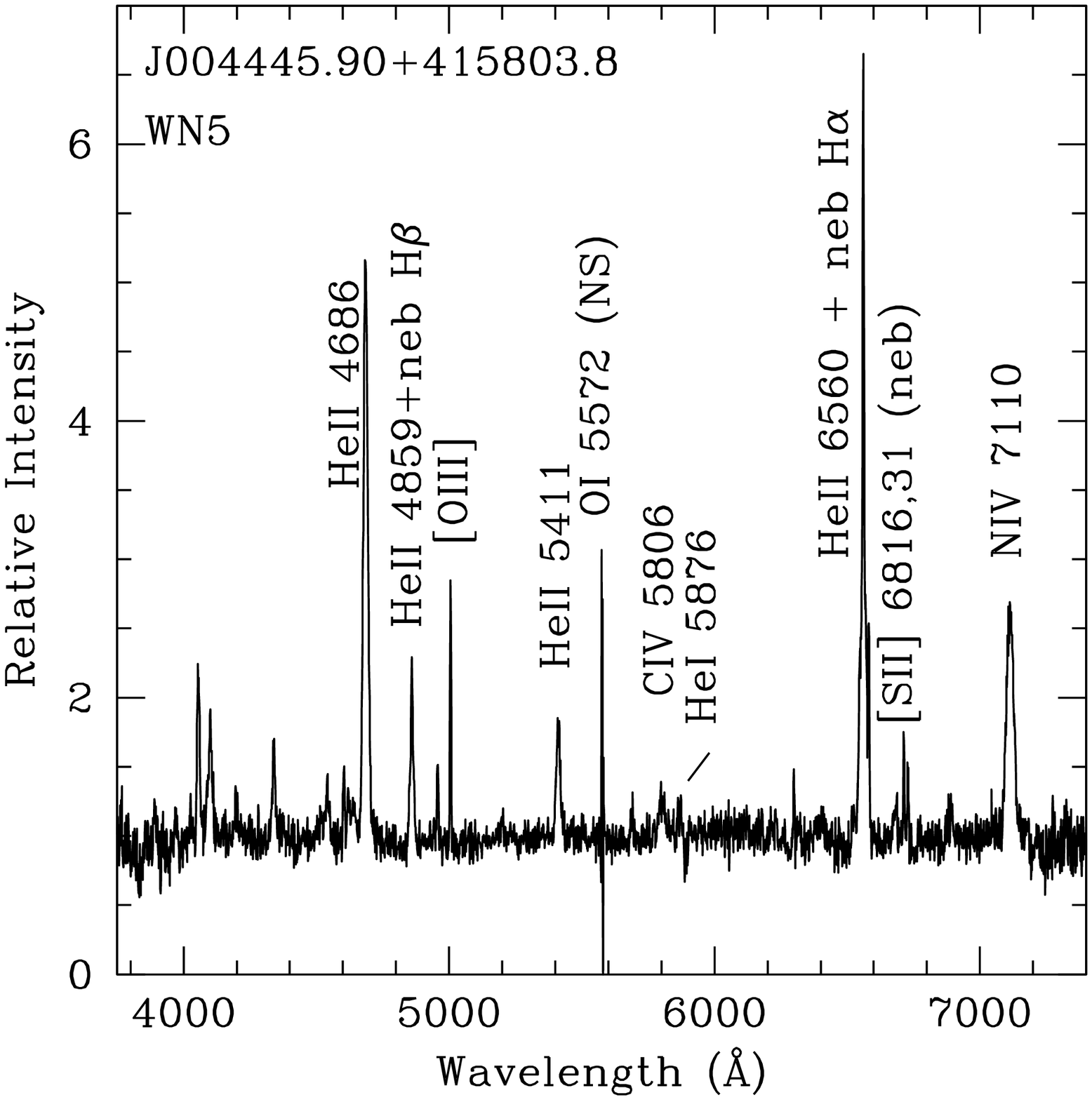}
\plotone{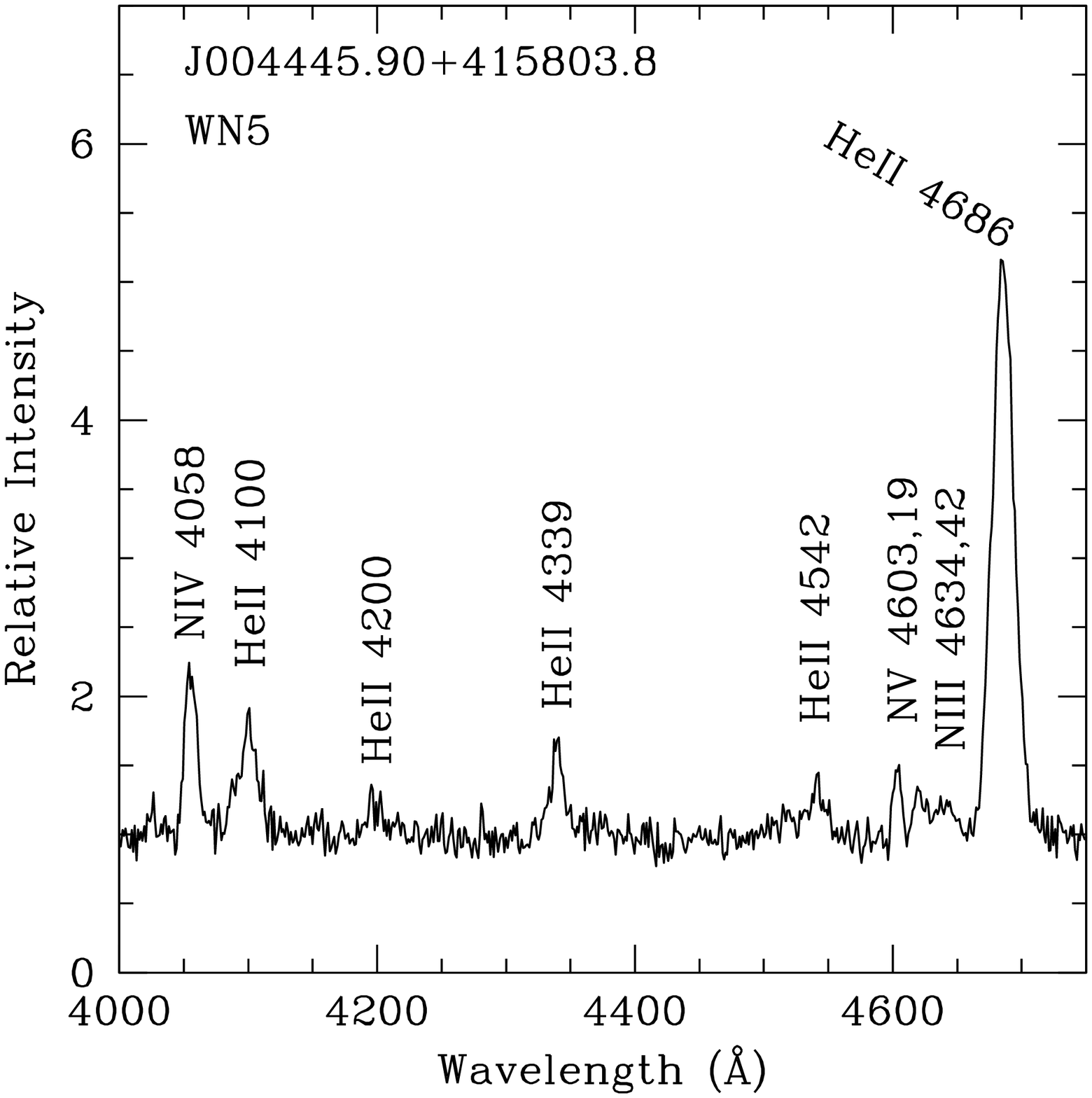}
\plotone{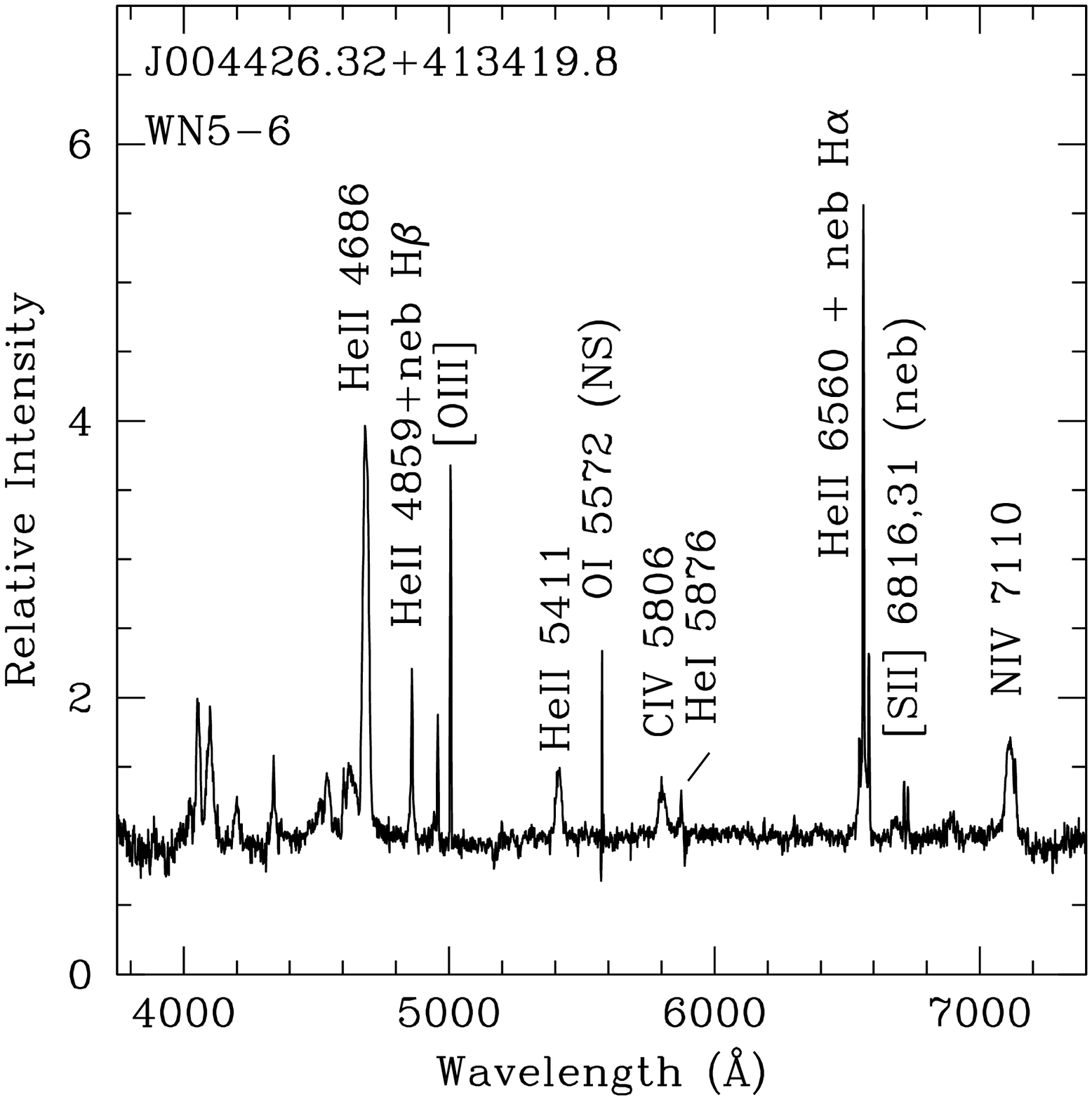}
\plotone{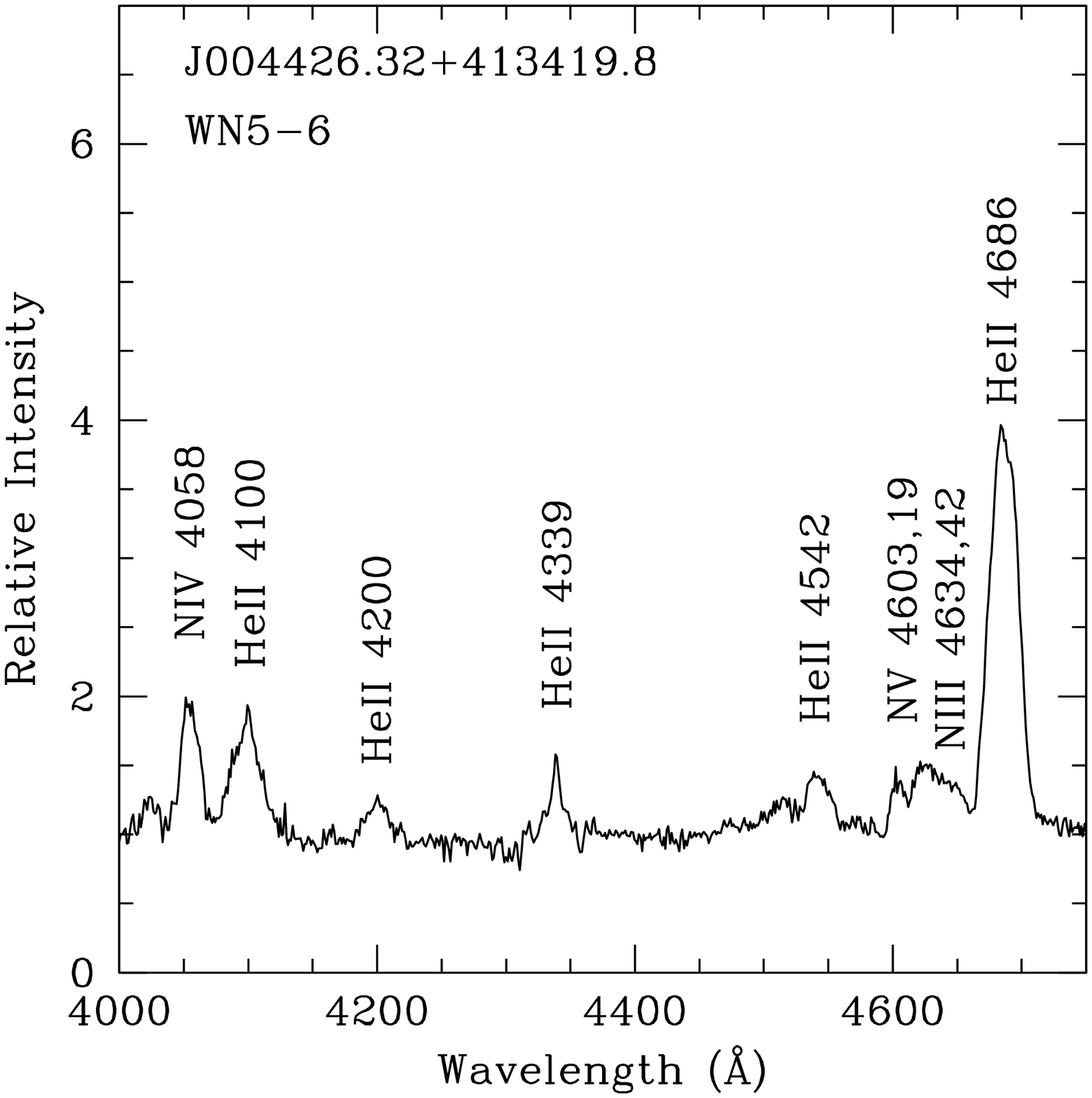}
\plotone{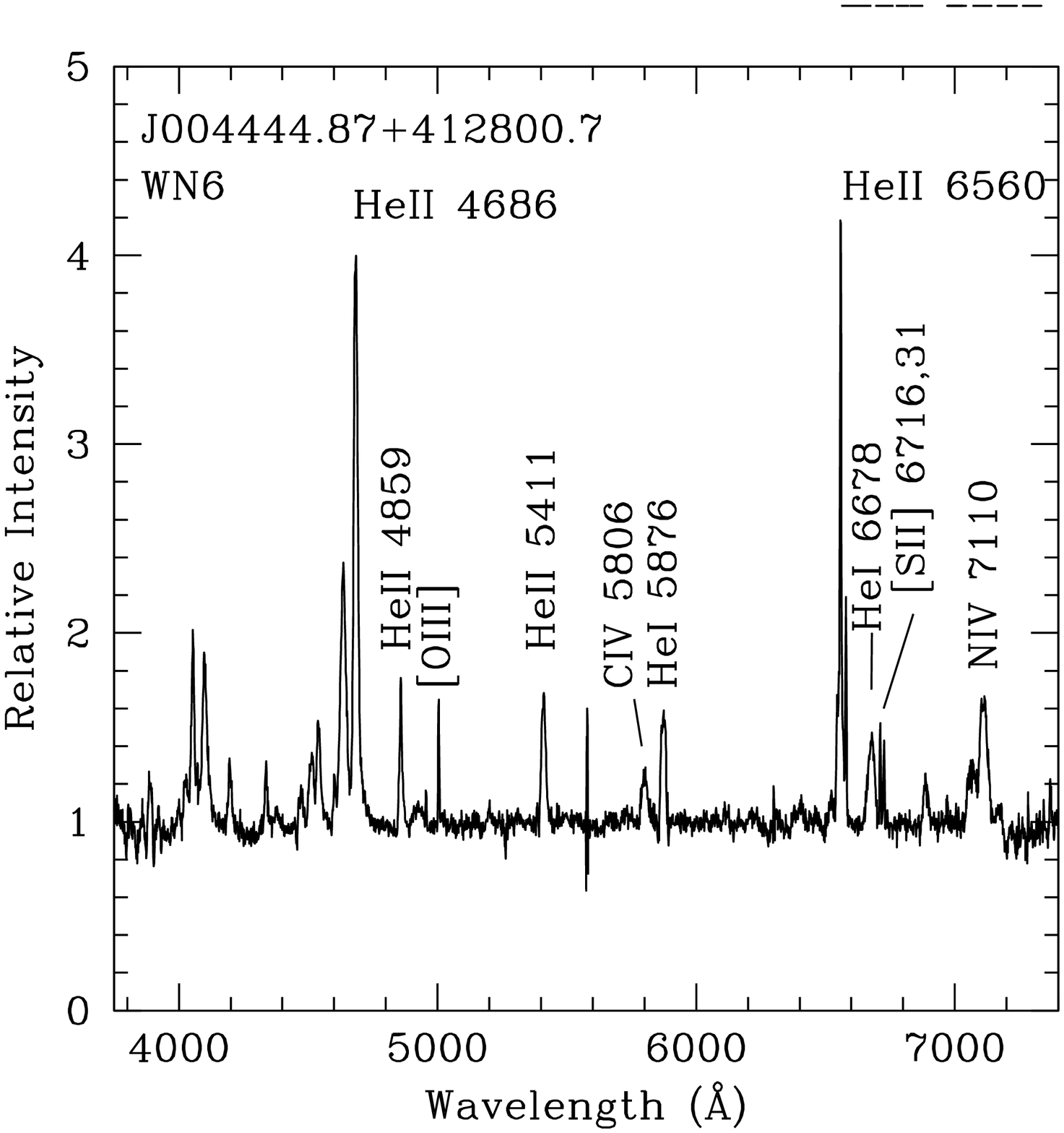}
\plotone{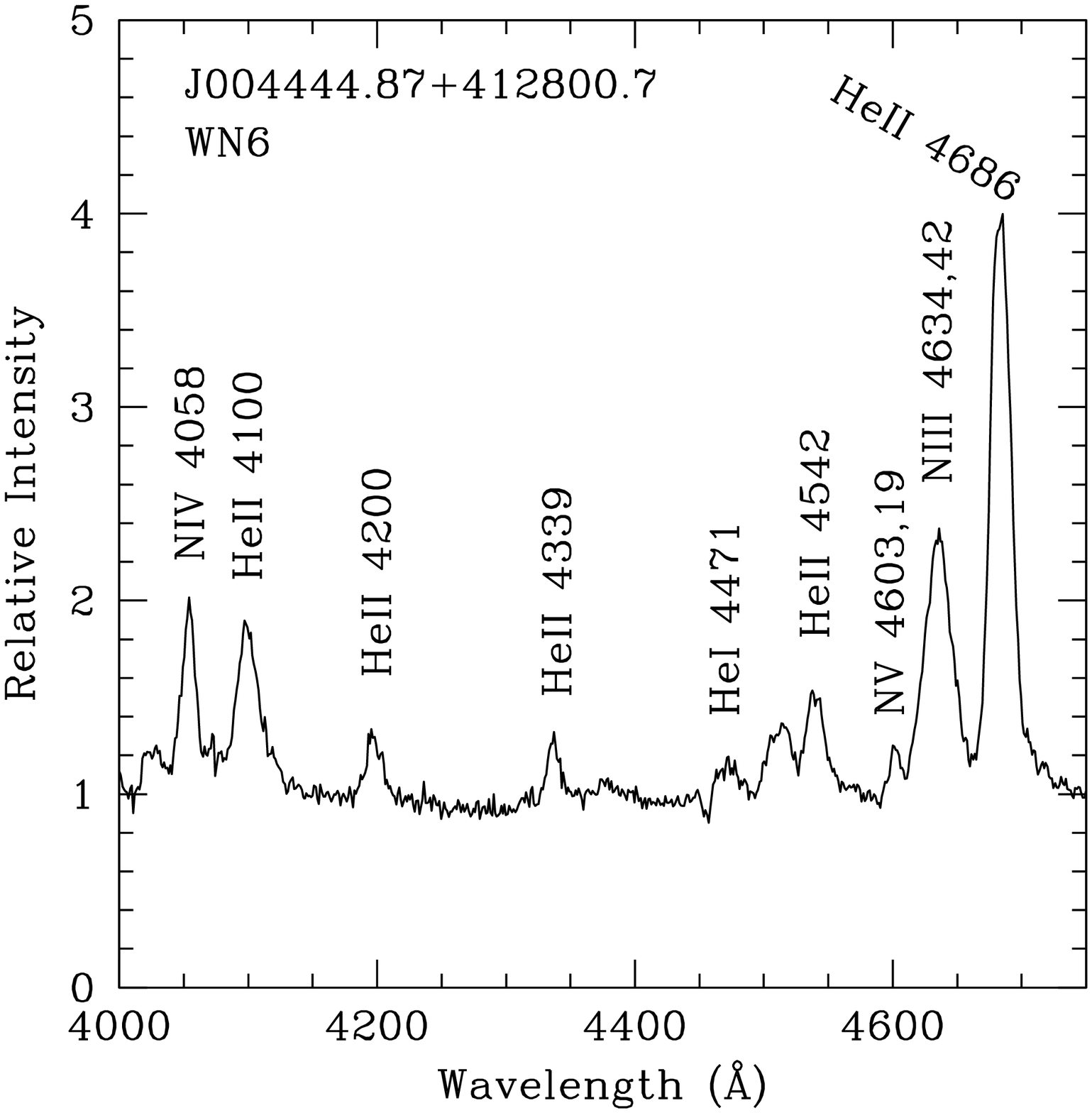}
\caption{Continued}
\end{figure}

\addtocounter{figure}{-1}
\begin{figure}
\epsscale{0.33}
\plotone{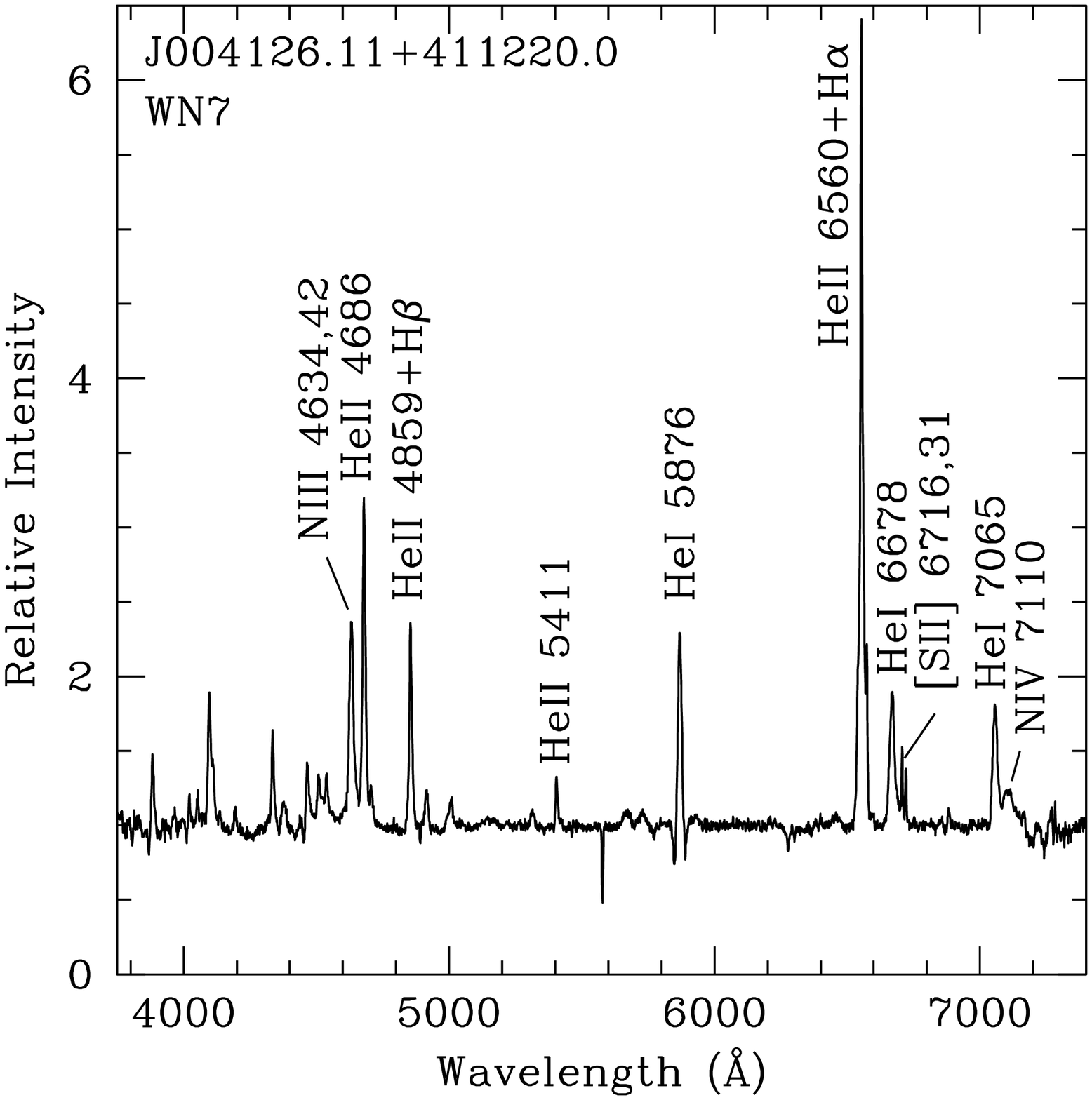}
\plotone{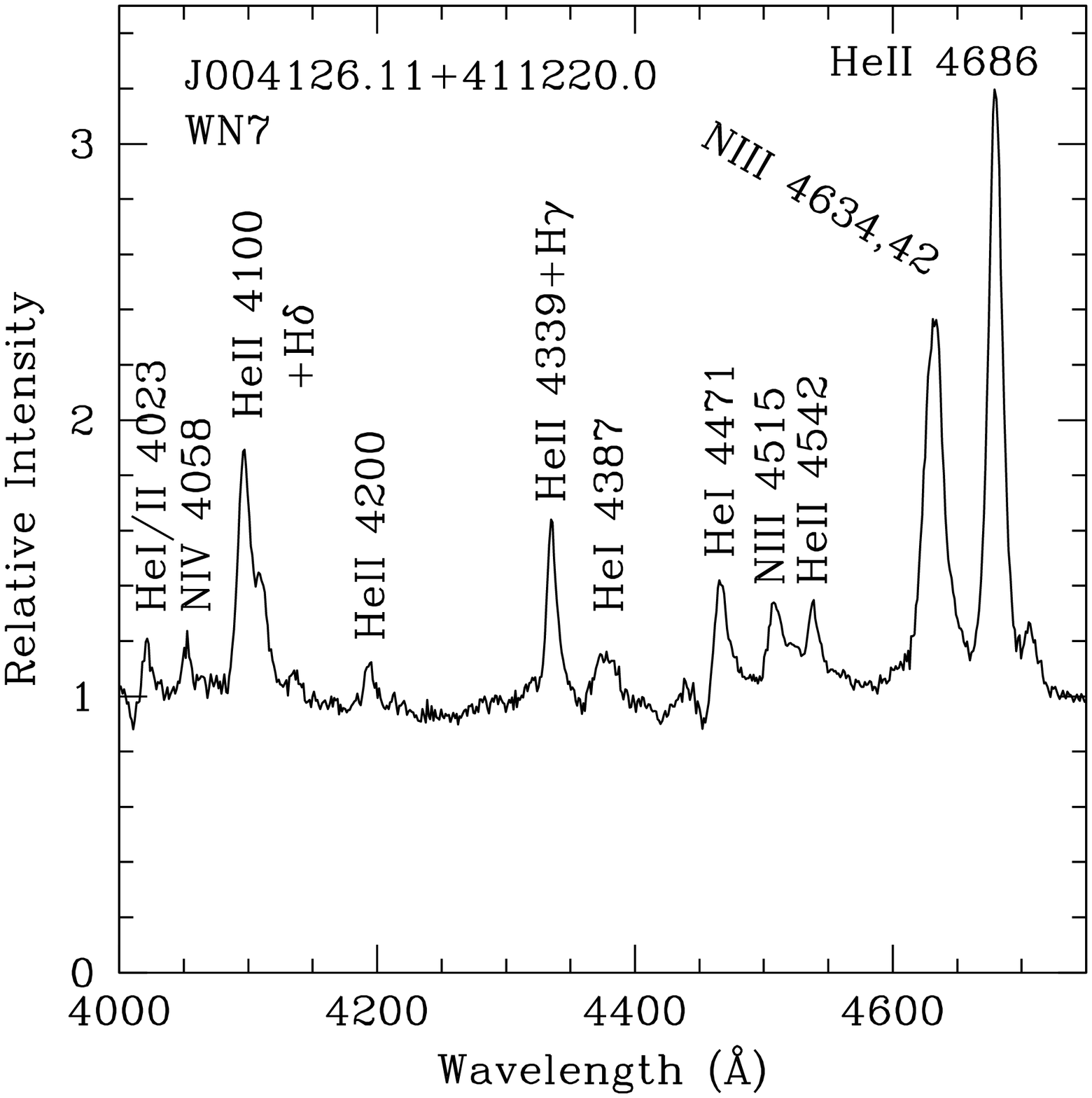}
\plotone{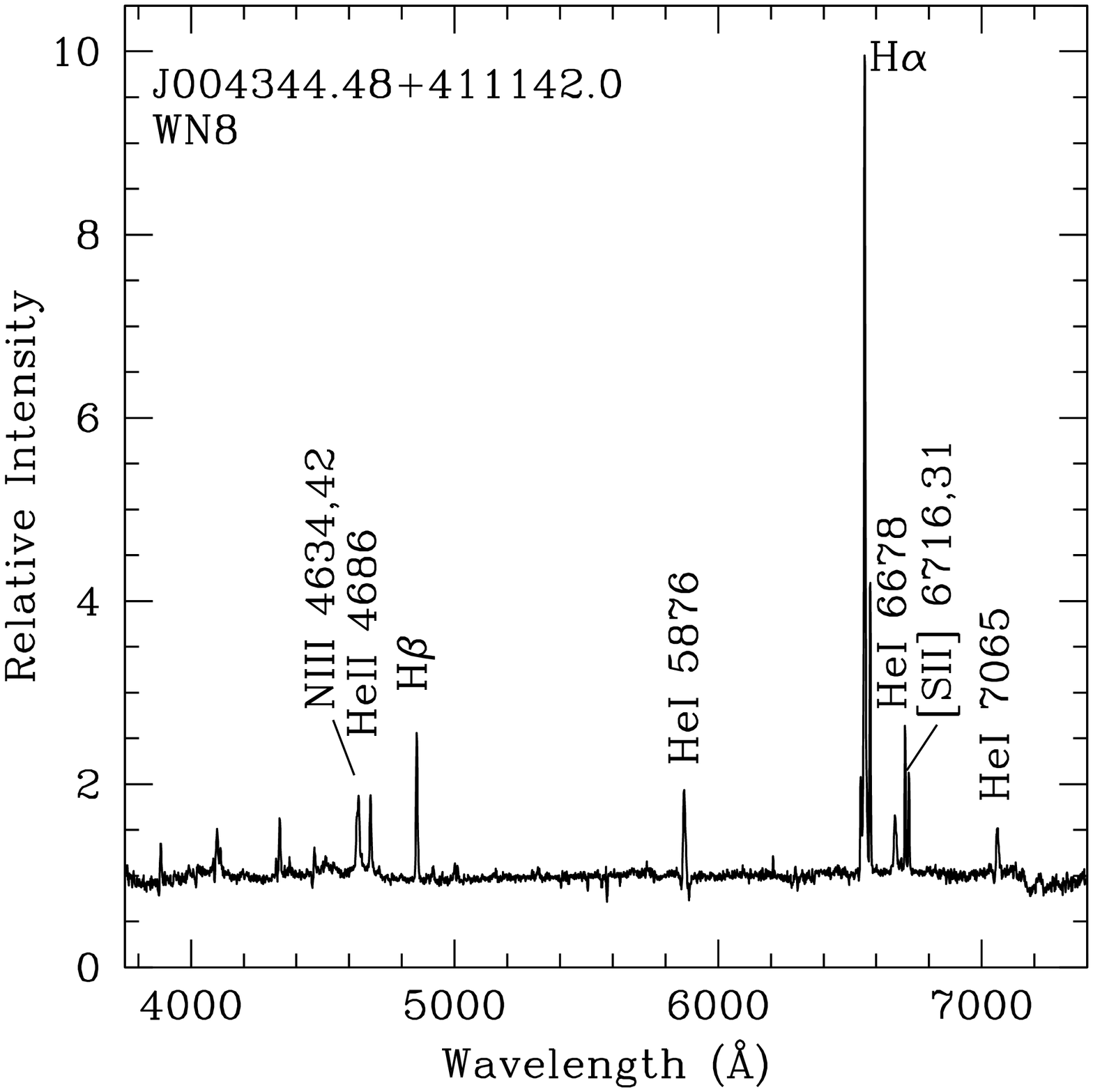}
\plotone{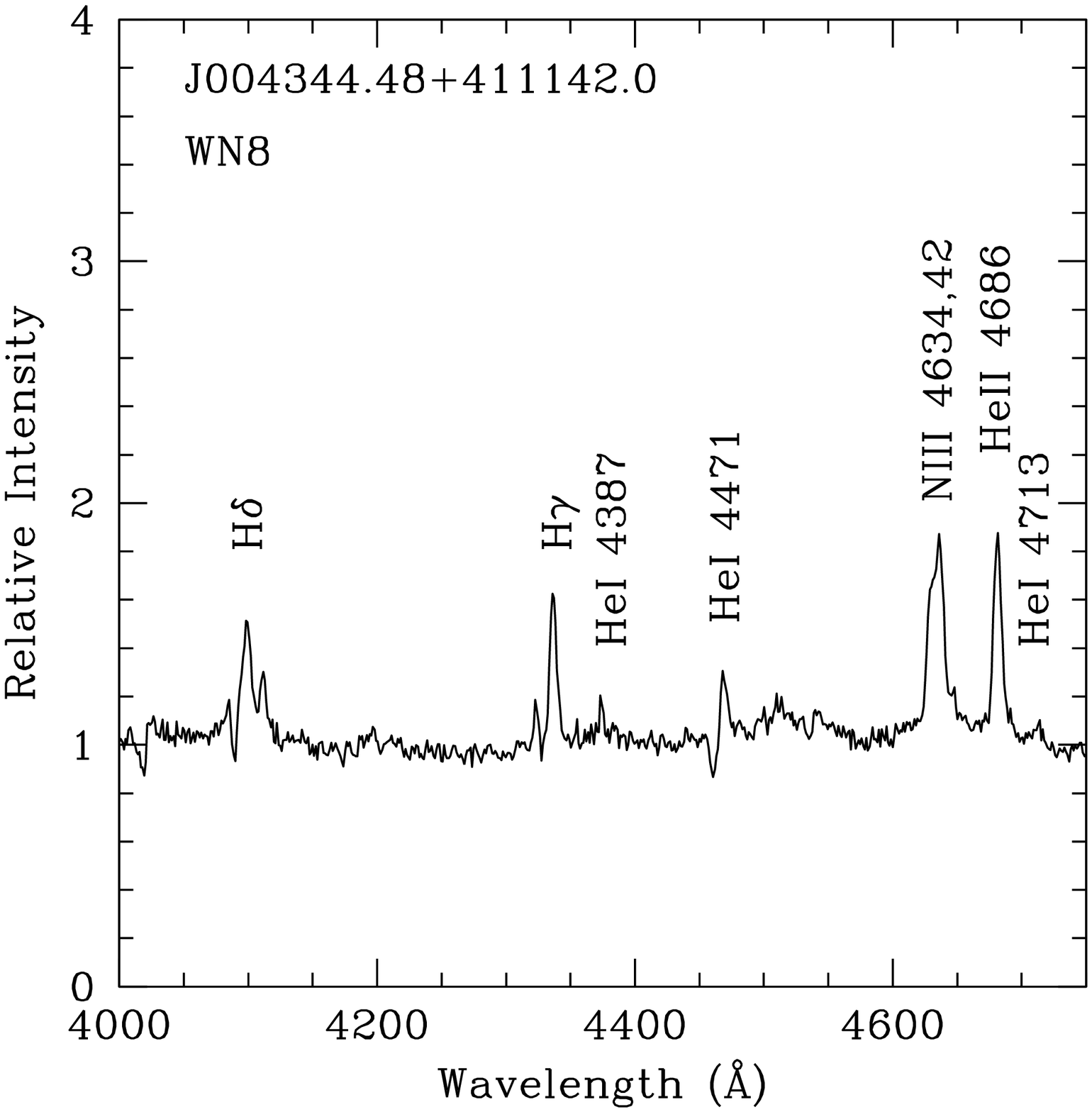}
\plotone{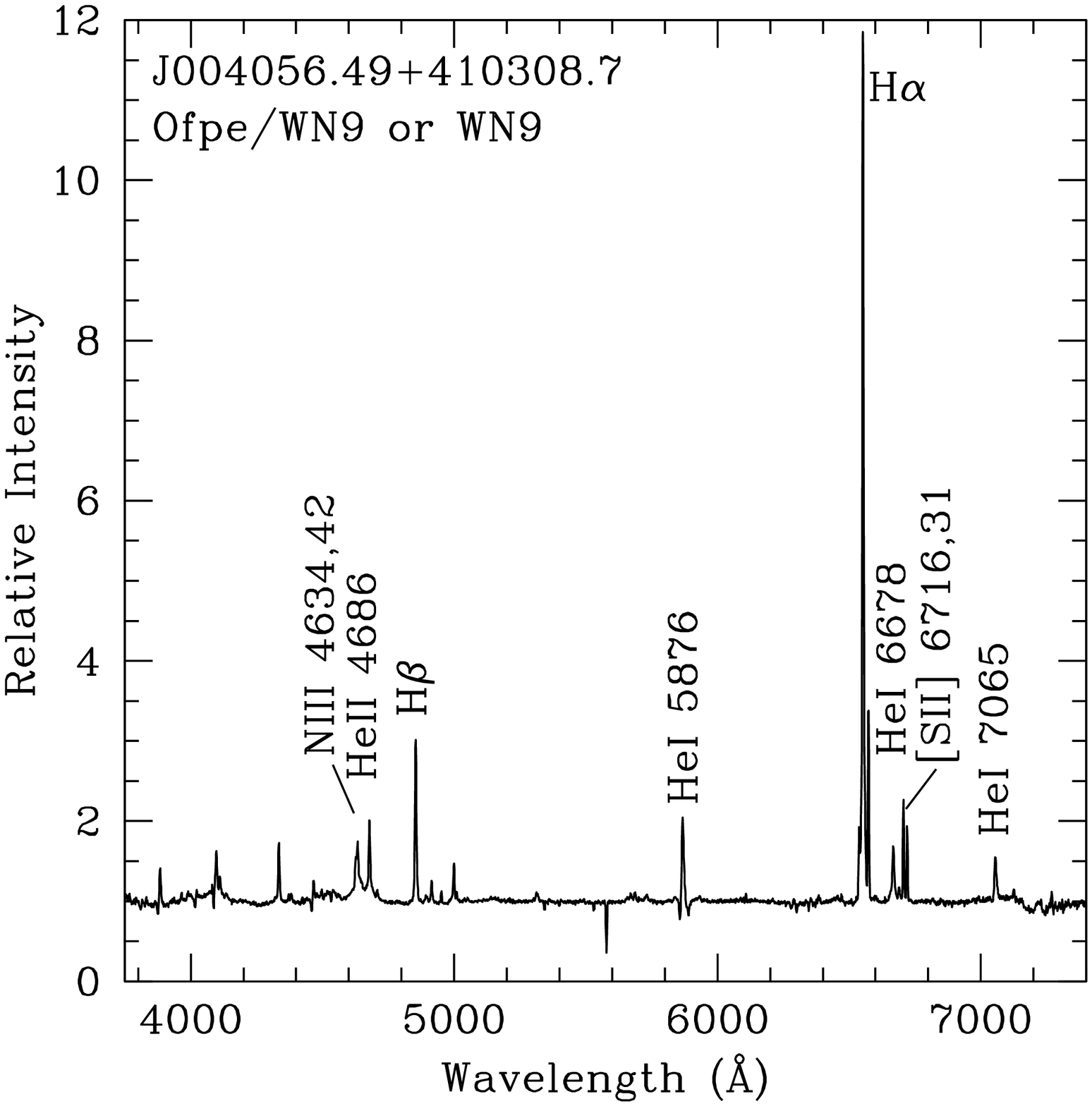}
\plotone{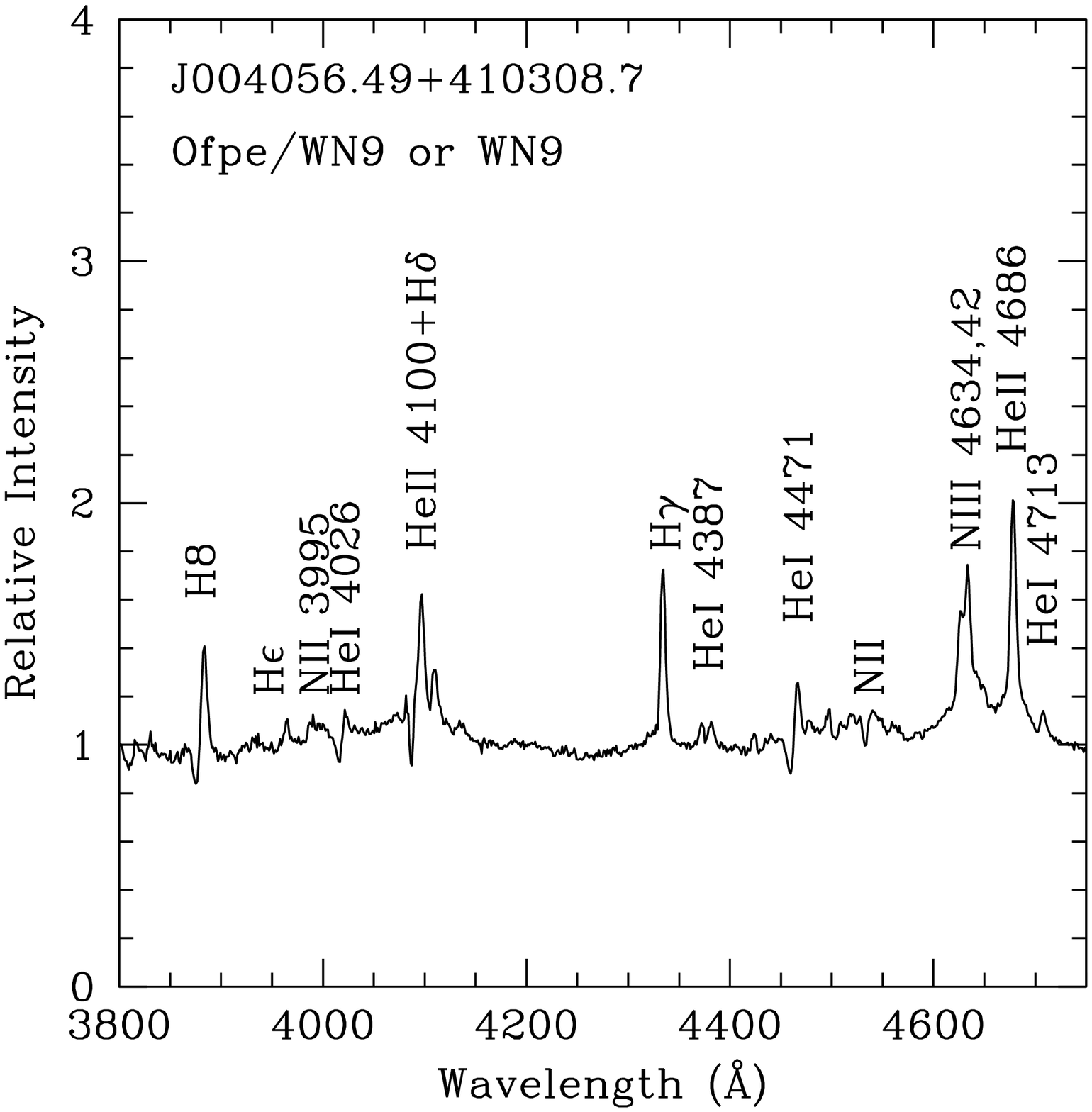}
\caption{Continued}
\end{figure}

\addtocounter{figure}{-1}
\begin{figure}
\epsscale{0.33}
\plotone{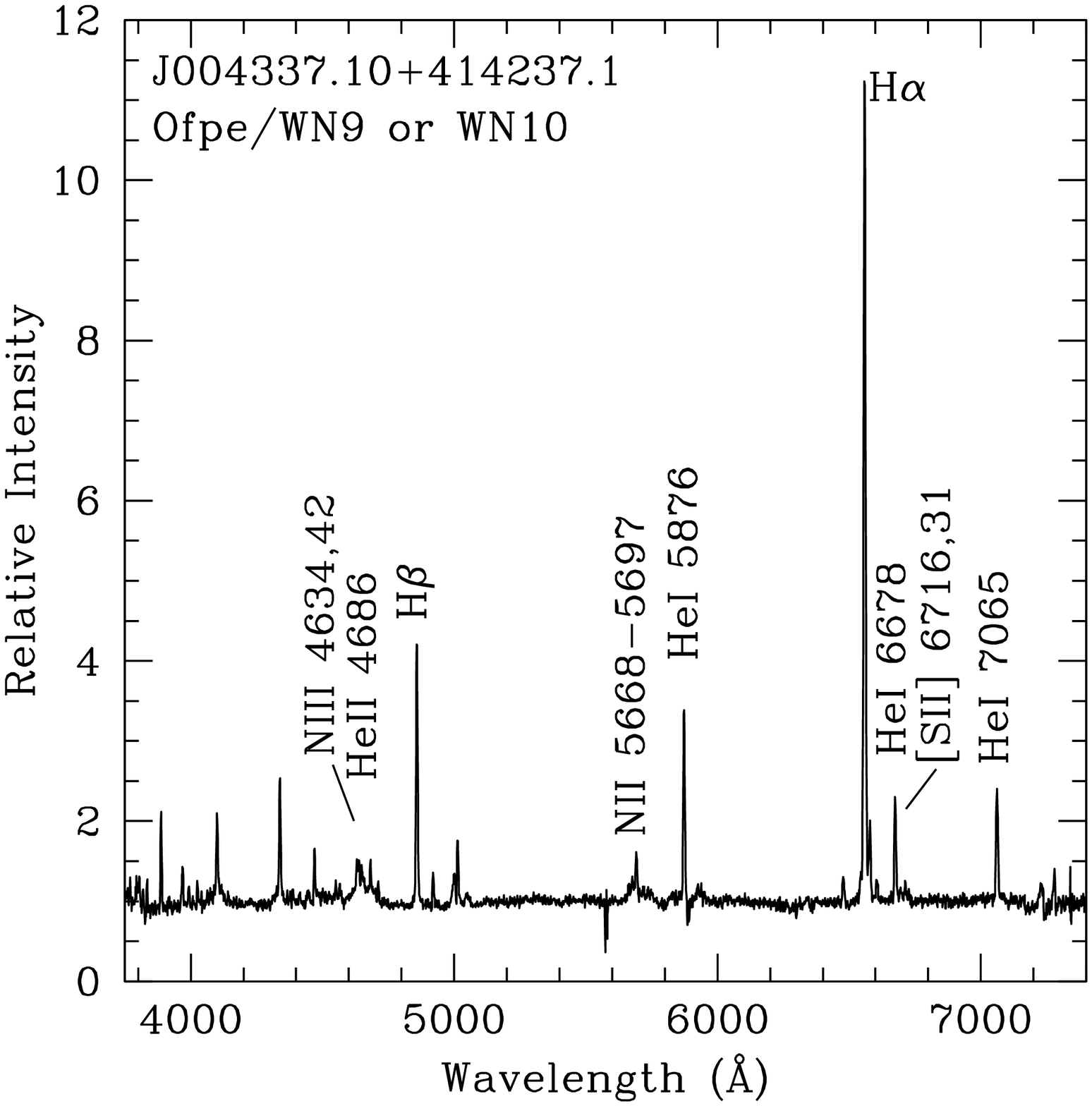}
\plotone{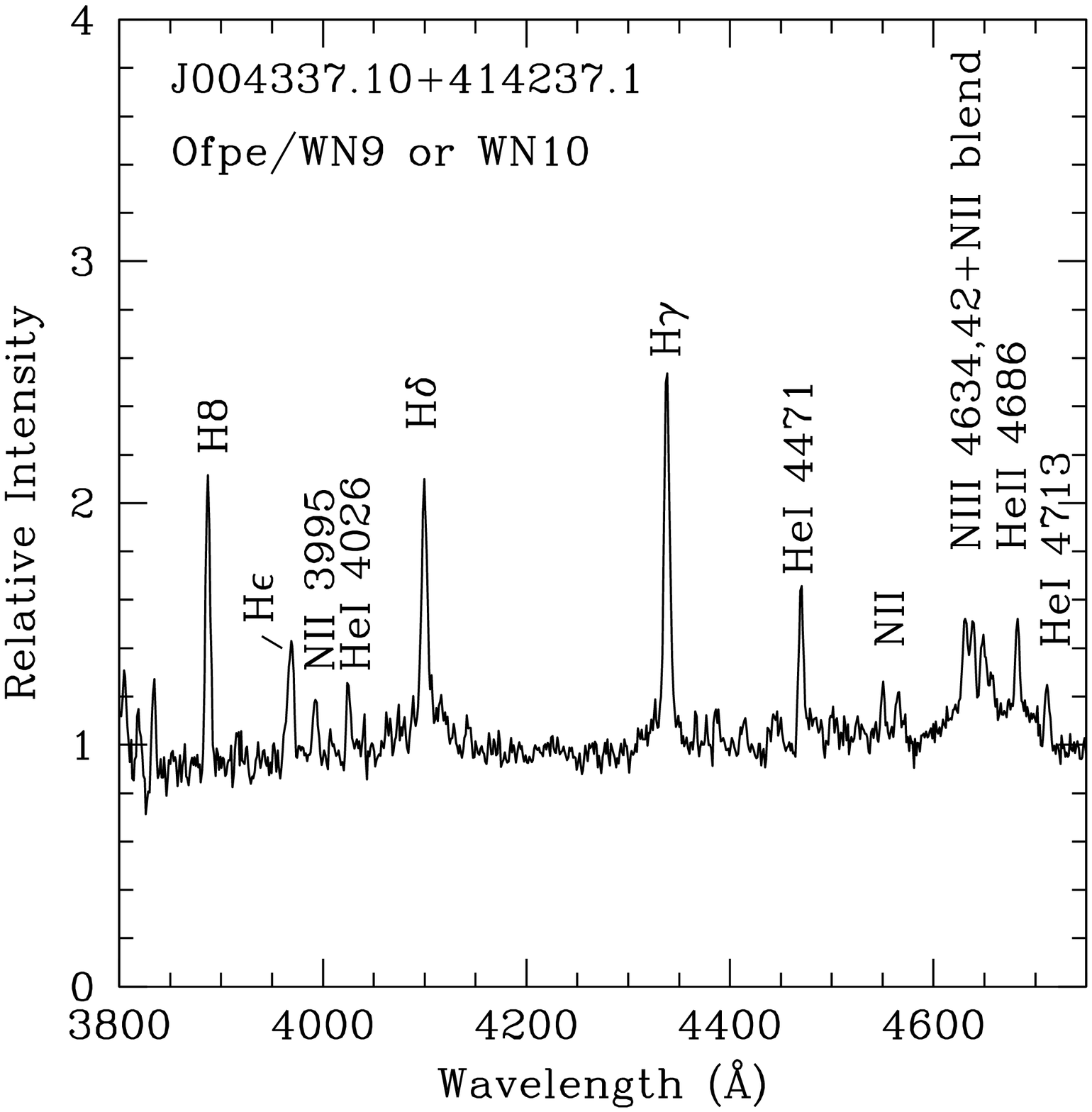}
\plotone{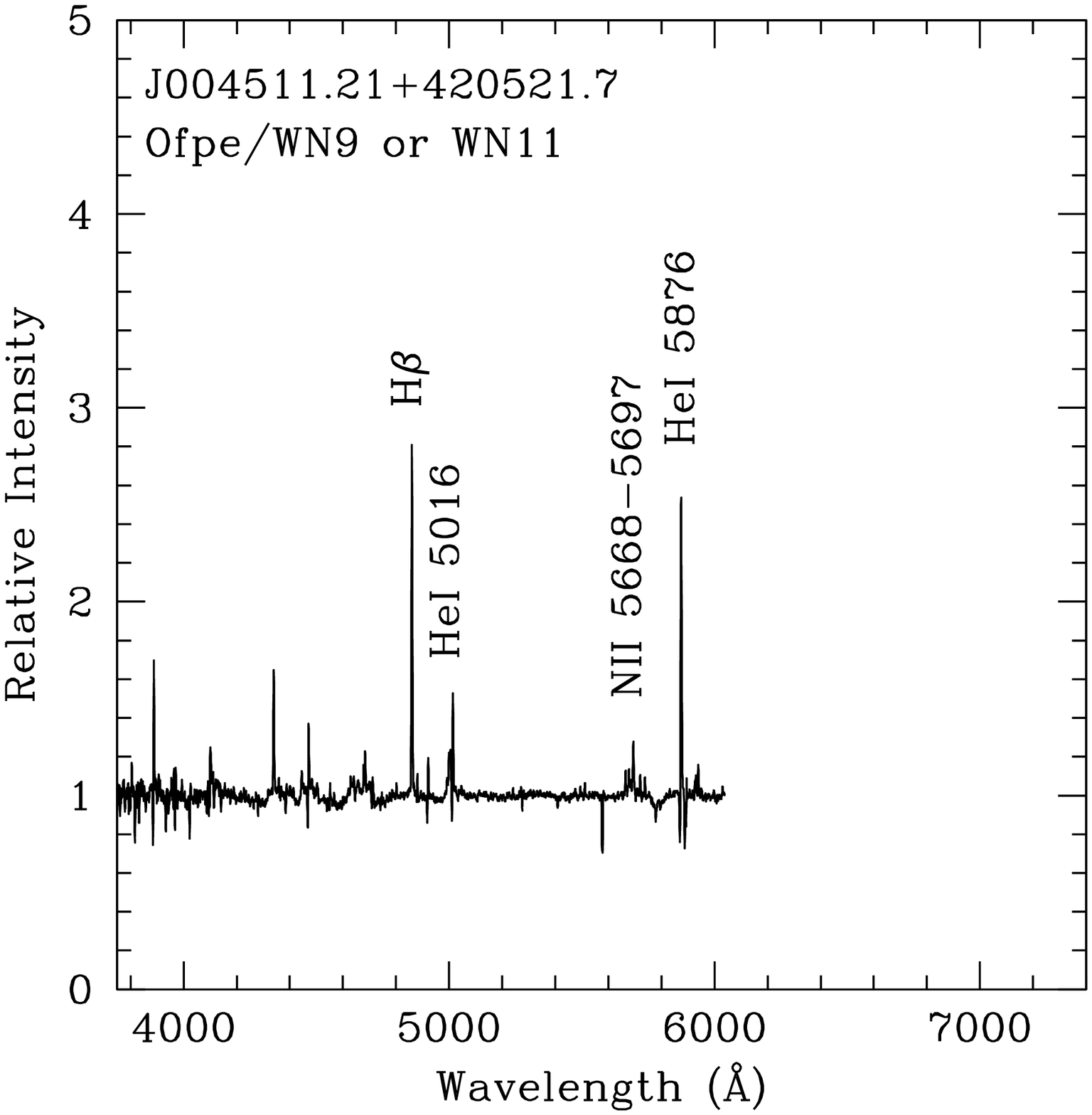}
\plotone{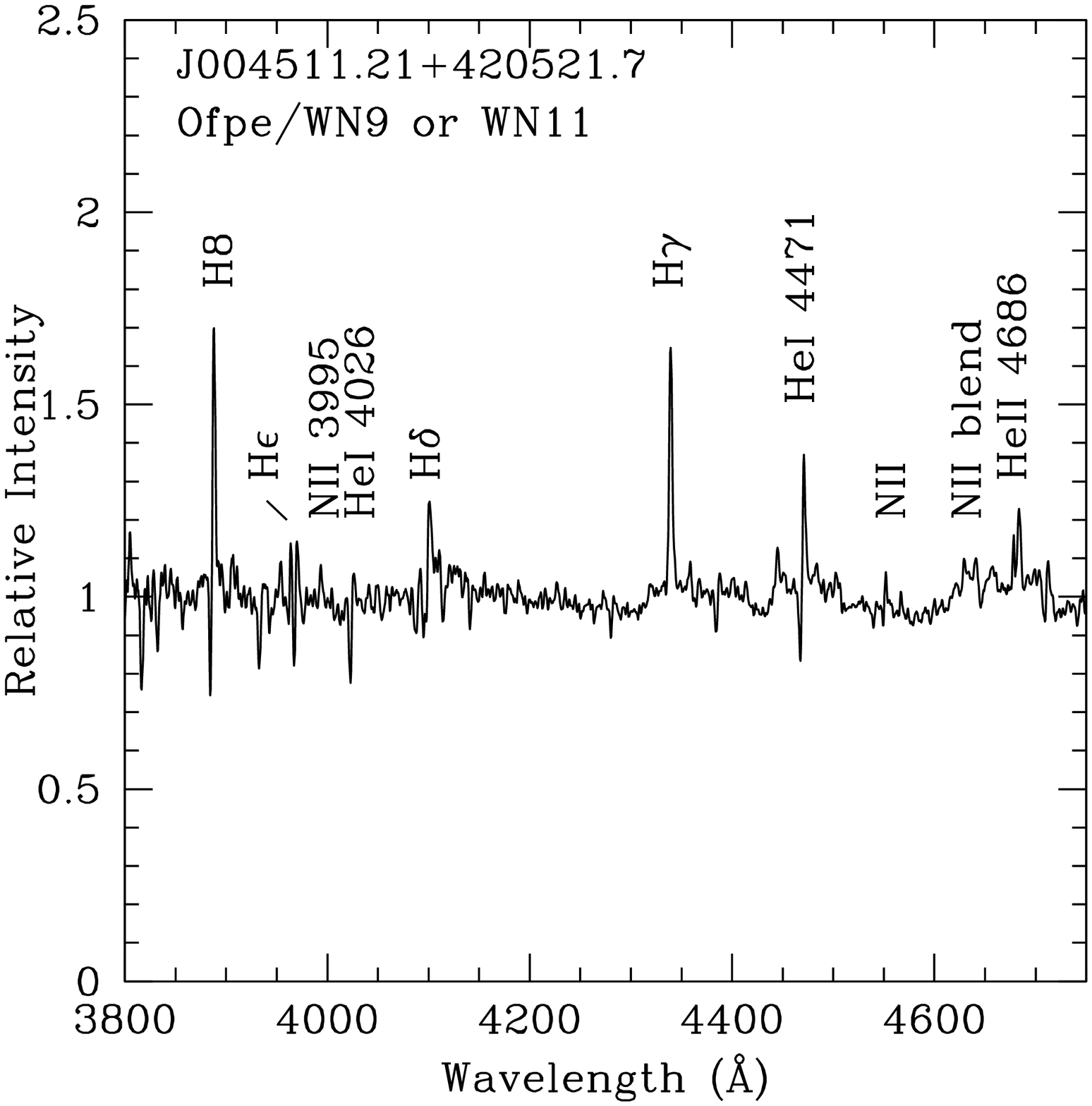}
\caption{Continued}
\end{figure}

\begin{figure}
\epsscale{0.33}
\plotone{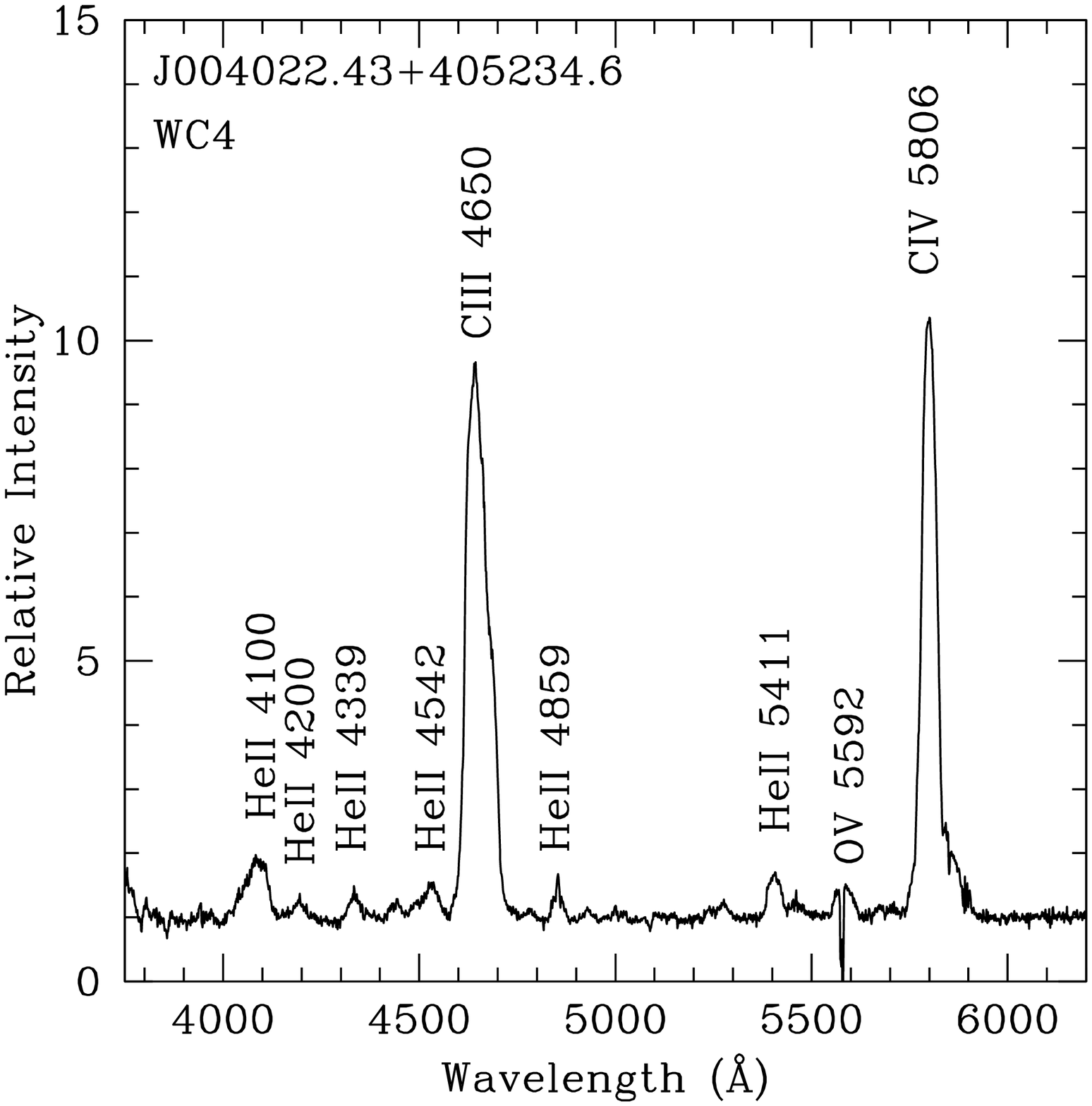}
\plotone{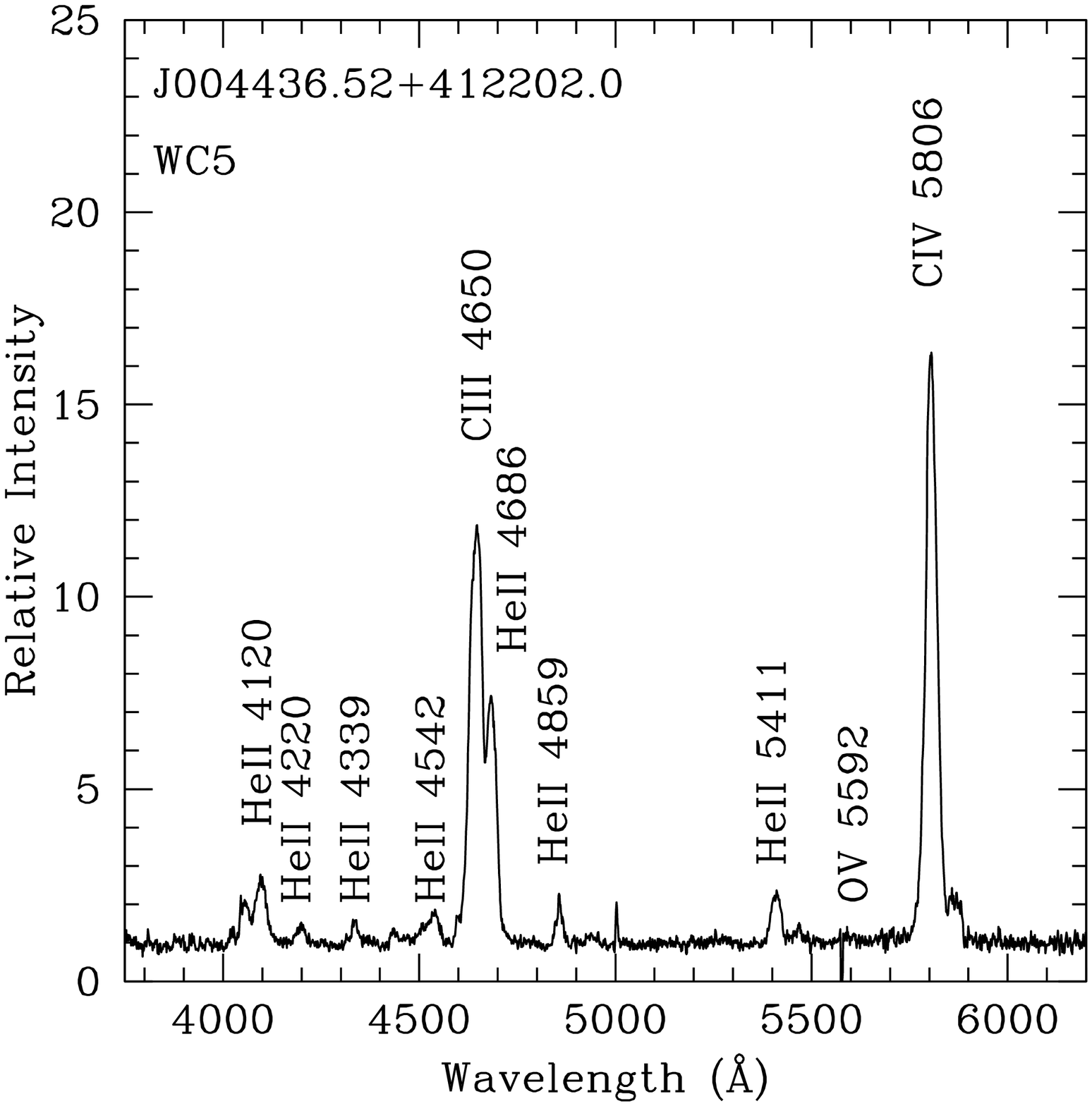}
\plotone{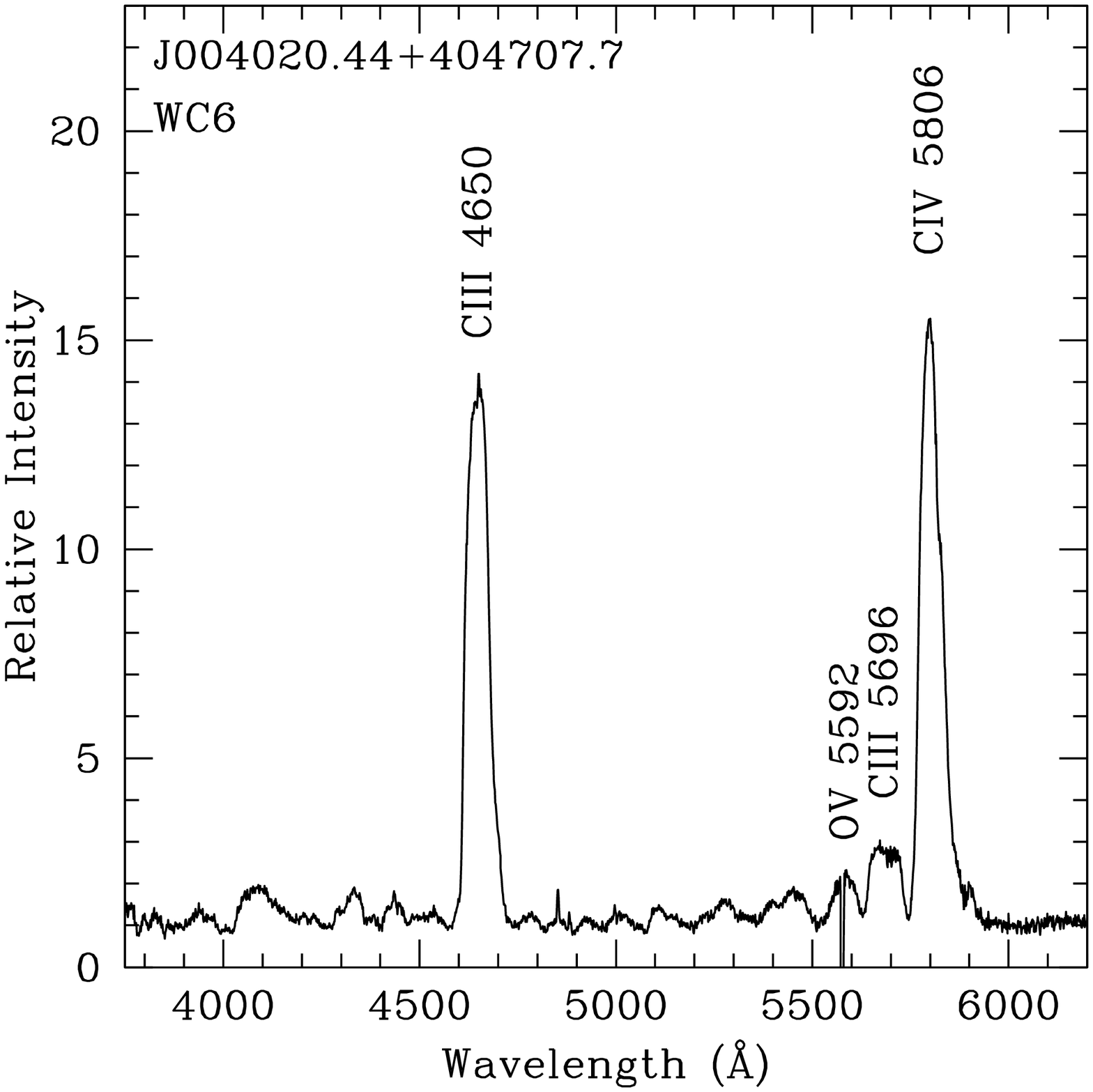}
\plotone{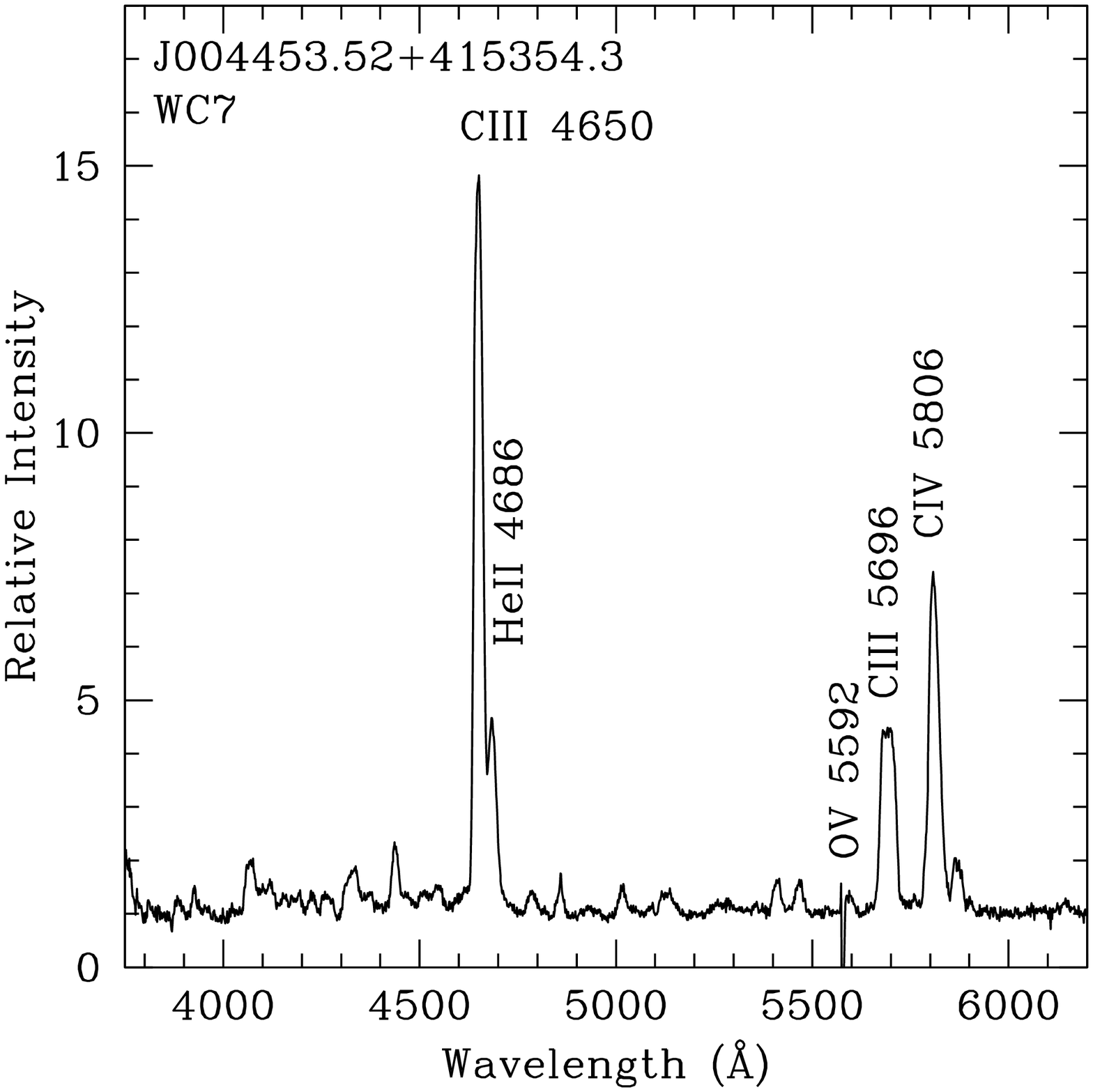}
\plotone{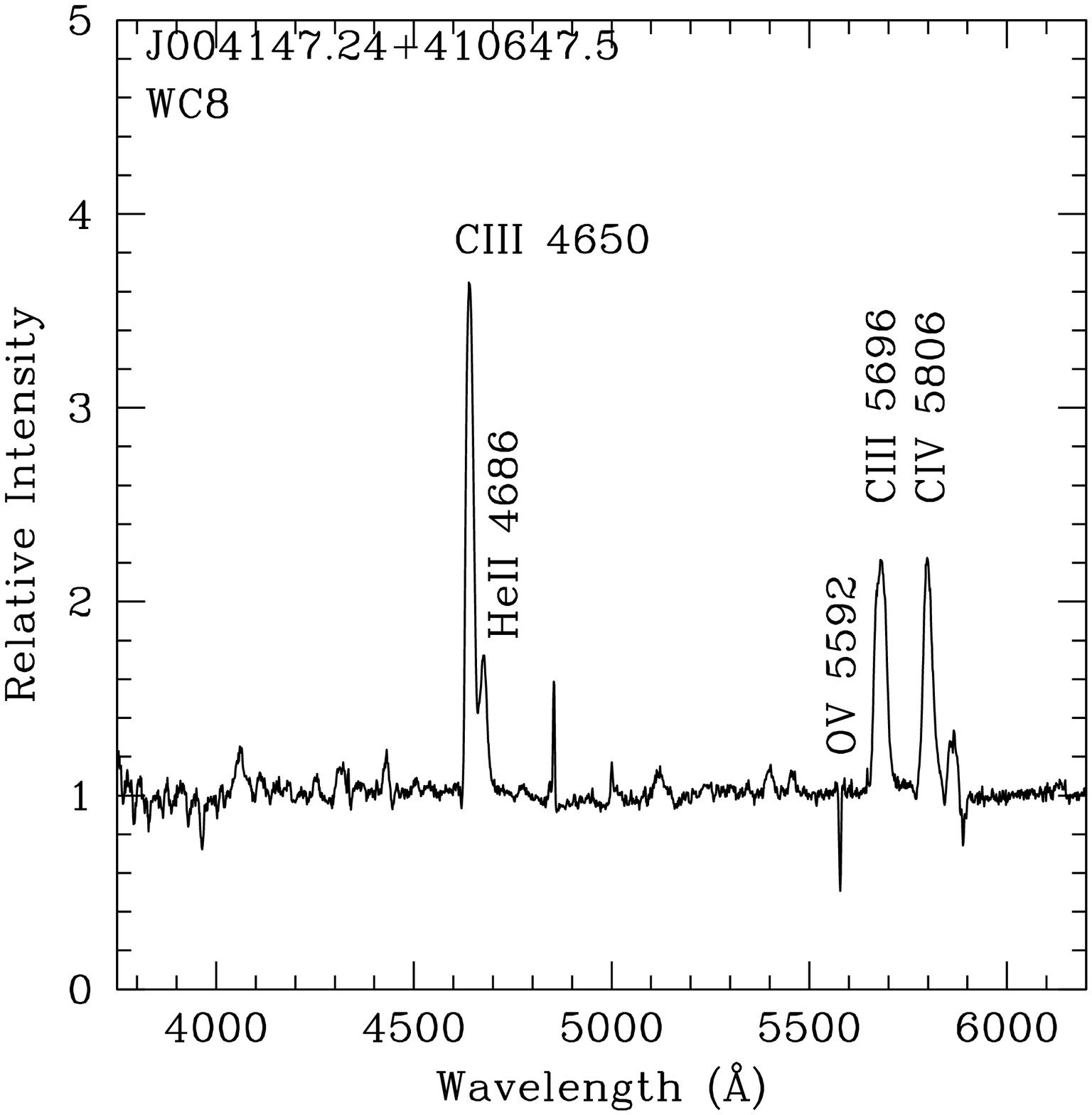}
\caption{\label{fig:wcs} The spectra of representative WC stars. The classification criteria are based primarily on the relative strengths of O V $\lambda 5592$, C III $\lambda 5696$, and C IV $\lambda 5806$.}
\end{figure}

\begin{figure}
\plotone{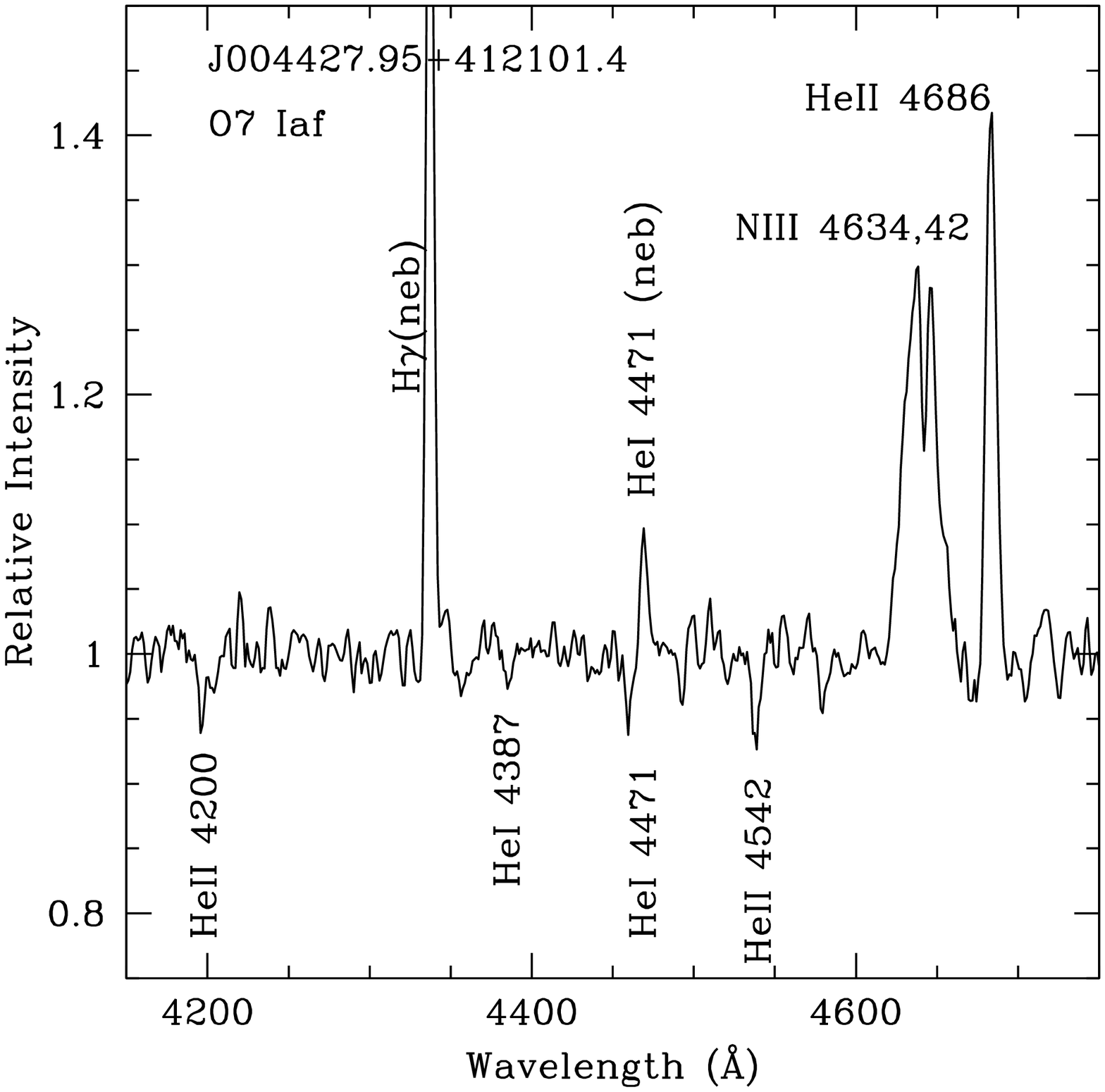}
\caption{\label{fig:Of} Spectrum of J004427.95+412101.4.  Previously called a WNL/Of star by Armandroff \& Massey (1991), we reclassify it here as O7 Iaf.  Only the primary classification region is shown, and the spectrum has been slightly smoothed by a 3 pixels boxcar function.  Strong nebular lines dominate the rest of the spectrum.}
\end{figure}

\begin{figure}
\plotone{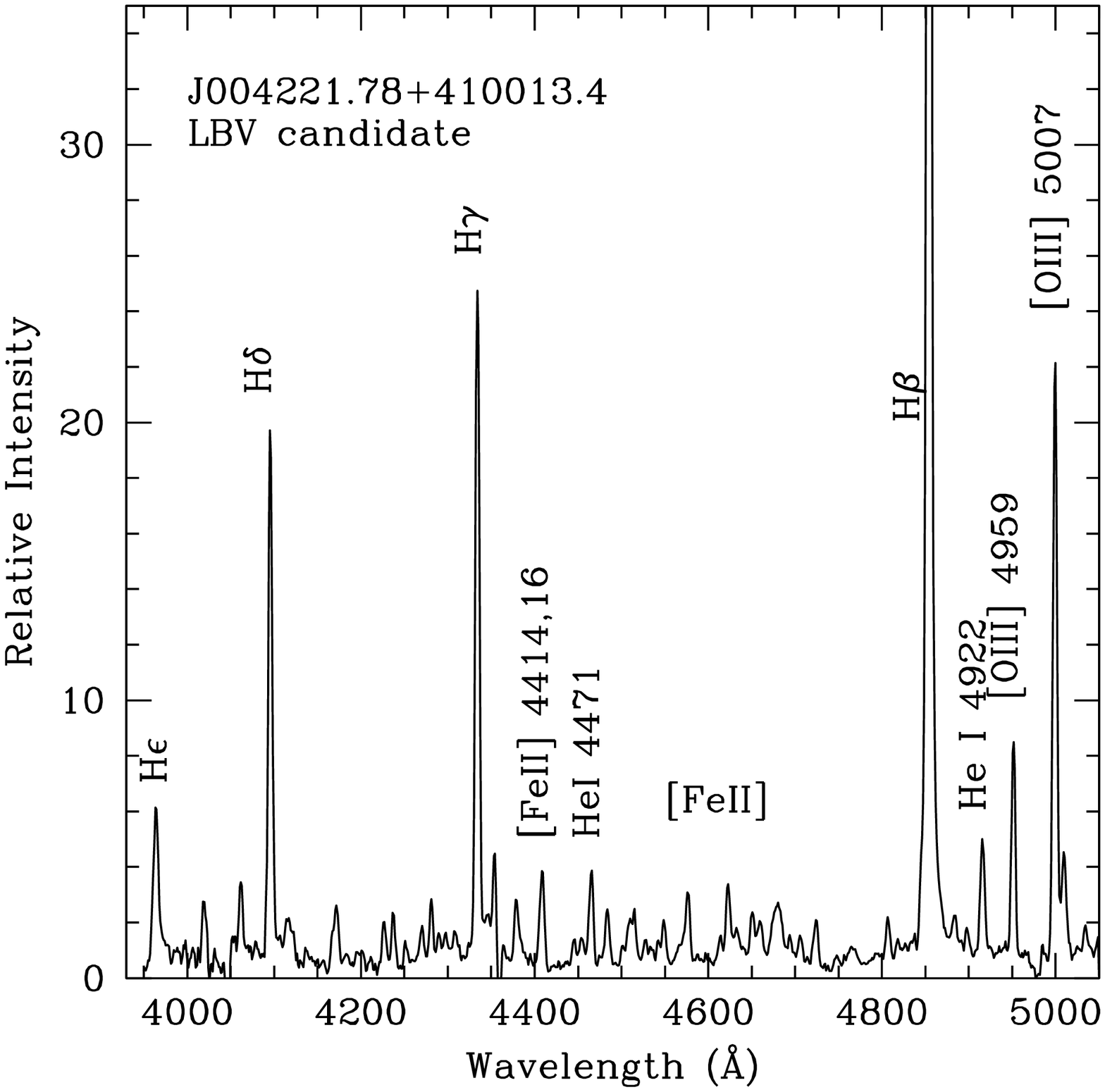}
\caption{\label{fig:LBV} Spectrum of J004221.78+410013.4.  Previously identified as a probable H$\alpha$ emission star by Massey et al.\ (2007a), our spectrum shows strong Balmer, He I, and [Fe II] emission.  The spectra is quite similar to several of the Hubble \& Sandage (1953) LBVs in their hot state (compare to Figure 10 of Massey et al.\ 2007a), and so we call this a hot LBV candidate.}
\end{figure}

\begin{figure}
\epsscale{0.48}
\plotone{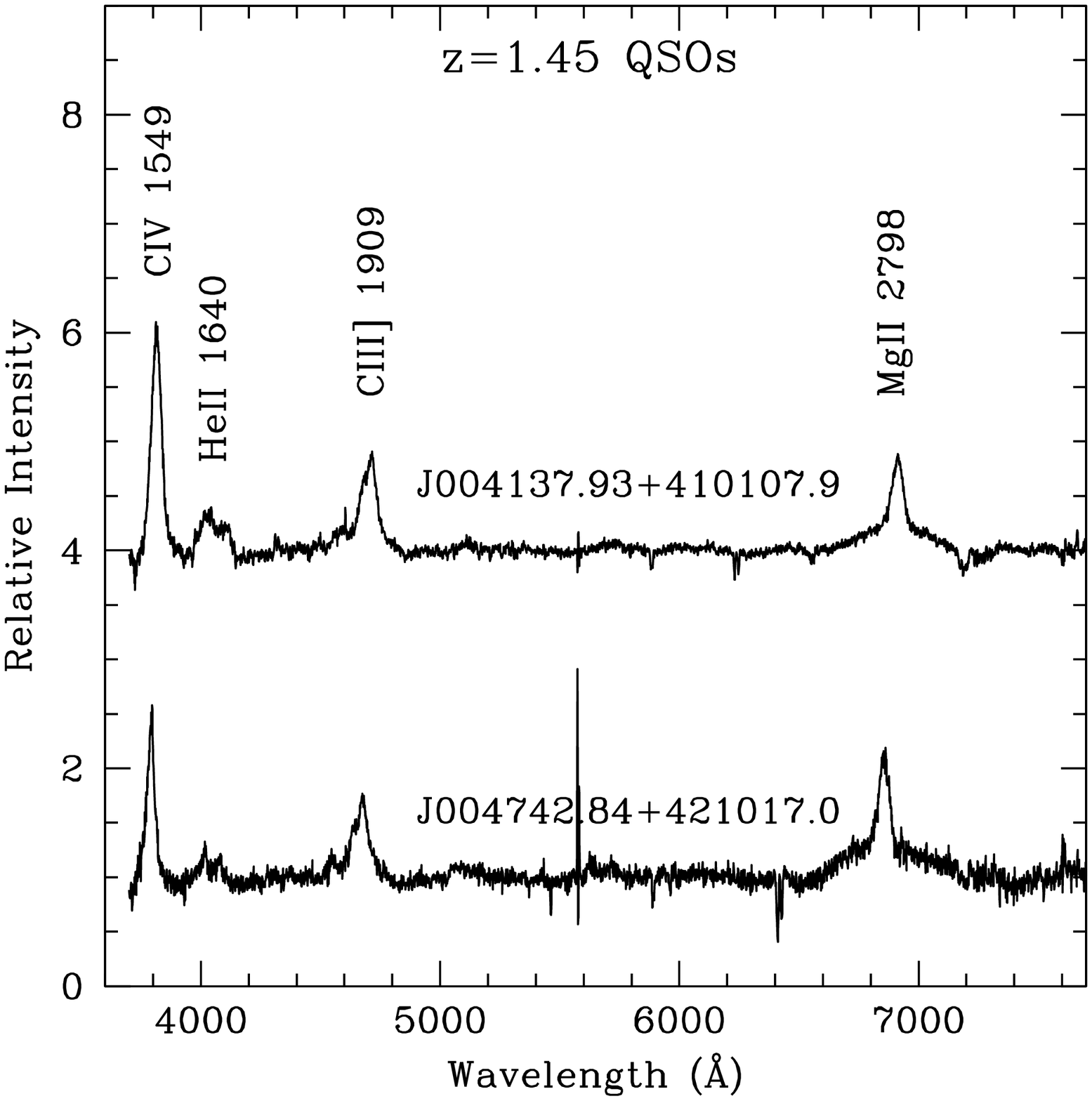}
\plotone{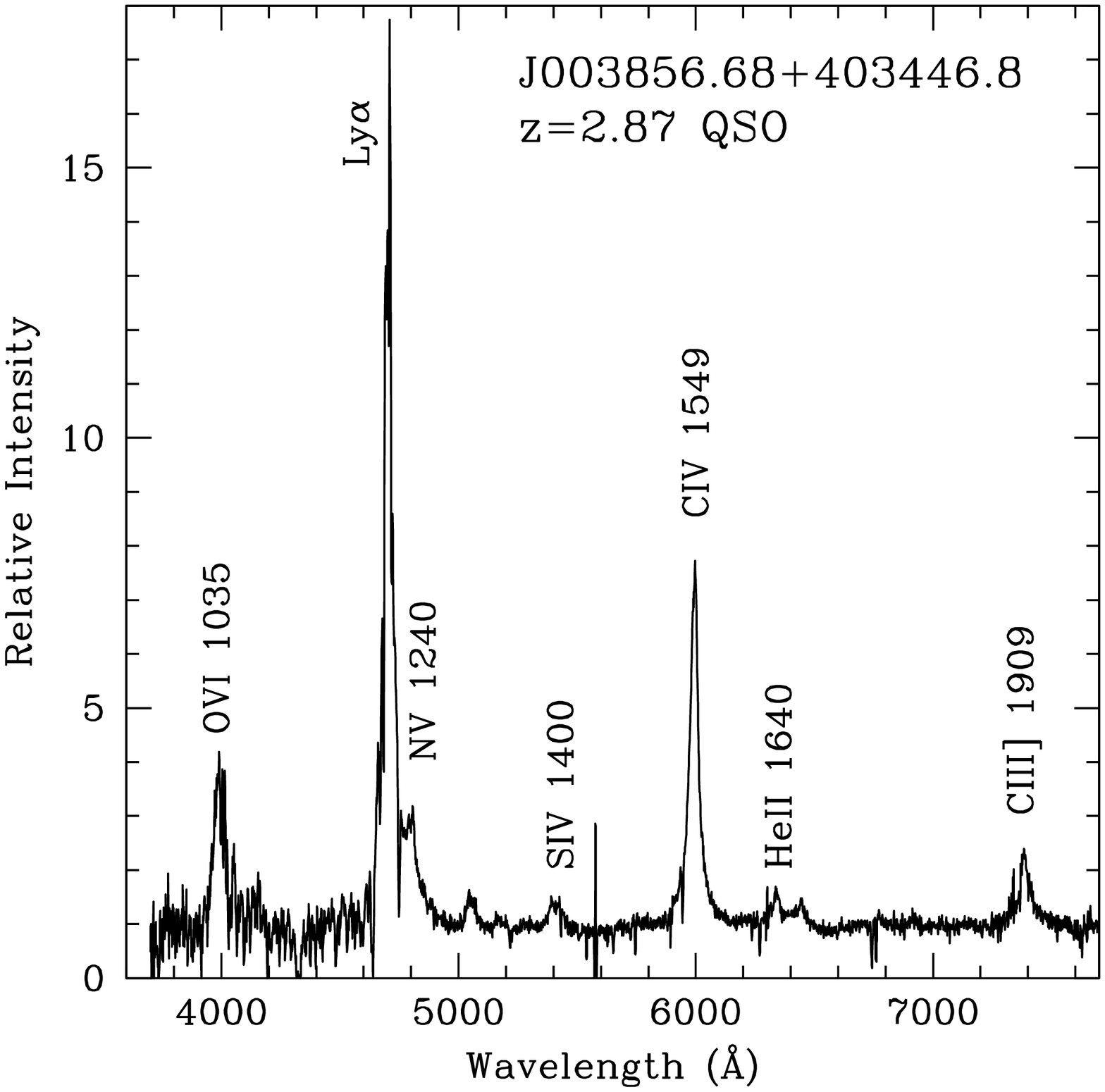}
\caption{\label{fig:qsos} The spectra of three QSOs.  The two on the left have a redshift of 1.45; the one on the right has a redshift of 2.87.  Note that in all three cases strong emission has been shifted in to the WR detection bands at 4650 -- 4690~\AA.}
\end{figure}

\begin{figure}
\epsscale{1}
\plotone{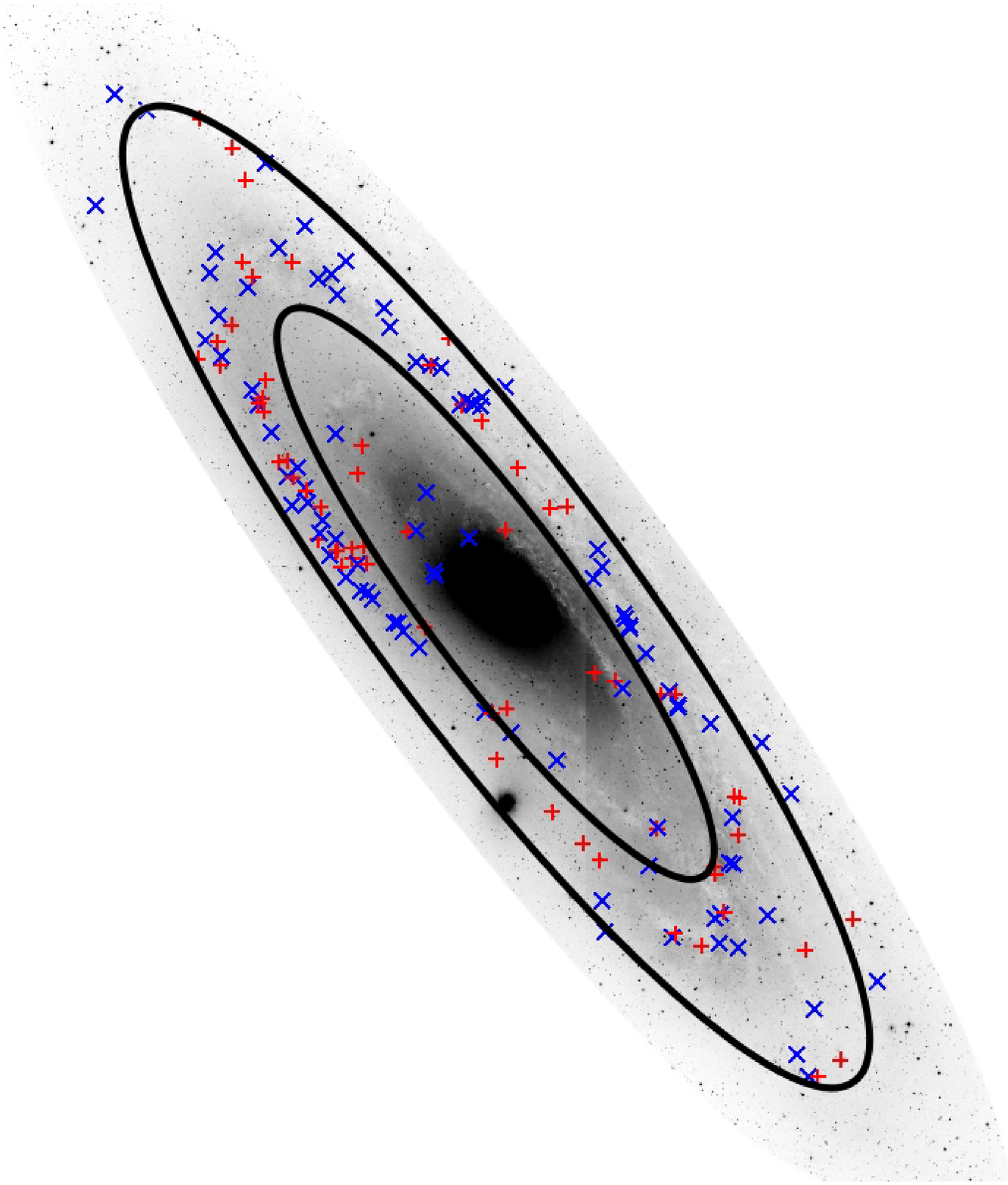}
\caption{\label{fig:WRlocations} Locations of all known WR stars in M31. The blue $\times$s represent WN stars while the red +s represent WC stars. The inner black ellipse is at 9 kpc ($\rho=0.43$) within the plane of M31, and the outer one is at 15 kpc ($\rho=0.71$). Our survey extended well beyond the region shown here (see Figure 1 in Massey et al.\ 2006), but no WRs were found in these outer regions.}
\end{figure}

\begin{figure}
\epsscale{1}
\plotone{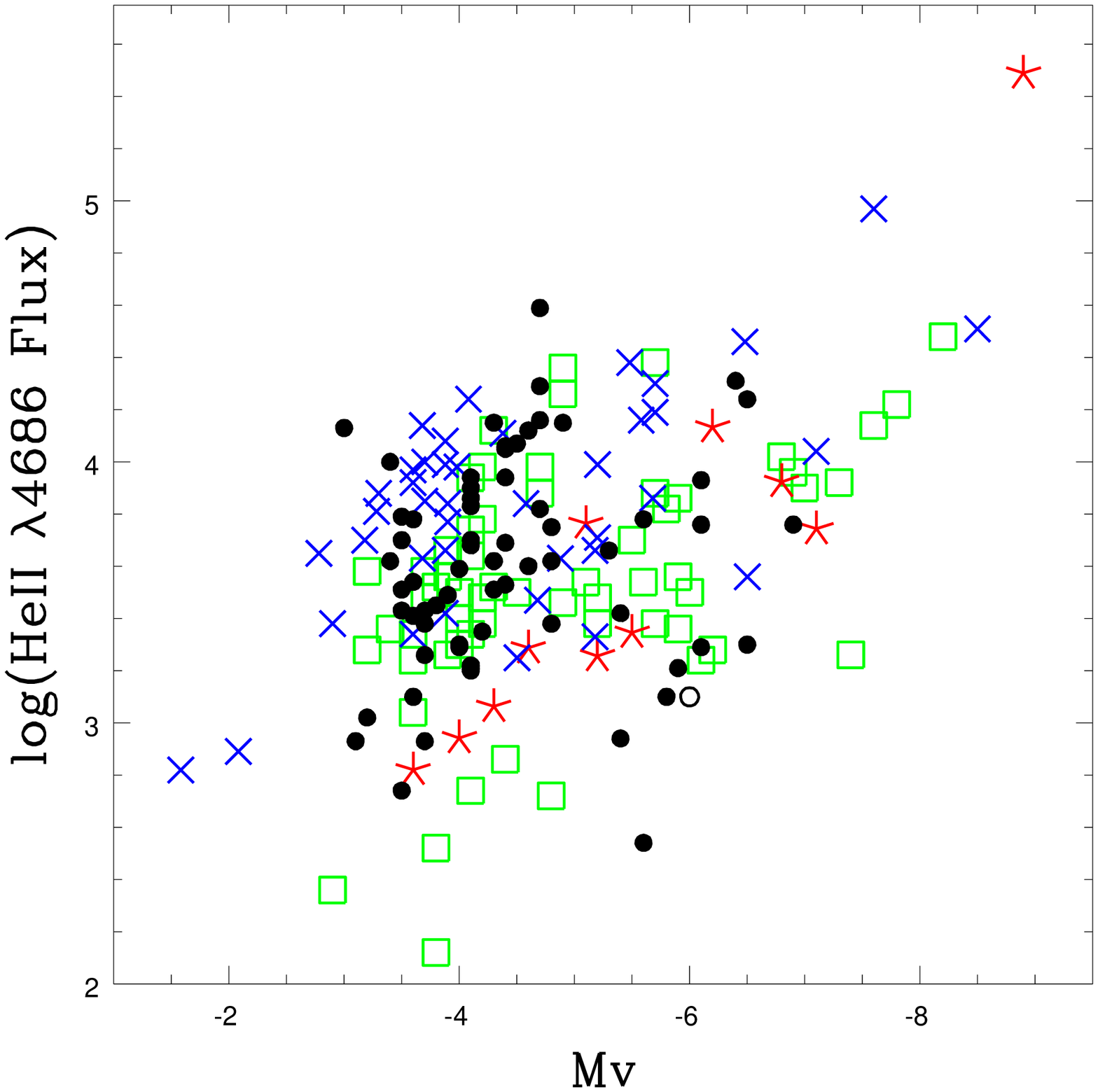}
\caption{\label{fig:complete} Sensitivity of the M31 survey. Here we plot the log of the flux in the He II $\lambda$4686 emission line against absolute visual magnitude $M_V$ for WNs. We have approximated the line flux as log(-EW) $- M_V/2.5.$ The observed M31 WNs are denoted by solid circles and the one Of-type star is denoted by an open circle. The SMC data are plotted as red asterisks, the LMC as blue $\times$s and M31 as green boxes. These values are shown for comparison and come from Massey et al.\ (2003), Conti \& Massey (1989) and Neugent \& Massey (2011), respectively. The locations of the solid circles show that this survey detected WNs with line fluxes as weak or weaker than the WNs known in M33 and the Magellanic Clouds.}
\end{figure}

\begin{figure}
\epsscale{0.4}
\plotone{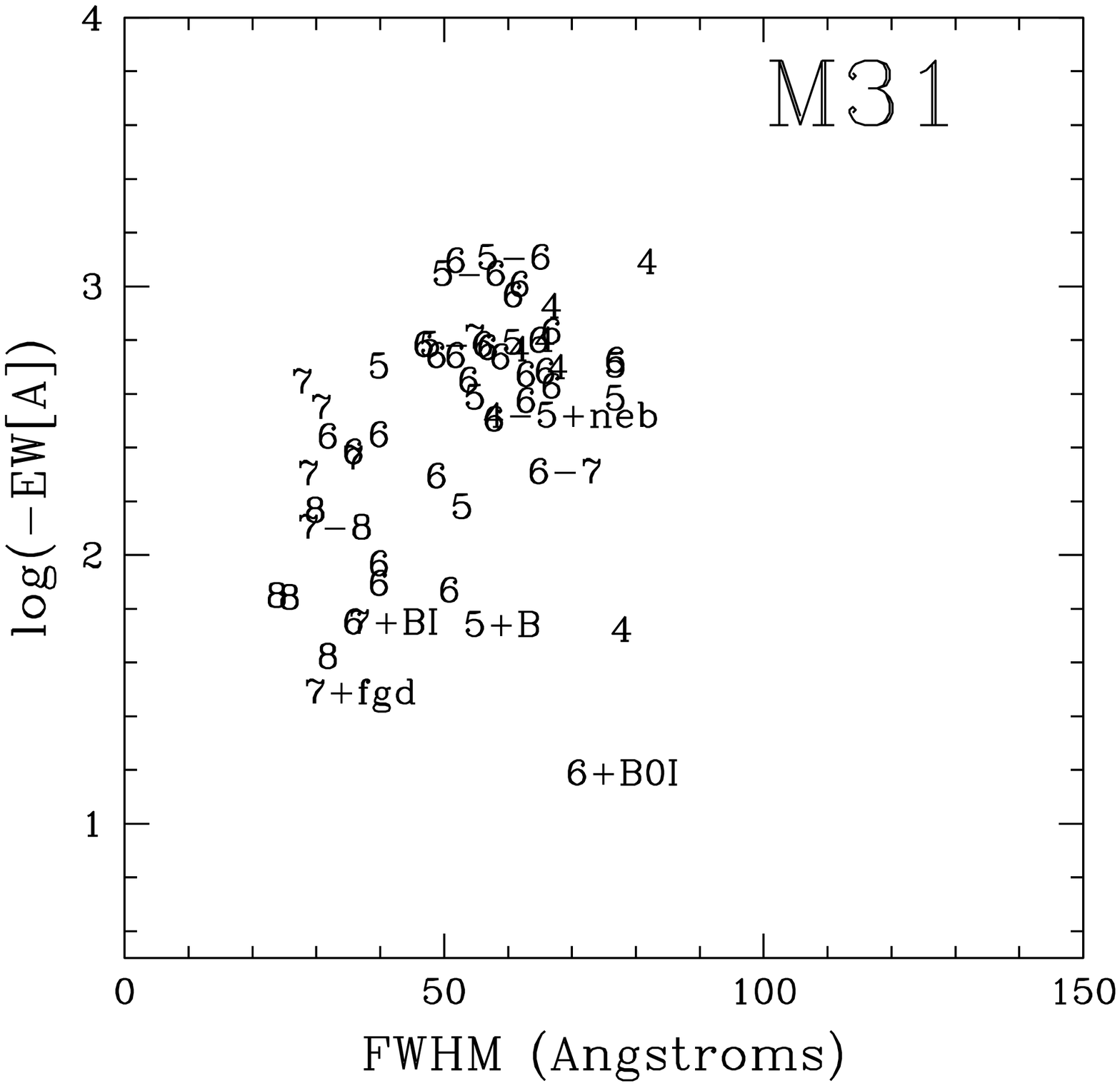}
\plotone{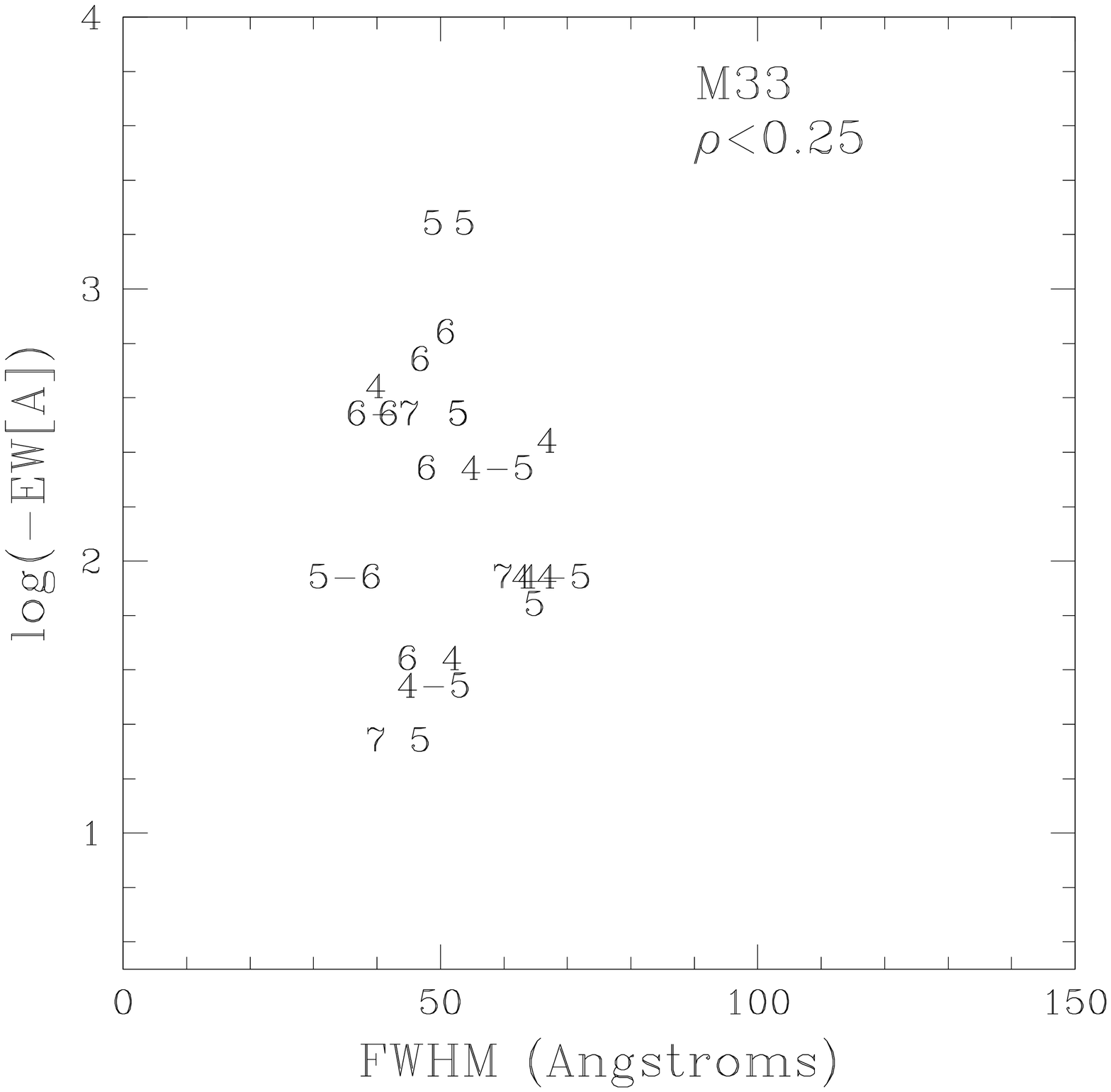}
\plotone{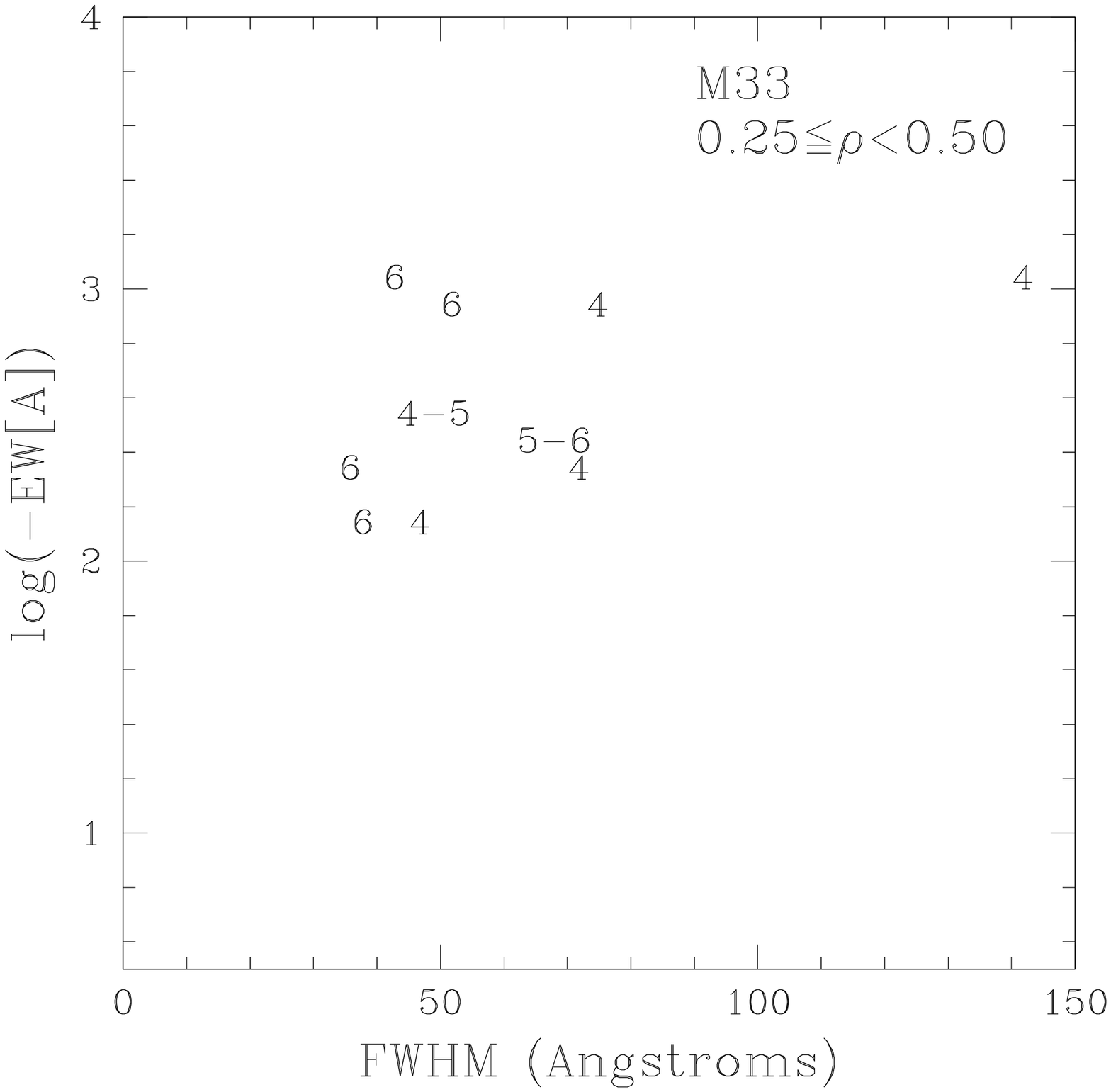}
\plotone{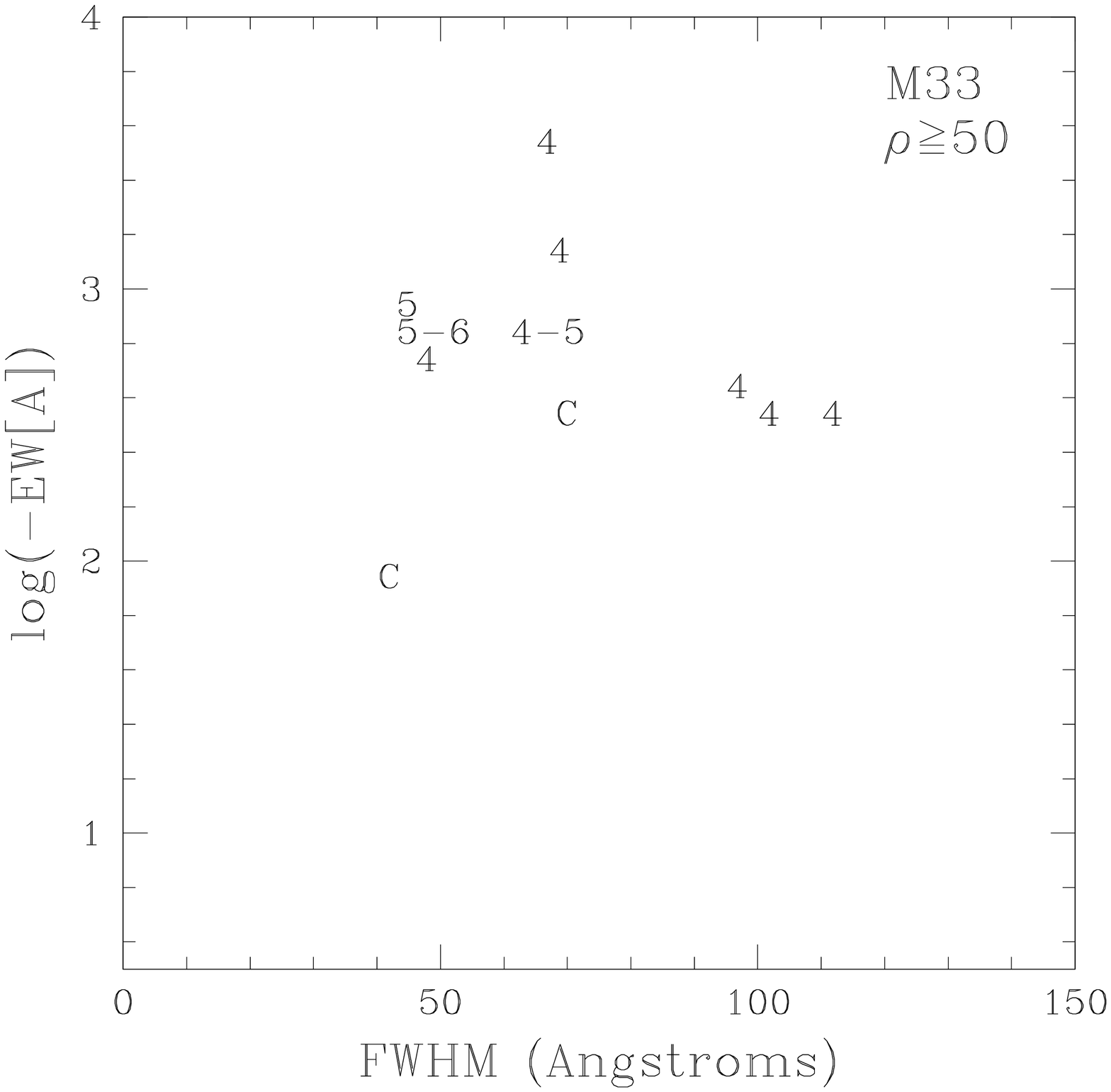}
\caption{\label{fig:hockeystick} Line strength vs.\ line width for M31 and M33 WCs. The C~IV $\lambda$5806 line strength is plotted against its line width for all observed M31 WC stars. The stars are represented in the figure by their subtypes. The M31 diagram shows points further to the left and of later types than in the inner portion of M33, suggesting its metallicity is higher than in the inner region of M33, thought to be solar.}
\end{figure}

\begin{figure}
\epsscale{0.3}
\plotone{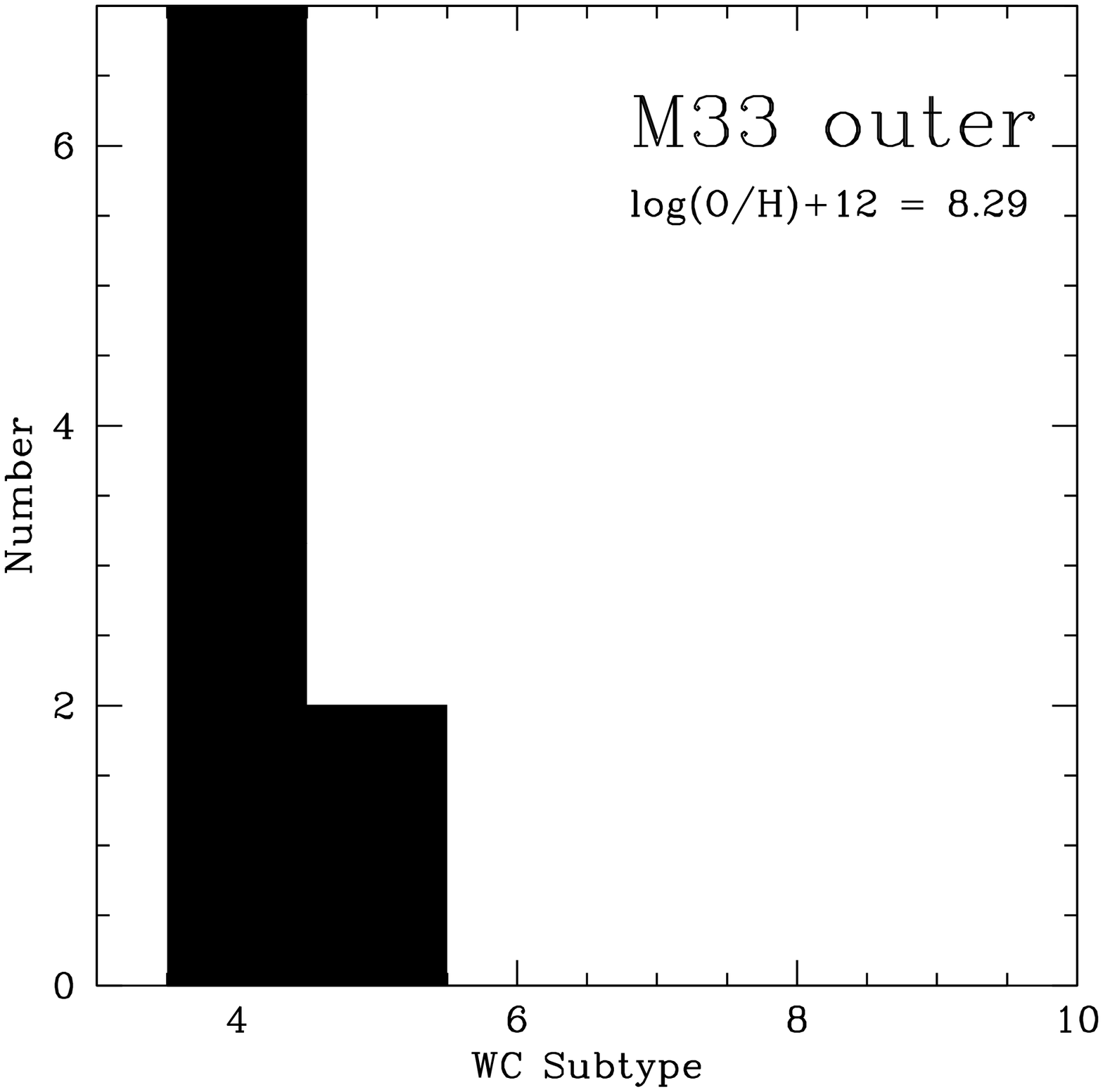}
\plotone{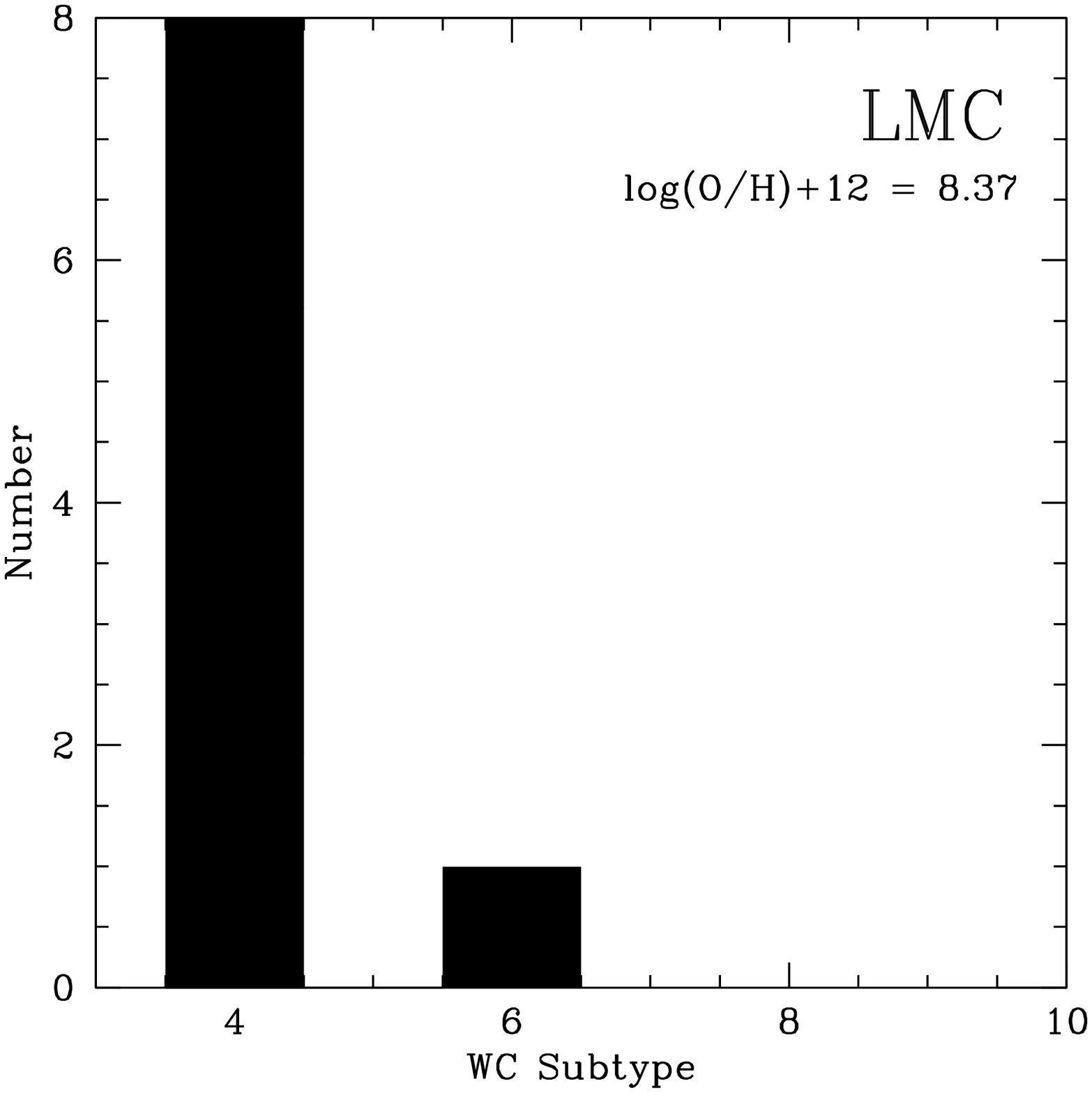}
\plotone{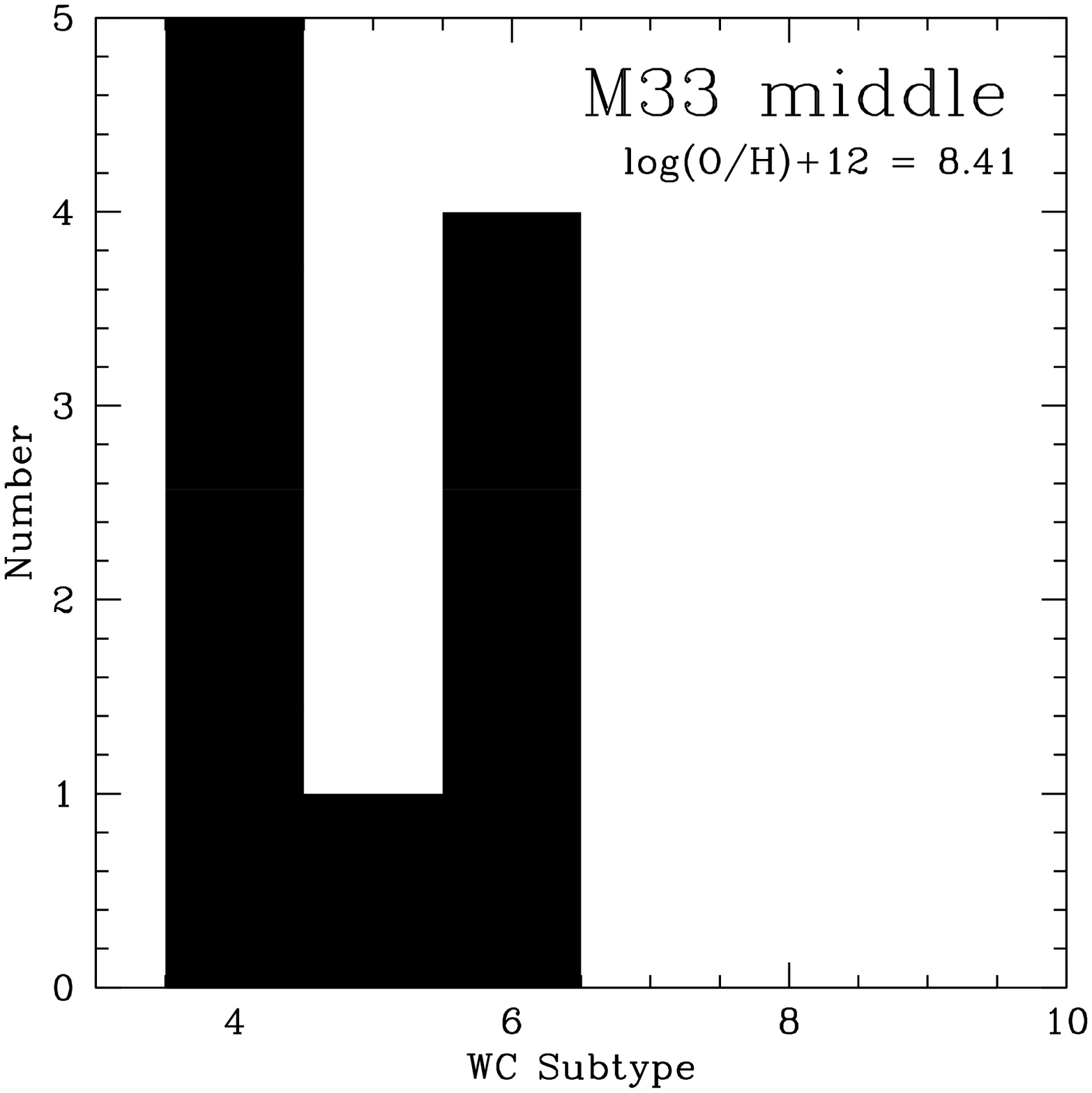}
\plotone{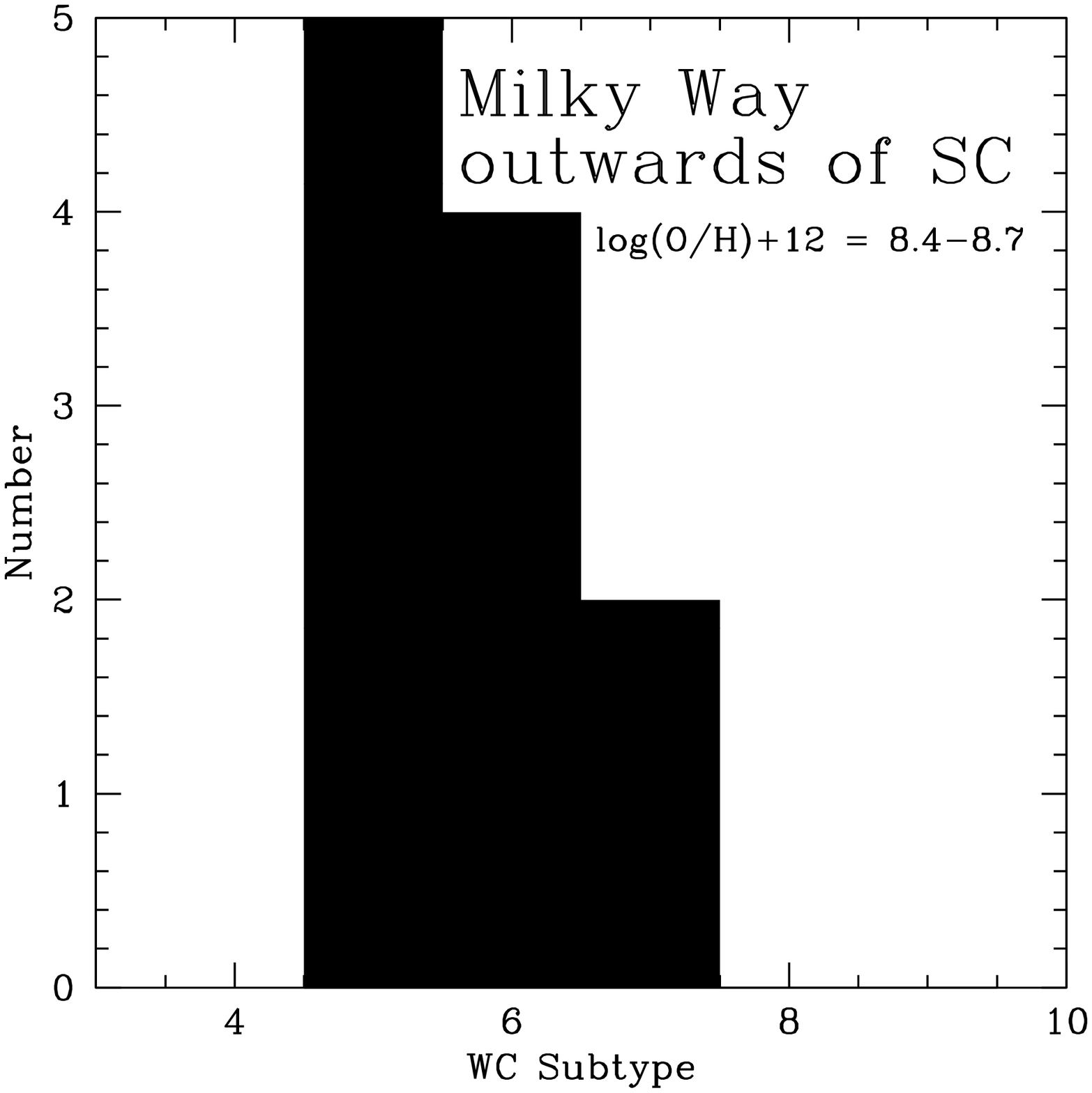}
\plotone{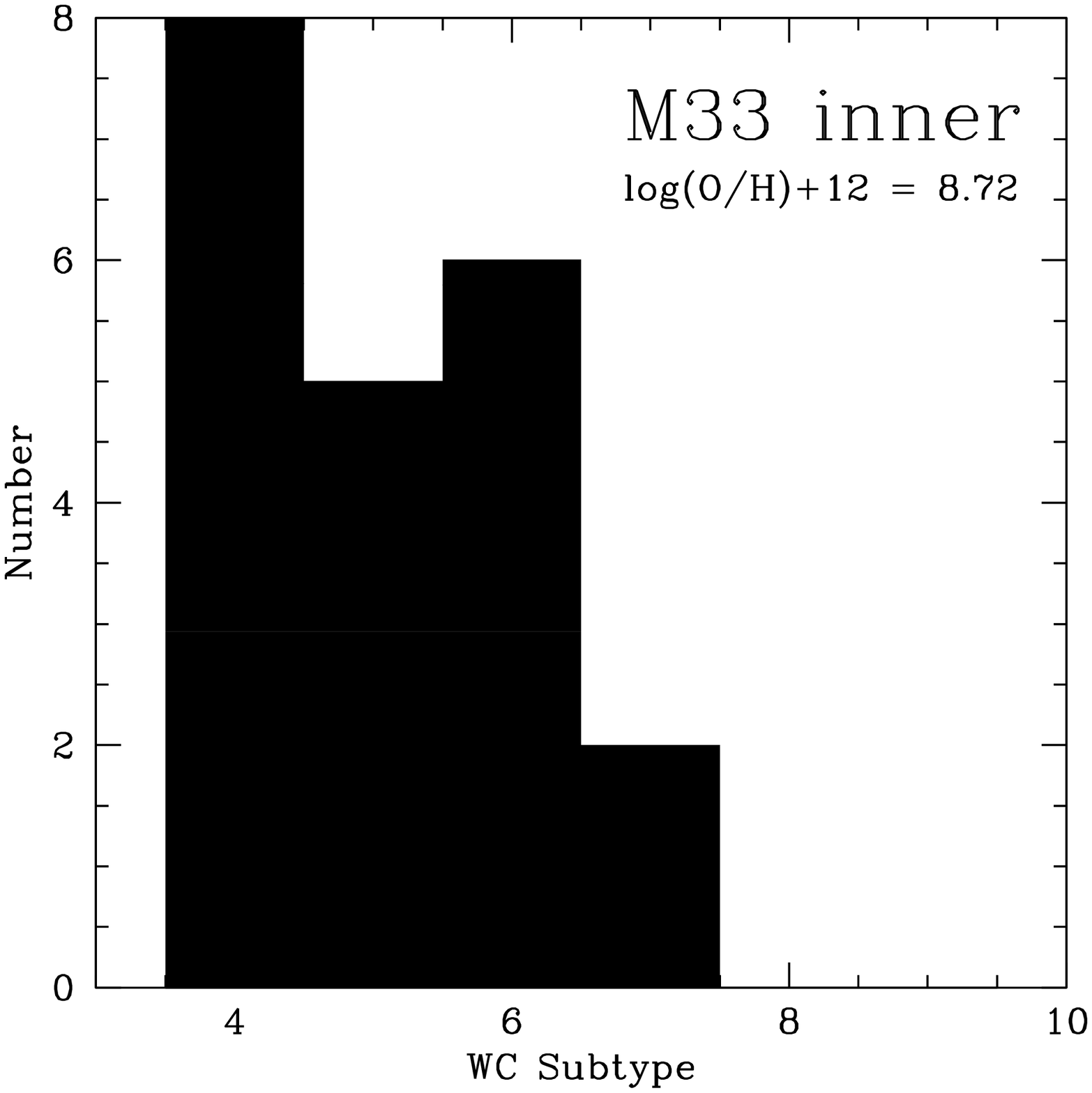}
\plotone{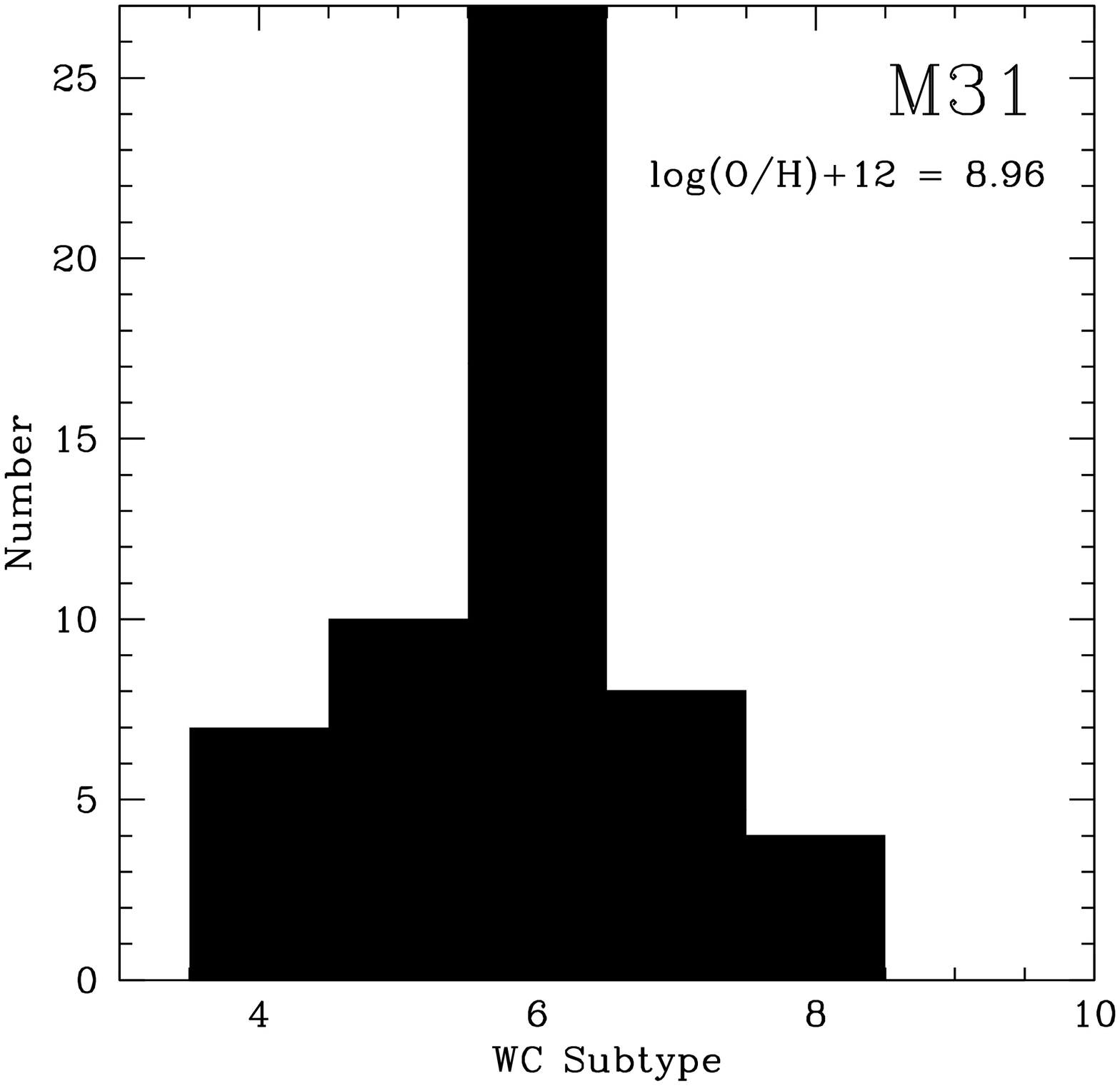}
\plotone{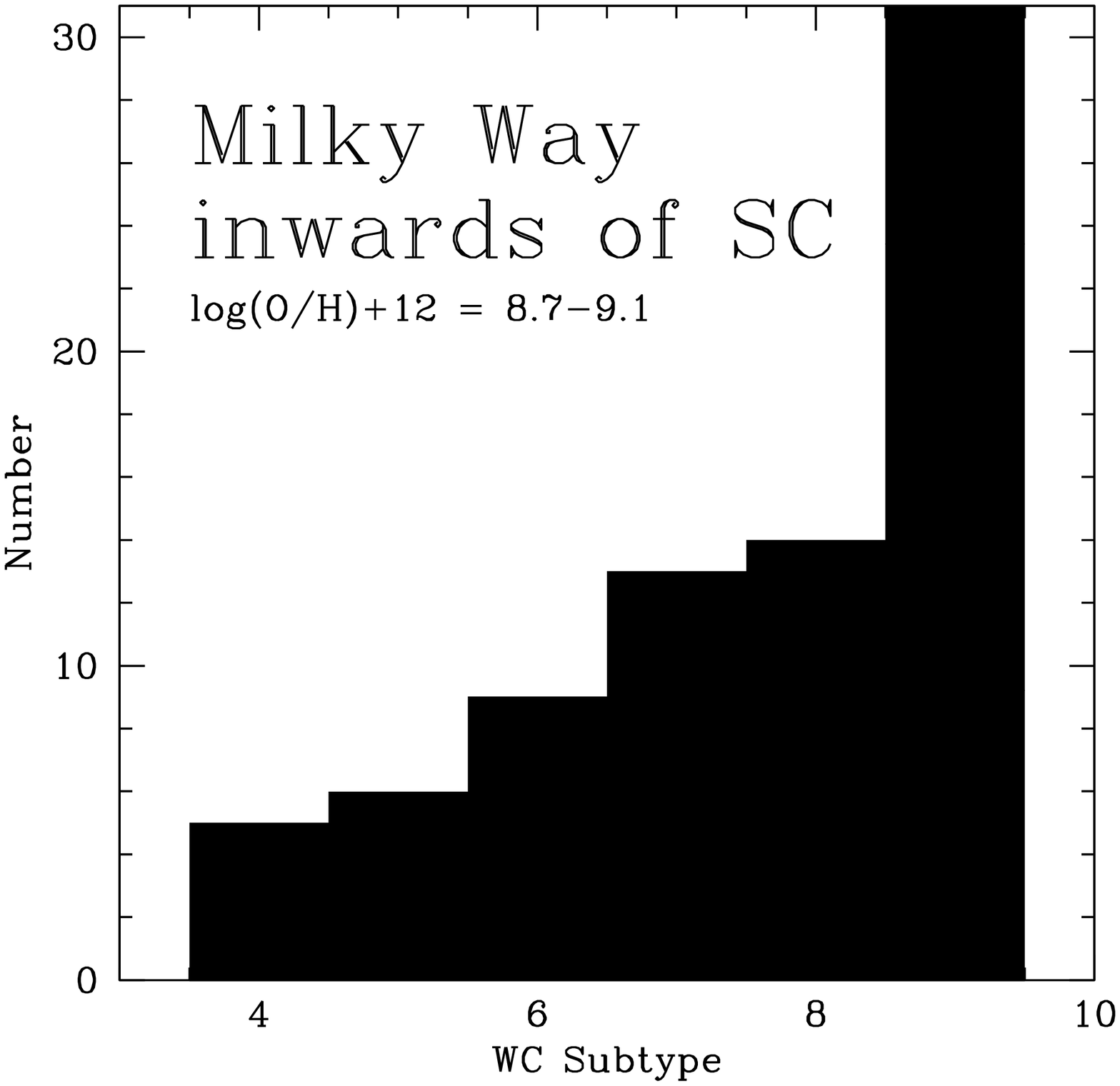}
\caption{\label{fig:histograms} WC subtype versus metallicity. As the metallicity increases, the relative number of early-type WCs decreases and the relative number of later-type WCs increases. Note: M33 inner ($\rho<0.25$); M33 middle ($0.25\le \rho <0.50$); M33 outer ($\rho\ge 0.5$); SC = Solar Circle.}
\end{figure}

\begin{figure}
\epsscale{0.8}
\plotone{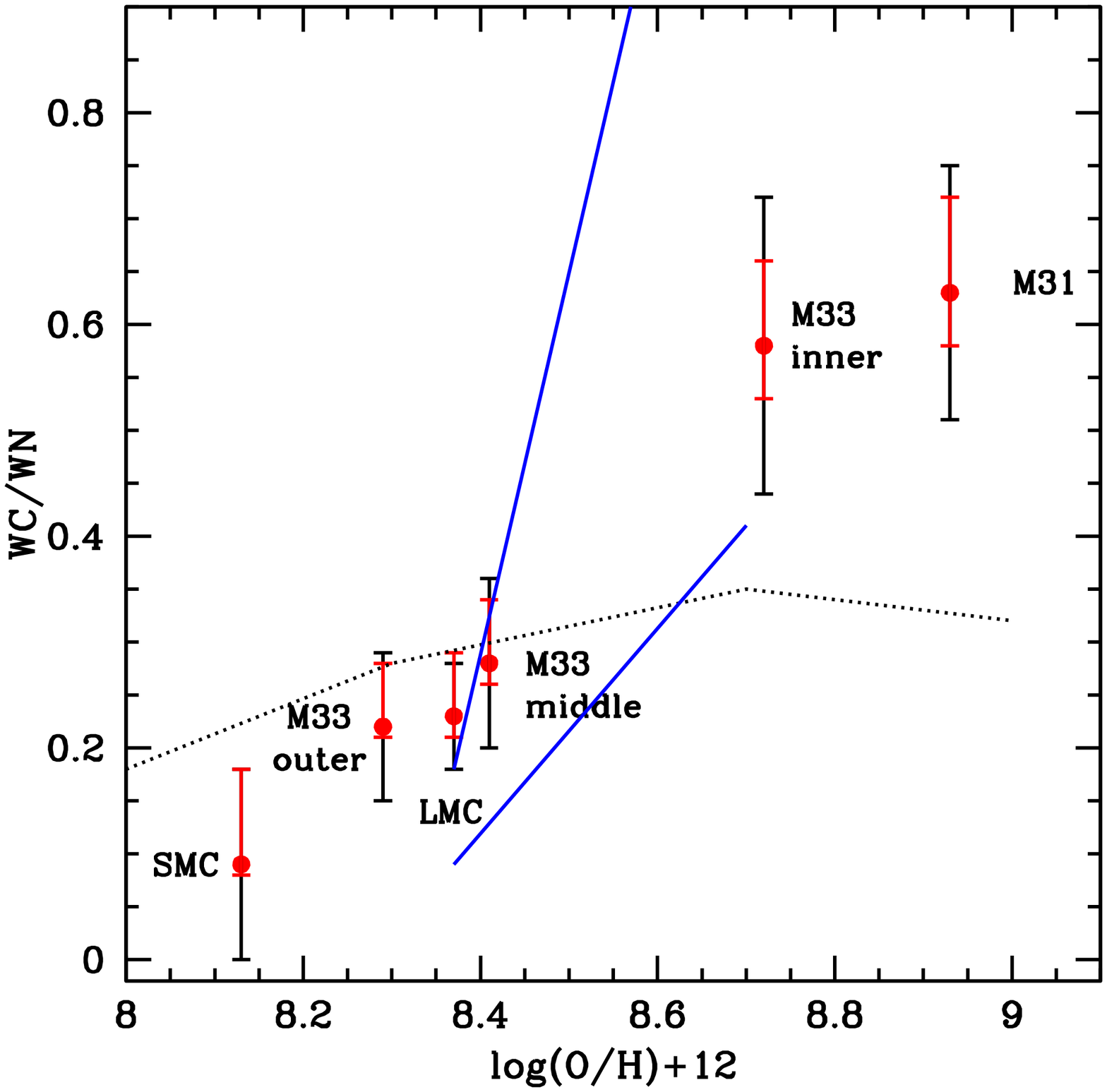}
\caption{\label{fig:wcwn} The WC to WN ratio plotted against the metal abundance log$\frac{O}{H} + 12$, which we define to correspond to 8.7 at $z=0.014$ (solar). The red error bars indicate the uncertainties in the measured WC/WN ratios, while the black error bars represent $\sqrt{N}$ uncertainties, which we argue are more appropriate for comparing with the predictions of the evolutionary models. The model results for an initial rotational velocity of 300 km s$^{-1}$ are shown by the dotted black line for the older Geneva models (Meynet \& Maeder 2005). The two blue solid lines are the predictions from the $z=0.006$ and 0.014 new Geneva models, with the lower one corresponding to a (realistic) initial rotation of 40\% of the critical break up speed, while the upper line corresponds to the (unrealistic) case of no initial rotation.}
\end{figure}

\clearpage

\begin{deluxetable}{l l l l}
\tablecaption{\label{tab:Whinners} WR Stars and Candidates Not Observed Spectroscopically}
\tablewidth{0pt}
\tablehead{
\colhead{Star\tablenotemark{a}}
&\colhead{ID\tablenotemark{b}}
&\colhead{Type}
&\colhead{Ref.\tablenotemark{c}}
}
\startdata
J004109.32+404902.6 &New      &\nodata & \nodata\\
J004303.60+413752.1 &New      &\nodata & \nodata \\
J004406.39+411921.0 &MS3      &WC     & 1 \\
J004425.10+412046.4 &New      &\nodata \\
J004425.32+412115.7 &OB33 WR1 &Candidate & 2 \\
J004425.43+412044.7 &OB33 WR2 &WC6-7 & 2 \\
J004513.68+413742.5 &OB48 WR3 &WC+abs & 3 \\
J004551.87+421014.0 &New      &\nodata  & \nodata\\
J004628.51+421127.6 & OB102 WR1 & WN & 2 \\
\enddata
\tablenotetext{a}{Designation from the LGGS.}
\tablenotetext{b}{MS3 from Moffat \& Shara 1983 with revised coordinates by MJ98; OB 33 stars identified in Armandroff \& Massey 1991; OB 48 star identified in Massey et al.\ 1986.}
\tablenotetext{c}{References for type: (1) Moffat \& Shara 1983; (2) Armandroff \& Massey 1991; (3) Massey et al.\ 1987.}
\end{deluxetable}

\begin{deluxetable}{l l l l l l r r r r r l}
\tabletypesize{\tiny}
\rotate
\tablecaption{\label{tab:Winners} M31 WR Stars with New Spectroscopic Data}
\tablewidth{0pt}
\tablehead{
\colhead{Star\tablenotemark{a}}
&\colhead{New}
&\multicolumn{3}{c}{Previous Data}
&
&\multicolumn{2}{c}{CIII $\lambda$4650/HeII $\lambda$4686}
&
&\multicolumn{2}{c}{CIV $\lambda$5806} \\ \cline{3-5} \cline{7-8} \cline{10-11} \\
&\colhead{Type}
&\colhead{ID\tablenotemark{b}}
&\colhead{Type}
&\colhead{Ref\tablenotemark{c}}
&
&\colhead{-EW(\AA)}
&\colhead{FWHM(\AA)}
&
&\colhead{-EW(\AA)}
&\colhead{FWHM(\AA)}}
\startdata
J003857.19+403132.2 &WN7:+neb                & New\tablenotemark{e}       &\nodata      & \nodata && 40 & 18 && \nodata & \nodata &               \\
J003911.04+403817.5 &WC6                         & MS 21     &WCE          & 1 &&    1127&     50&&   1127&     50&                    \\
J003919.53+402211.6 &WC4                         & OB136 WR 1&WN/C    & 2 &&      48&     76&&     48&     76&                    \\
J003933.46+402018.9 &WC5                         & OB138 WR 1&WC           & 2 &&     460&     75&&    460&     75&                    \\
J003935.63+402811.5 &WN5                         & New       &\nodata      & \nodata &&     190&     21&&    190&     21&                    \\
J003939.61+402031.1 &WN3                         & New       &\nodata      & \nodata &&     300&     29&&\nodata&\nodata&                    \\
J003939.97+403450.4 &WC6                         & New       &\nodata      & \nodata &&     500&     47&&    500&     47&                    \\
J003945.82+402303.2 &WN3+O4\tablenotemark{d}     & OB135 WR 1&WNE          & 2,3 &&      44&     40&&     44&     40&                    \\
J003948.84+405256.3 &WN4                         & New       &\nodata      & \nodata &&      86&     30&&     86&     30&                    \\
J004002.84+403852.8 &WN6-7                       & New       &\nodata      & \nodata &&      86&     21&&     86&     21&                    \\
J004005.65+405848.1 &WN4\tablenotemark{d}        & New       &\nodata      & \nodata &&     125&     45&&    125&     45&                    \\
J004019.47+405224.9 &WC4\tablenotemark{d}        & MS 12     &WC           & 2 &&    1120&     80&&   1120&     80&                    \\
J004020.44+404807.7 &WC6                         & MS 14     &WC           & 2 &&     920&     60&&    920&     60&                    \\
J004021.13+403520.6 &WN5                         & MS 20     &WN           & 2   &&     200&     25&&    200&     25&                    \\
J004022.43+405234.6 &WC4                         & MS 11     &WC4-5        & 4  &&     575&     64&&    575&     64&                    \\
J004023.02+404454.1 &WN4                         & OB78 WR 5 &WN/C         &2 &&     150&     26&&    150&     26&   MS 15                 \\
J004024.33+405016.2 &WN9                         & New\tablenotemark{e}  &\nodata      & \nodata &&       7&      9&&\nodata&\nodata&                 \\
J004026.23+404459.6 &WN6+abs\tablenotemark{d}    & OB78 WR 2 &WNL          & 2,3,5 &&      21&     23&&\nodata&\nodata&  IT 5-25               \\
J004029.27+403916.6 &WC5                         & MS 18     &WC           & 6 &&     560&     59&&    560&     59&                    \\
J004031.67+403909.0 &WN6                         & MS 17     &WN           & 6 &&      36&     28&&     36&     28&                    \\
J004032.58+403550.1 &WN6-7                       & New       &\nodata      & \nodata &&      35&     19&&     35&     19&                    \\
J004034.17+404340.4 &WC7+fgd                   & OB78 WR3  &WC6-7        & 7   &&      28&     28&&     28&     28&A9 V fgd comp         \\
J004034.69+404432.9 &WC4                         & OB78 WR 4 &WC5          & 2 &&     766&     65&&    766&     65&   MS 16                 \\
J004035.18+403838.5 &WN5                         & New       &\nodata      & \nodata &&      53&     20&&     53&     20&                    \\
J004036.76+410104.3 &WN8                         & New       &\nodata      & \nodata &&      35&     11&&\nodata&\nodata&                    \\
J004043.28+403525.2 &WC6                         & New       &\nodata      & \nodata &&     495&     57&&    495&     57&                    \\
J004056.49+410308.7 &WN9                         & OB69 WR 2 &Ofpe/WN9     & 2,3,8 &&      10&     10&&\nodata&\nodata&\\
J004056.72+410255.4 &WN3                         & OB69 WR 1 &WNE          &  6 &&     510&     51&&    510&     51&                    \\
J004058.49+410414.8 &WC6                         & OB69 WR 3 &WC           & 2 && \nodata&\nodata&&\nodata&\nodata&Low continuum       \\
J004059.13+403652.0 &WC6\tablenotemark{d}        & New       &\nodata      & \nodata &&      67&     49&&     67&     49&                    \\
J004100.97+403633.0 &WN3                         & New       &\nodata      & \nodata &&     220&     35&&    220&     35&                    \\
J004102.04+410446.0 &WN4                         & OB69 WR 4 &WNL          & 2  &&      49&     27&&     49&     27&                    \\
J004107.31+410417.0 &WC6                         & OB69 F1   &WC6-7        & 2   &&     290&     56&&    290&     56&                    \\
J004109.46+404907.8 &WN6\tablenotemark{d}        & New       &\nodata      & \nodata &&      10&     20&&     10&     20&                    \\
J004109.59+404856.2 &WC5+B                       & New       &\nodata      & \nodata &&      50&     53&&     50&     53&                    \\
J004114.95+404448.2 &WN6                         & New       &\nodata      & \nodata &&     185&     25&&    185&     25&                    \\
J004115.95+410906.8 &WN4                         & New       &\nodata      & \nodata &&     847&     29&&    847&     29&                 \\
J004126.11+411220.0 &WN7                         & New       &\nodata      & \nodata &&      33&     15&&\nodata&\nodata& Fig.~\ref{fig:wns}                 \\
J004126.39+411203.5 &WN8                         & New       &\nodata      & \nodata &&      17&     11&&\nodata&\nodata&                    \\
J004129.11+411311.4 &WN4                         & New       &\nodata      & \nodata &&      36&     31&&\nodata&\nodata&                    \\
J004129.17+411335.0 &WN7pec                      & New       &\nodata      & \nodata && \nodata&\nodata&&\nodata&\nodata&Low continuum, broad\\
J004130.37+410500.9 &WNL/Of+neb                  & New\tablenotemark{e}  &\nodata      & \nodata &&       5&      7&&\nodata&\nodata&                    \\
J004134.99+410552.3 &WC7                         & MS 8      &WCL          & 1   &&     586&     53&&    586&     53&                    \\
J004141.81+403711.5 &WN3                         & New       &\nodata      & \nodata &&     245&     25&&    245&     25&                    \\
J004143.00+411859.7 &WN6                         & New       &\nodata      & \nodata &&      93&     21&&     93&     21&                    \\
J004143.09+404045.9 &WN3                         & New       &\nodata      & \nodata &&     200&     30&&    200&     30&                    \\
J004144.47+404517.1 &WC6-7                       & MS 10     &WC6          & 3, 1 &&     185&     63&&    185&     63&                    \\
J004145.33+412104.0 &WN6                         & New       &\nodata      & \nodata &&      97&     21&&     97&     21&                    \\
J004147.24+410647.5 &WC8\tablenotemark{d}  & New       &\nodata      & \nodata &&      64&     22&&     64&     22&Beautiful Dup in LGG\\
J004148.27+411739.9 &WN6                         & New       &\nodata      & \nodata &&      87&     23&&     87&     23&                    \\
J004154.62+404713.8 &WC6                         & New       &\nodata      & \nodata &&     437&     64&&    437&     64&                    \\
J004203.94+412554.5 &WC6                         & New       &\nodata      & \nodata &&      71&     38&&     71&     38&                    \\
J004211.16+405648.0 &WN6                         & New       &\nodata      & \nodata &&     109&     20&&    109&     20&                    \\
J004213.21+405051.8 &WC4\tablenotemark{d}        & New       &\nodata      & \nodata &&     530&     60&&    530&     60&                    \\
J004214.36+412542.3 &WC6                         & MS 6      &WCL          & 1  &&     535&     55&&    535&     55&                    \\
J004234.42+413024.2 &WC8                         & MS 5      &WC7-8        & 3   &&     133&     28&&    133&     28&                    \\
J004238.90+410002.0 &WN7\tablenotemark{d}        & New       &\nodata      & \nodata &&      25&     16&&\nodata&\nodata&                    \\
J004240.81+410241.6 &WC6                         & New       &\nodata      & \nodata &&     425&     61&&    425&     61&                    \\
J004242.03+412314.9 &WC7                         & MS 7      &WCL          & 1 &&     208&     34&&    208&     34&                    \\
J004242.33+413922.7 &WN10                        & New\tablenotemark{e}       &\nodata      & \nodata &&  3 & 11 && \nodata & \nodata &                \\
J004247.12+405657.1 &WC4                         & New       &\nodata      & \nodata &&     455&     66&&    455&     66&                    \\
J004249.84+410215.7 &WC5-6                       & New       &\nodata      & \nodata &&     550&     46&&    550&     46&                    \\
J004254.65+410226.1 &WN3+neb                     & New       &\nodata      & \nodata &&     220&     34&&    220&     34&                    \\
X004256.05+413543.7 &WC6\tablenotemark{d}        & New       &\nodata      & \nodata &&     219&     34&&    219&     34&                    \\
J004256.85+413837.2 &WN6                         & New       &\nodata      & \nodata &&     126&     20&&    126&     20&                    \\
J004257.62+413727.8 &WN3                         & New       &\nodata      & \nodata &&      47&     50&&\nodata&\nodata&                    \\
J004302.05+413746.7 &WN9                         & New       &\nodata      & \nodata &&      13&     11&&\nodata&\nodata&                  \\
J004302.76+413753.6 &WN6                         & New       &\nodata      & \nodata &&      60&     20&&     60&     20&                    \\
J004304.34+412223.4 &WN5                         & New       &\nodata      & \nodata &&      38&     23&&     38&     23&                    \\
J004306.11+413813.7 &WN4                         & New       &\nodata      & \nodata &&     200&     37&&    200&     37&                    \\
X004308.25+413736.3 &WC6                         & New       &\nodata      & \nodata &&     250&     30&&    250&     30&                    \\
J004310.19+413744.0 &WN4                         & New       &\nodata      & \nodata &&     165&     36&&    165&     36&                    \\
J004316.44+414512.4 &WC6\tablenotemark{d}        & New       &\nodata      & \nodata &&      84&     38&&     84&     38&                    \\
J004321.48+414155.5 &WN6\tablenotemark{d}        & New       &\nodata      & \nodata &&     183&     31&&    183&     31&                    \\
J004324.95+411803.2 &WN4.5                       & New       &\nodata      & \nodata &&      37&     23&&     37&     23&                    \\
J004325.68+411830.2 &WN4.5                       & New       &\nodata      & \nodata &&      28&     25&&     28&     25&                    \\
J004327.92+414207.3 &WC6                         & New       &\nodata      & \nodata &&      51&     34&&     51&     34&                    \\
J004328.06+414212.3 &WN6                         & New       &\nodata      & \nodata &&     126&     32&&    126&     32&                    \\
J004330.76+412734.7 &WN4                         & New       &\nodata      & \nodata &&      49&     25&&     49&     25&                    \\
J004331.17+411203.5 &WC8                         & MS 4      &WC7-8        & 4    &&      63&     24&&     63&     24&                    \\
J004334.92+410953.2 &WN3?+neb+abs        & New       &\nodata      & \nodata &&      90&     87&&     90&     87& Foreground comp?   \\
J004336.51+412315.1 &WN4                         & New       &\nodata      & \nodata &&      46&     25&&     46&     25&                    \\
J004337.10+414237.1 &WN10                   & New\tablenotemark{e}  &\nodata      & \nodata &&       2&      3&&\nodata&\nodata&            \\
J004341.72+412304.2 &WC6                         & MS 2      &WC           & 1 &&     254&     38&&    254&     38&                    \\
J004344.48+411142.0 &WN8+neb\tablenotemark{d}    & New       &\nodata      & \nodata &&       6&      7&&\nodata&\nodata&Fig.~\ref{fig:wns}      \\
J004347.01+411238.0 &WN7                         & New       &\nodata      & \nodata &&      29&     29&&     29&     29&                    \\
J004349.72+411243.4 &WN6                         & New       &\nodata      & \nodata &&      61&     20&&     61&     20&                    \\
J004353.34+414638.9 &WN7                         & New       &\nodata      & \nodata &&      31&     26&&     31&     26&HeI P Cyg           \\
J004357.31+414846.2 &WN8                         & New\tablenotemark{e}  &\nodata      & \nodata &&      36&     15&&     36&     15&                    \\
J004403.39+411518.8 &WN6                         & New       &\nodata      & \nodata &&      90&     21&&     90&     21&                    \\
J004406.68+411612.0 &WN4                         & New       &\nodata      & \nodata &&     107&     33&&    107&     33&                    \\
J004408.58+412121.2 &WC6                         & New       &\nodata      & \nodata &&     480&     75&&    480&     75&                    \\
J004410.17+413253.1 &WC6                         & OB10 WR 1 &WC6-7        & 7 &&     614&     65&&    614&     65&         IT5-19           \\
J004410.91+411623.2 &WN4                         & New       &\nodata      & \nodata &&     192&     42&&    192&     42&Fig.~\ref{fig:wns}                  \\
J004412.44+412941.7 &WC6                         & IT 5-15   &WC6          & 2 &&     535&     55&&    535&     55&                    \\
J004413.06+411920.5 &WN8                         & New       &\nodata      & \nodata &&       7&      8&&\nodata&\nodata&                    \\
J004415.77+411952.5 &WC5                         & New       &\nodata      & \nodata &&     137&     51&&    137&     51&                    \\
J004416.20+412103.5 &WC+B0I?                     & New       &\nodata      & \nodata &&       8&     67&&      8&     67&                    \\
J004419.92+411751.2 &WN3                         & New       &\nodata      & \nodata &&      59&     31&&     59&     31&                    \\
J004420.58+415412.5 &WN3\tablenotemark{d}        & IT 1-40   &WN           & 9   &&     190&     30&&    190&     30&                    \\
J004422.24+411858.4 &WC7-8                       & OB32 WR 1 &WC6-7        & 2    &&     115&     27&&    115&     27&                    \\
J004425.10+412050.0 &WC6+B0I                     & New       &\nodata      & \nodata &&      14&     69&&     14&     69&                    \\
J004425.83+415019.4 &WN5                         & New       &\nodata      & \nodata &&      58&     27&&     58&     27&                    \\
J004426.17+412210.8 &WN6                         & New       &\nodata      & \nodata &&      43&     26&&     43&     26&                    \\
J004426.32+413419.8 &WN5\tablenotemark{d}      & New       &\nodata      & \nodata &&      85&     27&&     85&     27& Fig.~\ref{fig:wns}\\
J004429.89+412021.6 &WN3                         & New       &\nodata      & \nodata &&      55&     27&&     55&     27&                    \\
J004430.04+415237.1 &WN8                         & New       &\nodata      & \nodata &&       6&     11&&\nodata&\nodata&                    \\
J004434.57+412424.4 &WN3                         & New       &\nodata      & \nodata &&     258&     41&&    258&     41&                    \\
J004435.15+412545.2 &WC5                         & New       &\nodata      & \nodata &&     347&     75&&    347&     75&                    \\
J004436.22+412257.3 &WN5                         & New       &\nodata      & \nodata &&     195&     30&&    195&     30&                    \\
J004436.52+412202.0 &WC5                         & New       &\nodata      & \nodata &&     457&     38&&    457&     38&CIII 4650 component \\
J004437.61+415203.3 &WN4pec                      & OB54 WR 1 &WN           & 2 &&     270&     55&&    270&     55&broad lines, IT1-38        \\
J004443.10+412619.6 &WN6                         & New       &\nodata      & \nodata &&      18&     17&&     18&     17&                    \\
J004444.00+412739.9 &WC6                         & IT 5-3    &WC           & 9      &&     384&     65&&    384&     65&                    \\
J004444.87+412800.7 &WN6                         & New       &\nodata      & \nodata &&      50&     17&&     50&     17& Fig.~\ref{fig:wns}                \\
J004445.61+412806.5 &WN5                         & New       &\nodata      & \nodata &&      47&     24&&     47&     24&                    \\
J004445.90+415803.8 &WN5                         & New       &\nodata      & \nodata &&      85&     20&&     85&     20& Fig.~\ref{fig:wns}                  \\
J004449.41+413020.7 &WN3                         & New       &\nodata      & \nodata &&     130&     48&&    130&     48&Fig.~\ref{fig:wns}                  \\
J004451.98+412911.6 &WC7                         & IT 5-2    &WC           & 2     &&     321&     29&&    321&     29&                    \\
J004453.06+412601.7 &WN3+TiO\tablenotemark{d}    & New       &\nodata      & \nodata &&      22&     30&&\nodata&\nodata&                    \\
J004453.52+415354.3 &WC7                         & IT 1-48   &WC           & 2     &&     400&     26&&    400&     26&                    \\
J004455.63+413105.1 &WC6                         & OB42 WR 1 &WC6-7        & 2 &&     405&     52&&    405&     52&  IT5-4                  \\
J004455.82+412919.2 &WN2pec+neb                  & New       &\nodata      & \nodata &&      49&     47&&     49&     47&  Fig.~\ref{fig:wns}                \\
J004500.96+413058.8 &WC6\tablenotemark{d}        & OB42 WR 2 &WC6-7        & 2 &&     178&     47&&    178&     47&      IT5-10             \\
J004502.78+415533.7 &WN5pec                      & New       &\nodata      & \nodata &&     155&     36&&    155&     36&NV/NIIIblend. Two stars?\\
J004506.50+413425.1 &WN4+neb                     & New       &\nodata      & \nodata &&      67&     32&&     67&     32&                    \\
J004509.18+414021.4 &WC7                         & New       &\nodata      & \nodata &&     182&     27&&    182&     27&                    \\
J004510.39+413646.6 &WC7+BI                      & OB48 WR 6 &WC6-7+abs    & 2 &&      51&     35&&     51&     35&                    \\
J004511.21+420521.7 &WN11                 & New\tablenotemark{e}  &\nodata      & \nodata &&       4&      3&&\nodata&\nodata&             \\
J004511.27+413815.3 &WC6                         & OB48 WR 1 &WC6-7        & 6 &&     500&     50&&    500&     50&        IT5-01            \\
J004514.10+413735.2 &WN4                         & OB48 WR 2 &WNE          & 2  &&     180&     35&&    180&     35&                    \\
J004517.56+413922.0 &WN4brd                      & OB48-527  &WN           & 2 &&      56&     45&&     56&     45&                    \\
J004517.89+415209.5 &WC8                         & New       &\nodata      & \nodata &&      38&     30&&     38&     30&                    \\
J004520.80+415100.0 &WN4                         & New       &\nodata      & \nodata &&      80&     25&&     80&     25&                    \\
J004522.78+420318.2 &WC5-6                       & New       &\nodata      & \nodata &&    1160&     55&&   1160&     55&                    \\
J004524.26+415352.5 &WC6                         & IT 4-01   &WC           & 9     &&     550&     45&&    550&     45&                    \\
J004530.61+414639.2 &WC4-5+neb                   & New       &\nodata      & \nodata &&     300&     56&&    300&     56&                    \\
J004531.16+420658.1 &WC6                         & New       &\nodata      & \nodata &&     570&     63&&    570&     63&                    \\
J004537.05+414302.4 &WN3                         & New       &\nodata      & \nodata &&     155&     46&&    155&     46&                    \\
J004537.10+414201.4 &WC6                         & IT 4-13   &WC           & 2   &&     340&     61&&    340&     61&                    \\
J004537.28+414305.2 &WN4+neb                     & New       &\nodata      & \nodata &&      45&     26&&     45&     26&                    \\
J004539.15+414754.0 &WN5                         & New       &\nodata      & \nodata &&      80&     24&&     80&     24&                    \\
J004539.35+414439.8 &WC5                         & New       &\nodata      & \nodata &&     350&     53&&    350&     53&                    \\
J004541.29+415504.0 &WN4                         & New       &\nodata      & \nodata &&     115&     25&&    115&     25&                    \\
J004542.26+414510.1 &WN10                        & New\tablenotemark{e}       &\nodata      & \nodata &&  4 &  8 && \nodata & \nodata &                \\
J004544.65+415241.0 &WN3                         & New       &\nodata      & \nodata &&      60&     27&&\nodata&\nodata&                    \\
J004551.12+421015.4 &WC5-6                       & New       &\nodata      & \nodata &&    1010&     48&&   1010&     48&                    \\
J004551.35+414242.0 &WC6\tablenotemark{d}        & IT 4-14   &WC           & 9   &&     840&     59&&    840&     59&                    \\
J004645.02+421301.3 &WN3                         & New       &\nodata      & \nodata &&     440&     30&&    440&     30&                    \\
J004655.89+420015.1 &WN3+abs?                    & New       &\nodata      & \nodata &&      50&     26&&\nodata&\nodata&                    \\
\enddata
\tablenotetext{a}{J00... designations from LGGS; X00... designations new here.}
\tablenotetext{b}{Previous identifications: OB designations from Massey et al.\ 1986 except for OB 32, from Armandroff \& Massey 1991; MS= Moffat \& Shara 1983 with revised coordinates by MJ98; IT= Moffat \& Shara 1987, with revised coordinates by MJ98. ``New" means the star is newly identified WR star as part of the present program.}
\tablenotetext{c}{References for previous spectral types: (1) Moffat \& Shara 1983; (2) Armandroff \& Massey 1991; (3) Schild et al.\ 1990; (4) Willis et al.\ 1992; (5) Massey et al.\ 1995; (6) Massey et al.\ 1987; (7) Massey et al.\ 1986; (8) Massey 1998; (9) Moffat \& Shara 1987.}
\tablenotetext{d}{Two new spectra were obtained and compared.}
\tablenotetext{e}{Newly confirmed Wolf-Rayet star but not found as part of our imaging survey.}
\end{deluxetable}

\begin{deluxetable}{l c c l}
\tablecaption{\label{tab:toored} Red WR Candidates}
\tablewidth{0pt}
\tablehead{
\colhead{Star\tablenotemark{a}}
&\colhead{$V$\tablenotemark{a}}
&\colhead{$B-V$\tablenotemark{a}}
&\colhead{OB Assoc.\tablenotemark{b}}
}
\startdata
J003944.96+402038.2 & 21.22 & 0.80 & 139 \\
J003945.05+403030.5 & 20.62 & 0.87 & no \\
J004301.53+411226.8 & 18.90 & 0.71 & I \\
J004315.95+412708.5 & 18.70 & 0.73 & (12) \\
J004830.34+422250.9 & 19.69 & 0.80 & no \\
\enddata
\tablenotetext{a}{Designation from the LGGS.}
\tablenotetext{b}{OB associations as identified by van den Bergh 1964. Parenthesis means that the star is close to, but not in, the association. An ``I" means that the star is located in the Population I ring but not in an identified OB association.} 
\end{deluxetable}

\begin{deluxetable}{l l l}
\tablecaption{\label{tab:Losers} Newly Classified Non-WR Emission Line Sources }
\tablewidth{0pt}
\tablehead{
\colhead{Star\tablenotemark{a}}
&\colhead{Type}
&\colhead{Comment}
}
\startdata
J003856.68+403446.8&QSO z=2.76&     $V$=20.8 Ly$\alpha$, CIV $\lambda$1549, CIII] $\lambda$1909 \\                                        
J004137.93+410107.9&QSO z=1.45&     $V$=19.8 CIV $\lambda$1549, CIII] $\lambda$1909, MgII $\lambda$2798 \\                                         
J004221.78+410013.4&hotLBVcand&     H$\alpha$ source (Massey et al.\ 2007) \\ 
J004427.95+412101.4&O7Iaf+neb &       EW He II $\lambda$4686=$-5$\AA, OB33-WR3, WNL/Of (MJ98) \\  
J004742.84+421017.0&QSO z=1.45&     $V$=20.3 CIV $\lambda$1549, CIII] $\lambda$1909, MgII $\lambda$2798   \\                                               
\enddata
\tablenotetext{a}{Designation from the LGGS.}
\end{deluxetable}

\begin{deluxetable}{l r l r r r c l}
\tabletypesize{\small}
\tablecaption{\label{tab:BigTable} Spectroscopically Confirmed M31 Wolf-Rayet Stars }
\tablewidth{0pt}
\tablehead{
\colhead{Star\tablenotemark{a}}
&\colhead{$\rho$\tablenotemark{b}}
&\colhead{Type}
&\colhead{$V$\tablenotemark{a}}
&\colhead{$B-V$\tablenotemark{a}}
&\colhead{$M_V$\tablenotemark{c}}
&\colhead{OB Assoc.\tablenotemark{d}}
&\colhead{Other ID\tablenotemark{e}}
}
\startdata   
J003857.19+403132.2&    0.79   &WN7:+neb                   &    20.85 &     0.31 & \nodata &  128    & New        \\                                                                             
J003911.04+403817.5&    0.78   &WC6                        &    20.59 &    -0.35 &   -4.4  &  127    & MS 21      \\
J003919.53+402211.6&    0.70   &WC4                        &    19.17 &     0.38 & \nodata &  136    & OB 136 WR 1 \\
J003933.46+402018.9&    0.70   &WC5                        &    19.15 &    -0.04 &   -5.9  &  138    & OB 138 WR 1 \\
J003935.63+402811.5&    0.63   &WN5                        &    20.46 &     0.02 &   -4.6  & I & New        \\
J003939.61+402031.1&    0.70   &WN3                        &    22.31 &     0.35 & \nodata &  138    & New        \\
J003939.97+403450.4&    0.61   &WC6                        &    20.67 &    -0.17 &   -4.4  &  (39)   & New        \\
J003945.82+402303.2&    0.67   &WN3+O4                     &    18.48 &    -0.06 &   -6.5  &  135    & OB 135 WR 1  \\
J003948.84+405256.3&    0.76   &WN4                        &    20.63 &     0.06 &   -4.4  &  122    & New        \\
J004002.84+403852.8&    0.53   &WN6-7                      &    21.14 &    -0.04 &   -3.9  & I & New        \\
J004005.65+405848.1&    0.76   &WN4                        &    20.19 &     0.21 & \nodata &  121    & New        \\
J004019.47+405224.9&    0.55   &WC4                        &    20.55 &    -0.23 &   -4.5  &  132    & MS 12      \\
J004020.44+404807.7&    0.50   &WC6                        &    20.19 &    -0.34 &   -4.8  &  (72)   & MS 14      \\
J004021.13+403520.6&    0.52   &WN5                        &    20.63 &    -0.21 &   -4.4  &  80     & MS 20      \\
J004022.43+405234.6&    0.54   &WC4                        &    20.18 &    -0.27 &   -4.8  &  132    & MS 11      \\
J004023.02+404454.1&    0.47   &WN4                        &    20.61 &    -0.26 &   -4.4  &  78     & OB 78 WR 5 \\
J004024.33+405016.2&    0.50   &WN9                        &    19.15 &    -0.16 &   -5.9  &  72     & New        \\
J004026.23+404459.6&    0.46   &WN6+abs                    &    18.93 &    -0.17 &   -6.1  &  78     & OB 78 WR 2 \\
J004029.27+403916.6&    0.47   &WC5                        &    21.21 &    -0.27 &   -3.8  &  80b    & MS 18      \\
J004031.67+403909.0&    0.47   &WN6                        &    19.45 &     0.22 & \nodata &  80b    & MS 17      \\
J004032.58+403550.1&    0.51   &WN6-7                      &    19.37 &    -0.16 &   -5.6  &  81     & New        \\
J004034.17+404340.4&    0.43   &WC7+fgd                    &    19.90 &     0.78 & \nodata &  78     & OB 78 WR 3  \\
J004034.69+404432.9&    0.43   &WC4                        &    20.71 &    -0.45 &   -4.3  &  78     & OB 78 WR 4 \\
J004035.18+403838.5&    0.47   &WN5                        &    21.77 &     0.22 & \nodata &  (80b)  & New        \\
J004036.76+410104.3&    0.59   &WN8                        &    19.74 &    -0.14 &   -5.3  &  130    & New        \\
J004043.28+403525.2&    0.53   &WC6                        &    22.09 &     0.52 & \nodata & I & New        \\
J004056.49+410308.7&    0.49   &WN9                        &    18.09 &    -0.08 &   -6.9  &  69     & OB 69 WR 2  \\
J004056.72+410255.4&    0.48   &WN3                        &    20.30 &    -0.37 &   -4.7  &  69     & OB 69 WR 1 \\
J004058.49+410414.8&    0.50   &WC6                        &    21.89 &     0.10 &   -3.1  &  69     & OB 69 WR 3 \\
J004059.13+403652.0&    0.54   &WC6                        &    20.73 &     0.43 & \nodata &  82     & New        \\
J004100.97+403633.0&    0.55   &WN3                        &    21.39 &     0.09 &   -3.6  &  82     & New        \\
J004102.04+410446.0&    0.48   &WN4                        &    21.88 &     0.14 &   -3.1  &  69     & OB 69 WR 4 \\
J004107.31+410417.0&    0.44   &WC6                        &    21.31 &     0.30 & \nodata &  (69)   & OB 69 F1   \\
J004109.46+404907.8&    0.34   &WN6                        &    19.31 &     0.27 & \nodata &  21     & New        \\
J004109.59+404856.2&    0.34   &WC5+B                      &    18.97 &    -0.03 &   -6.1  &  21     & New        \\
J004114.95+404448.2&    0.42   &WN6                        &    20.50 &    -0.39 &   -4.5  & \nodata & New        \\
J004115.95+410906.8&    0.48   &WN4                        &    22.03 &    -0.18 &   -3.0  &  67     & New        \\
J004126.11+411220.0&    0.48   &WN7                        &    18.93 &     0.26 & \nodata &  66     & New        \\
J004126.39+411203.5&    0.47   &WN8                        &    20.13 &     0.27 & \nodata &  66     & New        \\
J004129.11+411311.4&    0.48   &WN4                        &    20.95 &     0.13 &   -4.1  &  66     & New        \\
J004129.17+411335.0&    0.48   &WN7pec                     &    21.41 &     0.38 & \nodata &  66     & New        \\
J004130.37+410500.9&    0.30   &WNL/Of+neb                 &    18.50 &     0.05 &   -6.5  &  (19)   & New        \\
J004134.99+410552.3&    0.28   &WC7                        &    20.38 &    -0.22 &   -4.6  &  19     & MS 8       \\
J004141.81+403711.5&    0.71   &WN3                        &    21.48 &    -0.21 &   -3.5  &  (145)  & New        \\
J004143.00+411859.7&    0.52   &WN6                        &    21.42 &     0.15 &   -3.6  &  65     & New        \\
J004143.09+404045.9&    0.63   &WN3                        &    20.96 &    -0.19 &   -4.1  &  83     & New        \\
J004144.47+404517.1&    0.51   &WC6-7                      &    19.48 &     0.24 & \nodata &  76     & MS 10      \\
J004145.33+412104.0&    0.56   &WN6                        &    21.05 &    -0.04 &   -4.0  & I & New        \\
J004147.24+410647.5&    0.22   &WC8                        &    18.82 &    -0.17 &   -6.2  &  20     & New        \\
J004148.27+411739.9&    0.45   &WN6                        &    20.36 &     0.06 &   -4.7  &  (65)   & New        \\
J004154.62+404713.8&    0.51   &WC6                        &    21.70 &     0.58 & \nodata &  (84)   & New        \\
J004203.94+412554.5&    0.55   &WC6                        &    19.33 &     0.14 &   -5.7  &  (63)   & New        \\
J004211.16+405648.0&    0.34   &WN6                        &    20.92 &    -0.30 &   -4.1  &  (24)   & New        \\
J004213.21+405051.8&    0.52   &WC4                        &    21.10 &    -0.10 &   -3.9  &  (86)   & New        \\
J004214.36+412542.3&    0.47   &WC6                        &    20.71 &     0.22 & \nodata &  (63)   & MS 6       \\
J004234.42+413024.2&    0.47   &WC8                        &    19.65 &    -0.15 &   -5.4  &  61     & MS 5       \\
J004238.90+410002.0&    0.42   &WN7                        &    20.76 &     0.51 & \nodata &  25     & New        \\
J004240.81+410241.6&    0.35   &WC6                        &    20.42 &     0.40 & \nodata & \nodata & New        \\
J004242.03+412314.9&    0.22   &WC7                        &    20.94 &     0.04 &   -4.1  & \nodata & MS 7       \\
J004242.33+413922.7&    0.67   &WN10                       &    18.56 &     0.17 &   -6.5  &  (171)    & New        \\
J004247.12+405657.1&    0.56   &WC4                        &    20.69 &     0.06 &   -4.3  & I & New        \\
J004249.84+410215.7&    0.43   &WC5-6                      &    21.63 &     0.20 &   -3.4  &  (26)   & New        \\
J004254.65+410226.1&    0.45   &WN3+neb                    &    20.55 &     0.59 & \nodata &  26     & New        \\
X004256.05+413543.7&    0.48   &WC6                        & \nodata  & \nodata  & \nodata &  60     & New        \\
J004256.85+413837.2&    0.55   &WN6                        &    21.39 &    -0.11 &   -3.6  &  (59)   & New        \\
J004257.62+413727.8&    0.52   &WN3                        &    20.76 &     0.31 & \nodata &  59     & New        \\
J004302.05+413746.7&    0.50   &WN9                        &    20.15 &     0.36 & \nodata &  59     & New        \\
J004302.76+413753.6&    0.50   &WN6                        &    21.92 &     0.32 & \nodata &  59     & New        \\
J004304.34+412223.4&    0.08   &WN5                        &    20.91 &    -0.14 &   -4.1  & \nodata & New        \\
J004306.11+413813.7&    0.49   &WN4                        &    21.57 &    -0.10 &   -3.5  &  59     & New        \\
X004308.25+413736.3&    0.46   &WC6                        & \nodata  & \nodata  & \nodata &  59     & New        \\
J004310.19+413744.0&    0.45   &WN4                        &    20.89 &     0.00 &   -4.1  &  59     & New        \\
J004316.44+414512.4&    0.62   &WC6                        &    19.96 &     0.23 & \nodata & I & New        \\
J004321.48+414155.5&    0.50   &WN6                        &    21.60 &     0.14 &   -3.4  &  (58)   & New        \\
J004324.95+411803.2&    0.25   &WN4.5                      &    20.97 &    -0.26 &   -4.1  &  1      & New        \\
J004325.68+411830.2&    0.24   &WN4.5                      &    21.29 &    -0.21 &   -3.7  &  1      & New        \\
J004327.92+414207.3&    0.47   &WC6                        &    20.47 &     0.25 & \nodata &  58     & New        \\
J004328.06+414212.3&    0.47   &WN6                        &    22.02 &     0.35 & \nodata &  58     & New        \\
J004330.76+412734.7&    0.15   &WN4                        &    21.06 &    -0.25 &   -4.0  & \nodata & New        \\
J004331.17+411203.5&    0.45   &WC8                        &    19.93 &     0.13 &   -5.1  & I & MS 4       \\
J004334.92+410953.2&    0.53   &WN3?+neb                   &    19.71 &     0.48 & \nodata &  29     & New        \\
J004336.51+412315.1&    0.22   &WN4                        &    21.41 &    -0.21 &   -3.6  &  2      & New        \\
J004337.10+414237.1&    0.44   &WN10                   &    19.41 &     0.14 &   -5.6  &  58     & New        \\
J004341.72+412304.2&    0.26   &WC6                        &    20.64 &    -0.33 &   -4.4  &  2      & MS 2       \\
J004344.48+411142.0&    0.55   &WN8+neb                    &    19.63 &     0.16 &   -5.4  &  30     & New        \\
J004347.01+411238.0&    0.54   &WN7                        &    20.21 &    -0.07 &   -4.8  &  30     & New        \\
J004349.72+411243.4&    0.56   &WN6                        &    20.72 &     0.03 &   -4.3  &  30     & New        \\
J004353.34+414638.9&    0.47   &WN7                        &    18.97 &    -0.24 &   -6.1  &  56     & New        \\
J004357.31+414846.2&    0.51   &WN8                        &    20.54 &     0.30 & \nodata &  110    & New        \\
J004403.39+411518.8&    0.59   &WN6                        &    21.35 &    -0.03 &   -3.7  &  31     & New        \\
J004406.39+411921.0&    0.52   &WC\tablenotemark{f}        &    21.33 &    -0.31 &   -3.7  &  36     & MS 3       \\
J004406.68+411612.0&    0.60   &WN4                        &    21.51 &    -0.03 &   -3.5  &  (31)   & New        \\
J004408.58+412121.2&    0.48   &WC6                        &    20.00 &     0.01 &   -5.0  &  36     & New        \\
J004410.17+413253.1&    0.29   &WC6                        &    19.50 &    -0.41 &   -5.5  &  10     & OB 10 WR 1 \\
J004410.91+411623.2&    0.62   &WN4                        &    20.34 &    -0.20 &   -4.7  &  (31)   & New        \\
J004412.44+412941.7&    0.34   &WC6                        &    19.67 &    -0.42 &   -5.4  &  (3)    & IT 5-15    \\
J004413.06+411920.5&    0.56   &WN8                        &    18.91 &     0.08 &   -6.1  &  32     & New        \\
J004415.77+411952.5&    0.57   &WC5                        &    22.77 &     0.22 & \nodata &  32     & New        \\
J004416.20+412103.5&    0.55   &WC+B0I?                    &    19.47 &    -0.38 &   -5.6  &  37     & New        \\
J004419.92+411751.2&    0.65   &WN3                        &    20.65 &     0.09 &   -4.4  &  32     & New        \\
J004420.58+415412.5&    0.55   &WN3                        &    20.39 &    -0.31 &   -4.6  &  (111)  & IT 1-40    \\
J004422.24+411858.4&    0.64   &WC7-8                      &    20.74 &    -0.17 &   -4.3  &  32     & OB 32 WR 1 \\
J004425.10+412050.0&    0.61   &WC6+B0I                    &    17.55 &     0.11 &   -7.5  &  33     & New        \\
J004425.43+412044.7&    0.62   &WC6-7\tablenotemark{f}     &    19.49 &     0.07 &   -5.5  &  33     & OB 33 WR 2 \\
J004425.83+415019.4&    0.45   &WN5                        &    20.42 &     0.02 &   -4.6  &  55     & New        \\
J004426.17+412210.8&    0.59   &WN6                        &    21.08 &     0.59 & \nodata &  38     & New        \\
J004426.32+413419.8&    0.36   &WN5-6                      &    21.14 &    -0.17 &   -3.9  & \nodata & New        \\
J004429.89+412021.6&    0.66   &WN3                        &    21.82 &    -0.15 &   -3.2  &  33     & New        \\
J004430.04+415237.1&    0.49   &WN8                        &    19.27 &     0.20 &   -5.8  &  54     & New        \\
J004434.57+412424.4&    0.60   &WN3                        &    20.32 &    -0.37 &   -4.7  &  39     & New        \\
J004435.15+412545.2&    0.57   &WC5                        &    21.64 &     0.07 &   -3.4  &  (39)   & New        \\
J004436.22+412257.3&    0.64   &WN5                        &    20.65 &    -0.28 &   -4.4  &  (38)   & New        \\
J004436.52+412202.0&    0.67   &WC5                        &    20.93 &    -0.25 &   -4.1  &  (38)   & New        \\
J004437.61+415203.3&    0.46   &WN4pec                     &    20.70 &    -0.12 &   -4.3  &  54     & OB 54 WR 1 \\
J004443.10+412619.6&    0.62   &WN6                        &    19.65 &    -0.33 &   -5.4  &  40     & New        \\
J004444.00+412739.9&    0.59   &WC6                        &    20.82 &     0.14 &   -4.2  &  41     & IT 5-3     \\
J004444.87+412800.7&    0.59   &WN6                        &    20.27 &    -0.16 &   -4.8  &  41     & New        \\
J004445.61+412806.5&    0.60   &WN5                        &    20.81 &    -0.14 &   -4.2  &  41     & New        \\
J004445.90+415803.8&    0.56   &WN5                        &    21.20 &    -0.20 &   -3.8  & I & New        \\
J004449.41+413020.7&    0.58   &WN3                        &    21.57 &    -0.02 &   -3.5  &  (41)   & New        \\
J004451.98+412911.6&    0.62   &WC7                        &    20.56 &    -0.46 &   -4.5  &  41     & IT 5-2     \\
J004453.06+412601.7&    0.69   &WN3+TiO                    &    21.51 &    -0.10 &   -3.5  & I & New        \\
J004453.52+415354.3&    0.48   &WC7                        &    20.72 &    -0.28 &   -4.3  & I & IT 1-48    \\
J004455.63+413105.1&    0.60   &WC6                        &    20.34 &    -0.25 &   -4.7  &  42     & OB 42 WR 1 \\
J004455.82+412919.2&    0.64   &WN2pec+neb                 &    20.10 &     0.39 & \nodata &  (42)   & New        \\
J004500.96+413058.8&    0.64   &WC6                        &    19.77 &     0.05 &   -5.3  &  42     & OB 42 WR 2 \\
J004502.78+415533.7&    0.50   &WN5pec                     &    20.14 &    -0.22 &   -4.9  &  53     & New        \\
J004506.50+413425.1&    0.61   &WN4+neb                    &    20.24 &     0.15 &   -4.8  &  48     & New        \\
J004509.18+414021.4&    0.53   &WC7                        &    20.39 &     0.05 &   -4.6  &  48     & New        \\
J004510.39+413646.6&    0.60   &WC7+BI                     &    18.42 &    -0.12 &   -6.6  &  48     & OB 48 WR 6 \\
J004511.21+420521.7&    0.65   &WN11                   &    18.64 &     0.23 & \nodata &  (108)  & New        \\
J004511.27+413815.3&    0.58   &WC6                        &    20.30 &    -0.34 &   -4.7  &  48     & OB 48 WR 1 \\
J004513.68+413742.5&    0.60   &WC+abs\tablenotemark{f}    &    20.91 &     0.39 & \nodata &  48     & OB 48 WR 3 \\
J004514.10+413735.2&    0.61   &WN4                        &    20.92 &    -0.09 &   -4.1  &  48     & OB 48 WR 2 \\
J004517.56+413922.0&    0.60   &WN4brd                     &    18.62 &    -0.15 &   -6.4  &  48     & OB 48-527  \\
J004517.89+415209.5&    0.50   &WC8                        &    18.11 &    -0.02 &   -6.9  & I & New        \\
J004520.80+415100.0&    0.51   &WN4                        &    21.32 &     0.01 &   -3.7  &  (46)   & New        \\
J004522.78+420318.2&    0.60   &WC5-6                      &    21.94 &     0.22 & \nodata & I & New        \\
J004524.26+415352.5&    0.52   &WC6                        &    21.16 &    -0.37 &   -3.9  & I & IT 4-01    \\
J004530.61+414639.2&    0.58   &WC4-5+neb                  &    21.42 &     0.54 & \nodata &  49     & New        \\
J004531.16+420658.1&    0.65   &WC6                        &    20.01 &    -0.31 &   -5.0  &  107    & New        \\
J004537.05+414302.4&    0.66   &WN3                        &    20.91 &     0.05 &   -4.1  &  95     & New        \\
J004537.10+414201.4&    0.68   &WC6                        &    19.94 &    -0.01 &   -5.1  & I & IT 4-13    \\
J004537.28+414305.2&    0.66   &WN4+neb                    &    21.04 &     0.22 & \nodata &  95     & New        \\
J004539.15+414754.0&    0.61   &WN5                        &    20.68 &     0.02 &   -4.3  &  (49)   & New        \\
J004539.35+414439.8&    0.65   &WC5                        &    21.50 &     0.31 & \nodata &  95     & New        \\
J004541.29+415504.0&    0.57   &WN4                        &    20.92 &    -0.02 &   -4.1  &  (51)   & New        \\
J004542.26+414510.1&    0.66   &WN10                       &    19.35 &     0.55 & \nodata &  (49)    & New        \\
J004544.65+415241.0&    0.60   &WN3                        &    21.33 &     0.18 &   -3.7  &  (50)   & New        \\
J004551.12+421015.4&    0.69   &WC5-6                      &    20.89 &    -0.30 &   -4.1  &  106    & New        \\
J004551.35+414242.0&    0.76   &WC6                        &    20.63 &    -0.46 &   -4.4  & \nodata & IT 4-14    \\
J004628.51+421127.6&    0.73   &WN\tablenotemark{f}        &    20.67 &     0.04 &   -4.4  &  102    & OB 102 WR 1 \\
J004645.02+421301.3&    0.78   &WN3                        &    21.60 &     0.19 &   -3.4  &  (102)  & New        \\
J004655.89+420015.1&    0.89   &WN3+abs?                   &    21.00 &    -0.24 &   -4.0  &  150    & New        \\
\enddata                                                                                  
\tablenotetext{a}{From the LGGS, except for ``X00..." names, which are new here.}
\tablenotetext{b}{Normalized galactocentric distance within the plane of M31.}
\tablenotetext{c}{The absolute visual magnitude computed assuming $E(B-V)=0.2$ ($A_V=0.62$) for stars with $B-V\leq 0.2$. We assumed a true distance modulus of 24.4 (0.76 Mpc) using van den Bergh 2000.}
\tablenotetext{d}{OB associations as identified by van den Bergh 1964. Parenthesis means that the star is close to, but not in, the association. An ``I" means that the star is located in the Population I ring but not in an identified OB association.} 
\tablenotetext{e}{Previous IDs: OB designations from Massey et al.\ 1986 except for OB 32, from Armandroff \& Massey 1991; MS=Moffat \& Shara 1983 with revised coordinates by MJ98; IT= Moffat \& Shara 1987, with revised coordinates by MJ98. New = newly identified as part of this program.}
\tablenotetext{f}{Spectral type from MJ98 and references therein.}
\end{deluxetable}

\begin{deluxetable}{l c c c c c c c c c}
\tablecaption{\label{tab:WCWN} Numbers of WC and WN\tablenotemark{a}}
\tablewidth{0pt}
\tablehead{
\colhead{Region}
&\colhead{$\bar{\rho}$}
&\multicolumn{2}{c}{log$\frac{O}{H}+12$}
&\colhead{Number}
&\colhead{Number}
&\colhead{WC/WN}
&\multicolumn{2}{c}{Errors\tablenotemark{b}}  \\ \cline{3-4} \cline{8-9}
&
&\colhead{Value}
&\colhead{Ref.\tablenotemark{c}}
&\colhead{WCs} & \colhead{WNs} & 
& \colhead{5\%} 
& \colhead{$\sqrt{N}$} \\
}
\startdata
M31 (all)                       &0.53 & 8.93 & 1 & 62 & 92 & 0.67 &  $^{+0.09}_{-0.05}$ & 0.11 \\
M31 ($A_V< 1.5$)        &0.53 & 8.93 & 1 & 44 & 70 & 0.63 &  $^{+0.08}_{-0.05}$ & 0.12 \\
M33 ( $\rho<0.25$)         & 0.16 & 8.7 & 2 & 26 & 45 & 0.58 &  $^{+0.08}_{-0.05}$ & 0.14  \\
M33 ($0.25\le \rho <0.50$)        & 0.38  & 8.4 & 2 & 15 & 54 & 0.28 &  $^{+0.06}_{-0.02}$ & 0.08  \\
M33 ($\rho\ge 0.5$)               & 0.69  & 8.3 & 2 & 12 & 54 & 0.22 & $^{+0.06}_{-0.01}$ & 0.07  \\
LMC                                             & \nodata    & 8.4 & 3 & 25 & 109 & 0.23 & $^{+0.06}_{-0.02}$ & 0.05 \\
SMC                                             & \nodata    & 8.1 & 4 & 1   &   11 & 0.09 & $^{+0.09}_{-0.01}$ &  0.09 \\
\enddata
\tablenotetext{a}{Updated version of Table 9 of Neugent \& Massey 2011, now including M31 and improved error estimates.}
\tablenotetext{b}{The 5\% error estimate on the WC/WN ratio is computed assuming that 5\% of the total sample is missing (rounded to the nearest integer value) and that these are either all WCs (upper error) or all WNs (lower error). The $\sqrt{N}$ estimate of the error is based upon a stocastical argument that at a slightly different time the number of WNs and WCs would each differ by the square root of the number that we observed today, and hence the error will be WC/WN $\times$ [1/WC + 1/WN]$^{0.5}$.}
\tablenotetext{c}{References for oxygen abundances: 1--Zaritsky et al.\ 1994, adjusted using their gradient to the ring that contains most of the WR stars. 2--Magrini et al.\ 2007 two-component model adjusted to average $\rho$ values in the various bins. See discussion in Neugent \& Massey 2011. 3--Russell \& Dopita 1990.}
\end{deluxetable}

\end{document}